\documentclass[useAMS,usenatbib]{mnras_new}

\usepackage{enumitem}
\usepackage{calc}
\usepackage{chngcntr}
\usepackage{epsfig}
\usepackage{amsmath}
\usepackage{natbib}
\usepackage{graphicx}
\usepackage[caption=false]{subfig}
\usepackage{wrapfig}
\usepackage[T1]{fontenc}
\usepackage{enumitem}
\usepackage[normalem]{ulem}
\usepackage{textcomp}
\usepackage{longtable}
\usepackage{answers}
\usepackage{array}
\usepackage{tablefootnote}
\title[Southern Orion A: First Look]
  {The JCMT Gould Belt Survey: A First Look at Southern Orion A with SCUBA-2}
\author[S. Mairs et al.]
  {S. Mairs$^{1, 2}$, D. Johnstone$^{1,2}$, H. Kirk$^{2}$, J. Buckle$^{3,4}$, D.S. Berry$^{5}$,  H. Broekhoven-Fiene$^{1,2}$,  \newauthor M.J. Currie$^{5}$, M. Fich$^{6}$, S. Graves$^{5,7}$, J. Hatchell$^{8}$, T. Jenness$^{5,9}$,  J.C. Mottram$^{10,11}$,  \newauthor D. Nutter$^{12}$, K. Pattle$^{13}$, J.E. Pineda$^{14,15,16}$, C. Salji$^{3,4}$,   J. Di Francesco$^{1,2}$, M.R. Hogerheijde$^{10}$,  \newauthor D. Ward-Thompson$^{13}$, P. Bastien$^{17}$,  D. Bresnahan$^{13}$,  H. Butner$^{18}$, M. Chen$^{1,2}$,  \newauthor A. Chrysostomou$^{19}$, S. Coud\'e$^{17}$,  C.J. Davis$^{20}$,  E. Drabek-Maunder$^{21}$, A. Duarte-Cabral$^{8}$, \newauthor J. Fiege$^{22}$, P. Friberg$^{5}$,  R. Friesen$^{23}$, G.A. Fuller$^{15}$,  J. Greaves$^{24}$, J. Gregson$^{25,26}$, \newauthor W. Holland$^{27,28}$,  G. Joncas$^{29}$, J.M. Kirk$^{13}$, L.B.G. Knee$^{2}$, K. Marsh$^{12}$,  B.C. Matthews$^{1,2}$, \newauthor G. Moriarty-Schieven$^{2}$, C. Mowat$^{8}$, J. Rawlings$^{30}$,  J. Richer$^{3,4}$, D. Robertson$^{31}$, \newauthor E. Rosolowsky$^{32}$, D. Rumble$^{8}$, S. Sadavoy$^{11}$,  H. Thomas$^{5}$, N. Tothill$^{33}$, S. Viti$^{30}$, \newauthor G.J. White$^{25,26}$, J. Wouterloot$^{5}$, J. Yates$^{30}$, M. Zhu$^{34}$
\\\\\\\\
Affiliations are listed at the end of the paper}

\date{Released 2016 Xxxxx XX}

\pagerange{\pageref{firstpage}--\pageref{lastpage}} \pubyear{2016}

\def\LaTeX{L\kern-.36em\raise.3ex\hbox{a}\kern-.15em
    T\kern-.1667em\lower.7ex\hbox{E}\kern-.125emX}

\begin{document}

\label{firstpage}

\maketitle


\begin{abstract}

We present the JCMT Gould Belt Survey's first look results of the southern extent of the Orion A Molecular Cloud ($\delta \leq -5\mathrm{:}31\mathrm{:}27.5$). Employing a two-step structure identification process, we construct individual catalogues for large-scale regions of significant emission labelled as \textit{islands} and smaller-scale subregions called \textit{fragments} using the \mbox{850 $\mu$m} continuum maps obtained using SCUBA-2. We calculate object masses, sizes, column densities, and concentrations. We discuss fragmentation in terms of a Jeans instability analysis and highlight interesting structures as candidates for follow-up studies. Furthermore, we associate the detected emission with young stellar objects (YSOs) identified by \textit{Spitzer} and \textit{Herschel}. 
We find that although the population of active star-forming regions contains a wide variety of sizes and morphologies, there is a strong positive correlation between the concentration of an emission region and its calculated Jeans instability. There are, however, a number of highly unstable subregions in dense areas of the map that show no evidence of star formation. We find that only $\sim$72\% of the YSOs defined as Class 0+I and flat-spectrum protostars coincide with dense \mbox{850 $\mu$m} emission structures (column densities $>3.7\times10^{21}\mathrm{\:cm}^{-2}$). The remaining 28\% of these objects, which are expected to be embedded in dust and gas, may be misclassified. Finally, we suggest that there is an evolution in the velocity dispersion of young stellar objects such that sources which are more evolved are associated with higher velocities.  

\end{abstract}


\section{Introduction}
\label{introductionsec}

The James Clerk Maxwell Telescope's (JCMT) Gould Belt Legacy Survey (GBS, \citealt{wardthompson2007}) is a large-scale project which has mapped the notable star-forming regions within 500 pc of the Sun such as Orion A \citep{salji2015} and Orion B \citep{kirkfirstlook2016}, Taurus \citep{buckle2015}, Ophiuchus \citep{pattle2015}, Serpens \citep{rumble2015}, Auriga-California (Broekhoven-Fiene et al., submitted), and Perseus (Chen et al., accepted),   in \mbox{450 $\mu$m} and \mbox{850 $\mu$m} continuum emission as well as $^{12}$CO, $^{13}$CO, and C$^{18}$O spectral lines (see \citealt{buckle2012} and references therein). In this paper, we present the first results from the Southern Orion A region observed at \mbox{\mbox{850 $\mu$m}} with the Submillimetre Common-User Bolometer Array 2 (SCUBA-2) instrument \citep{holland2013}. 

Southern Orion A is a 2.8\textdegree  \hspace{0.3mm}  x 3.9\textdegree  \hspace{0.3mm}  region within the Orion cloud complex, predominantly composed of the L1641 cloud, which is an active star-formation site approximately 450 pc (see \citealt{muench2008} for a detailed review of the distance to Orion) from the Sun. The southern tip of the L1640 cloud to the north, however, is also included (i.e., the region south of $\delta \leq -5\mathrm{:}31\mathrm{:}27.5\arcsec$). Northern Orion A is arguably the most well-studied nearby star-forming region, as it is home to the Orion Nebula and the integral shaped filament (ISF; \citealt{bally1987}, \citealt{johnstone1999}; also see \citealt{salji2015cores} and \citealt{salji2015} for a GBS analysis of Orion A North). The Southern Orion A region, however, is also an area of interest, showing several different stages of low- and intermediate-mass star and cluster formation (see Chapter 20 of \citealt{reipurth2008}). 

The most southern declinations observed in this study (-7\textdegree:00$\arcmin$ to -9\textdegree:25$\arcmin$) have received less focus in previous literature than the northern section of the cloud. There is, however, still a wealth of data available. For example, \cite{bally1987} analysed extensive $^{13}$CO maps observed with the AT\&T Bell Laboratories \mbox{7~m} antenna and noted that the L1641 cloud was concentrated into a filamentary structure down to -9\textdegree \hspace{0.3mm} in declination with a north-south velocity gradient (see \citealt{allen2008} and references therein for a thorough review of L1641).   

The detected emission in Southern Orion A includes OMC-4, OMC-5, and L1641N, several active sites of Galactic star formation close to the Sun. It contains dozens of embedded sources (\citealt{johnstone2006}; \citealt{chen1996}; \citealt{ali2004}), the NGC 1999 reflection nebula and its associated A0e star V380 Ori (\citealt{stanke2010}; \citealt{johnstone2006}), as well as the famous  Herbig-Haro objects \citep{herbigpaper} HH 34, HH 1/2, and HH 222 with their sources and their prominent, young outflows (\citealt{johnstone2006}; \citealt{stanke2002}; \citealt{reipurth2002}; \citealt{reipurth2013}). Observations of the cold dust emission from (sub)millimeter detectors, however, are generally limited at the lower declinations in Southern Orion A.  Facilities, such as the Caltech Submillimetre Observatory (CSO) or the IRAM \mbox{30~m}  Telescope, have mainly focused on the Orion BN-KL complex or the Orion Bar, and have thus only sparsely sampled these lower declinations (see, for examples, \citealt{CSOref1}, \citealt{CSOref2}, \citealt{iramref1}, \citealt{iramref2}, and references therein).   As such, most of the early submillimeter continuum observations of Southern Orion A were made with SCUBA-2's predecessor, SCUBA (\citealt{difrancesco2008}; \citealt{nutter2007}; \citealt{johnstone2006}).   Indeed, these SCUBA observations revealed many clumps toward Southern Orion A for the first time.

The SCUBA-2 observations presented here, however, have a sensitivity which is an order of magnitude deeper than the maps presented in \citealt{johnstone2006} along with a much wider spatial coverage (8100 arcmin$^{2}$ compared to 2300 arcmin$^{2}$ in the original Southern Orion A SCUBA data).  Thus, we have a much better diagnostic to characterize the dense, cold dust in Southern Orion A.  To complement these new continuum observations of dense, often gravitationally unstable gas, we use extinction data taken in the \textit{J}, \textit{H} and \textit{K} bands that were determined from the Near-infrared Color Excess ({\sc{NICE}}) team (M. Lombardi, private communication, July 18$^{th}$, 2015), and the young stellar object (YSO) catalogues of \cite{megeath2012} and \cite{stutz2013} obtained using the \textit{Spitzer Space Telescope} and the \textit{Herschel Space Observatory}, respectively. The correlation between YSOs of different classes and the observed gas and dust structure is a powerful tool that can be used to help discern the dominant physical processes which influence star formation. 
Analysing the locations of protostars and their more-evolved counterparts with respect to the gas and dust in a molecular cloud is imperative for studying a variety of topics including cluster formation and the effect of feedback on the star-formation process.  

In Section \ref{drsec}, we summarise the observations and data reduction methods employed in this study. In Section \ref{cataloguessec}, we display the \mbox{450 $\mu$m} and \mbox{850 $\mu$m} SCUBA-2 maps of Southern Orion A, present our structure identification procedure, and discuss the population of objects in terms of larger-scale extinction, Jeans stability, and concentration. In Section \ref{ysosec}, we examine the associations between YSOs and dense continuum structure. We also investigate fragmentation as observed in the continuum data in terms of its effect on star formation and note interesting candidates for follow-up studies. We conclude this section with a discussion on the spatial distribution of young stellar objects, and we construct a simple model to understand the widespread locations of young stars across Southern Orion A. Finally, in Section \ref{conclusionsec}, we summarise our main results.

\section{Observations and Data Reduction}
\label{drsec}

The observations presented throughout this paper were performed using the SCUBA-2 instrument \citep{holland2013} as part of the JCMT Gould Belt Survey \citep{wardthompson2007}. This instrument has provided continuum coverage at both \mbox{850 $\mu$m} and \mbox{450 $\mu$m} simultaneously at effective beam sizes of 14.1$\arcsec$ and 9.6$\arcsec$, respectively \citep{dempsey2013}. In this work, we present Southern Orion A in both wavelengths, but focus mainly on the \mbox{850 $\mu$m} data for analysis. All of the observations were taken in the PONG1800 mapping mode \citep{kackley2010}, yielding circular maps (``PONGs'') of $\sim$0.5\textdegree  \hspace{0.3mm} in diameter. There are seventeen 0.5\textdegree  \hspace{0.3mm} subregions across the Orion A Molecular Cloud, thirteen of which cover Southern Orion A. These locations were individually observed four to six times throughout February 2012 to January 2015, and were then co-added (once co-added, these structures are referred to as ``tiles'') and mosaicked to form the final map. The tiles slightly overlap to provide a more uniform noise level throughout the whole of the Orion A Molecular Cloud. For a summary of the typical noise present in each tile after contamination from \mbox{CO(J=3-2)} has been removed (see the discussion below and the Appendix), see Table \ref{noisetable}. All observations were taken in dry weather ($\tau_{225\mathrm{\:GHz}}<0.08$) and two PONGs were taken in very dry weather ($\tau_{225\mathrm{\:GHz}}<0.05$).  To define the northern boundaries of the Southern Orion A region, a cut-off was then applied at $\delta = -5\mathrm{:}31\mathrm{:}27.5$ so that the northern half of integral shaped filament, including the Orion Nebula Cluster (ONC), was not included in this analysis. For analyses performed on Orion A North, which slightly overlaps with this region (OMC-4 is in both the Orion A North map as well as the Southern Orion A map), see \cite{salji2015cores} and \cite{salji2015}. 

\begin{table*}
\caption{A summary of the typical noise present in each of the seventeen publicly available tiles which comprise the Orion A Molecular Cloud. Contamination from CO has been removed in the \mbox{850 $\mu$m} images.}
\label{noisetable}
\begin{tabular}{|c|c|c|c|c|}
\hline
Tile Name & \multicolumn{1}{|p{1.75cm}|}{\centering Central R.A. \\ (J2000)} & \multicolumn{1}{|p{1.5cm}|}{\centering Central Dec \\ (J2000)} & \multicolumn{1}{|p{2.25cm}|}{\centering \mbox{850 $\mu$m} Noise\\ (mJy beam$^{-1}$)} &  \multicolumn{1}{|p{2.25cm}|}{\centering 450$\mu$m Noise\\ (mJy beam$^{-1}$)}\\
\hline\hline
OMC1\_TILE1   & 5:34:18   & -5:09:58 & 4.0 &  58\\ 
OMC1\_TILE2   & 5:34:57  &  -5:40:00 & 3.7 &  39\\
OMC1\_TILE3   & 5:36:22  &  -5:16:56 & 3.7  & 34\\ 
OMC1\_TILE4   & 5:35:50  &  -4:46:06 & 3.6  & 39\\  
OMC1\_TILE56 & 5:35:44  &  -6:07:25 & 3.7  & 53\\ 
OMC1\_TILE7   & 5:36:12  & -6:31:30 & 3.1  & 34\\  
OMC1\_TILE8   & 5:36:45  & -7:02:26 & 3.5  & 63\\  
OMC1\_TILE9   & 5:38:16  & -6:39:56 & 3.2  & 67\\  
OMC1\_TILE10 &  5:38:48 &  -7:10:27 & 3.4 & 63\\  
OMC1\_TILE11 & 5:40:06   & -7:33:22 & 3.0  & 43\\  
OMC1\_TILE12 & 5:40:58   & -8:00:26 & 3.3  & 67\\  
OMC1\_TILE13 & 5:42:48   & -8:16:14 & 3.3  & 63\\  
OMC1\_TILE14 & 5:40:58   & -8:32:13 & 3.4  & 58\\  
OMC1\_TILE15 & 5:42:49   & -8:47:54  & 3.4  & 53\\ 
OMC1\_TILE16 & 5:40:57   &  -9:03:53 & 3.3  & 53\\  
OMC1\_TILE17 & 5:33:09   &   -5:37:46 & 3.3  & 43\\ 
\hline
\end{tabular}
\begin{flushleft}
These measurements of the \mbox{850 $\mu$m} and \mbox{450 $\mu$m} noise levels are based a point source detection using pixel sizes of 3$\arcsec$ and 2$\arcsec$, respectively, and beam FWHM values of 14.1$\arcsec$ and 9.6$\arcsec$, respectively. \\
Note that four of the observations were taken during SCUBA-2 science verification. They can be found in CADC under the project code 'MJLSG22'
%
%
%
%
\end{flushleft}
\end{table*}

The data reduction procedure was performed using the iterative map-making technique {\sc{makemap}} (explained in detail by \citealt{chapin2013}) in the {\sc{SMURF}} package (\citealt{jenness2013}) found within Starlink (\citealt{currie2014}). The \mbox{850 $\mu$m} continuum image studied here is part of the GBS LR1 release (see \citealt{mairs2015}, for an overview). In this data release, after the iterative map-making procedure was performed for each observation, the individual maps were co-added for a higher signal-to-noise ratio (SNR) and the resulting image was used to define regions of genuine emission. A mask was constructed with boundaries defined by an SNR of at least 2. This mask was used to highlight emission regions and perform a second round of data reduction to recover better any faint and extended structure\footnote{Note that the boundaries employed in this paper are more conservative than those used in \cite{mairs2015}. The same SNR was used to identify significant structure, but in this analysis, no smoothing was applied to the boundaries whereas in the analysis of \cite{mairs2015}, the boundaries were smoothed to incorporate more diffuse structure.}. The map is gridded to 3$\arcsec$ pixels (as opposed to the GBS Internal Release 1 (IR1) reduction method where the pixels were 6$\arcsec$) and the iterative solution converged when the difference in individual pixels changed on average by <0.1\% of the rms noise present in the map. The final mosaic was originally in units of picowatts (pW) but was converted to \mbox{mJy arcsec$^{-2}$} using the \mbox{850 $\mu$m} aperture flux conversion factor \mbox{2.34 Jy pW$^{-1}$ arcsec$^{-2}$} and \mbox{4.71 Jy pW$^{-1}$ arcsec$^{-2}$} at \mbox{450 $\mu$m} \citep{dempsey2013}. 

The \mbox{CO(J=3-2)} emission line contributes to the flux measured in these \mbox{850 $\mu$m} continuum observations (\citealt{johnstone1999}, \citealt{drabek2012}). As \cite{drabek2012} and \cite{coude2016} discuss, however, this line generally contributes low-level emission to continuum observations ($\leq 20\%$) with only a few sources associated with stellar outflows having anomalously high contamination ($\sim80\%$); see the Appendix for our own analysis of the \mbox{CO(J=3-2)} contamination in Southern Orion A. After the \mbox{850 $\mu$m} map was produced, therefore, we subtracted the \mbox{CO(J=3-2)} emission line from the continuum map using ancillary GBS data. 

In the following, the \mbox{850 $\mu$m} map refers to the data from which the \mbox{CO(J=3-2)} emission line has been subtracted. The final SCUBA-2 maps are not sensitive to large-scale structures as these are filtered out during the data reduction \citep{chapin2013}. For an overview of the GBS LR1 filtering parameters as well as results from testing the completeness of this method using artificial sources, see \cite{mairs2015}. Briefly, a spatial filtering scale of 10$\arcmin$ is applied to all the data residing outside the SNR-defined mask. This means that small-scale sources (<5$\arcmin$) are confidently recovered but larger-scale structures between 5$\arcmin$ and $10\arcmin$ will have missing flux. The severity of this problem depends on the emission structure of the source, the size of the SNR boundary drawn around it 
during the data reduction, and the inherent background structure of the map. The filter will subtract out of the map any large, faint modes causing the total, observed flux of sizeable objects that have compact, bright components to be underestimated. 


\section{Structure Within Southern Orion A}
\label{cataloguessec}

\begin{figure*}	
\centering
\includegraphics[width=19.5cm,height=17.5cm]{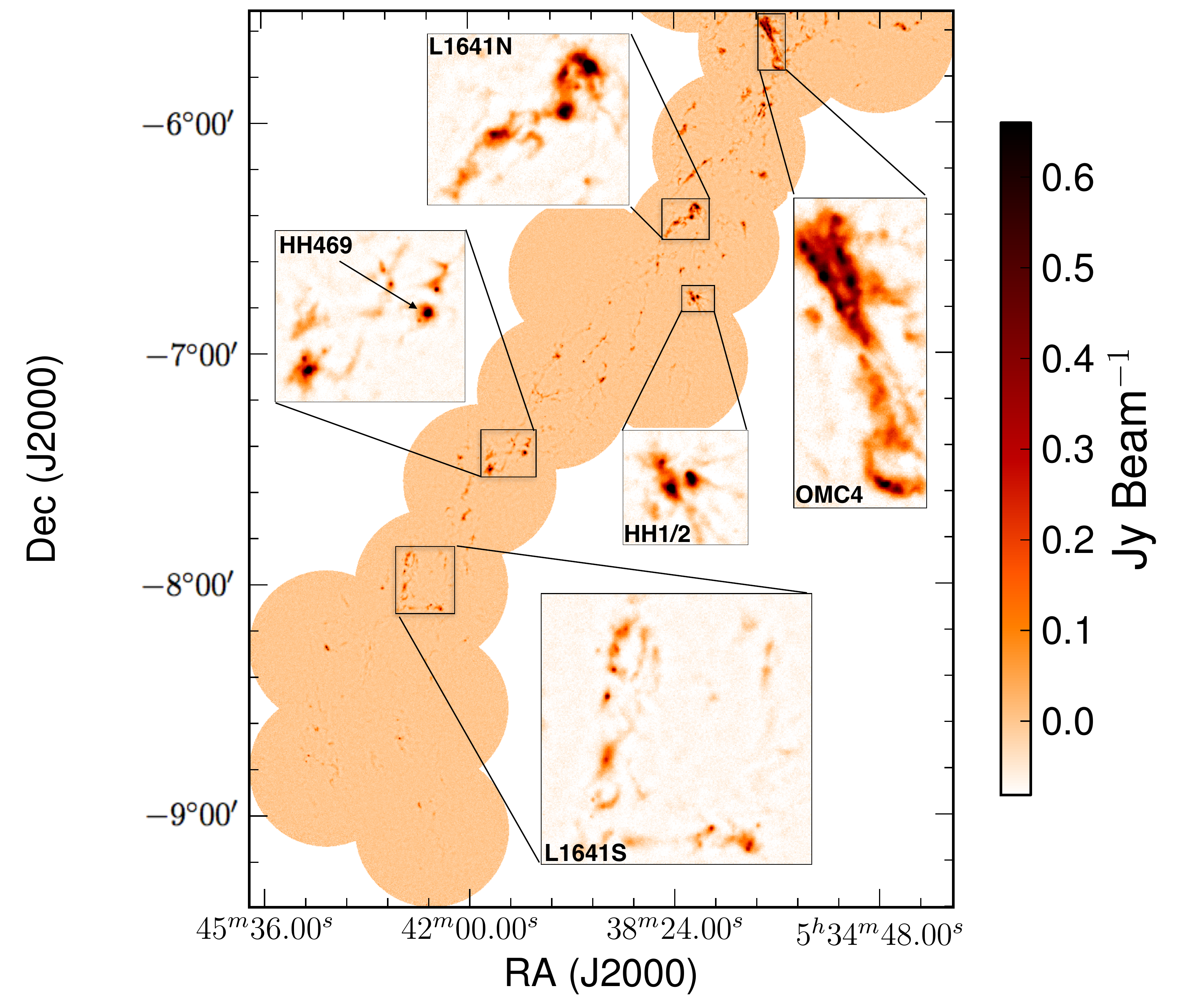}
\caption{The \mbox{850 $\mu$m} SCUBA-2 map of the GBS-defined Southern Orion A region. Several areas of significant emission are highlighted as insets in the main image. These include the ``V-shaped'' OMC-4 structure at the northern tip of the map \citep{johnstone1999}, HH 1/2 (\citealt{johnstone2006}; also see \citealt{herbig1951}, \citealt{haro1952}, and \citealt{haro1953}), HH469 \citep{aspin2000}, L1641-N, and L1641-S \citep{fukui1986}.}
\label{bigmap}
\end{figure*}

\begin{figure*}	
\centering
\includegraphics[width=19.5cm,height=17.5cm]{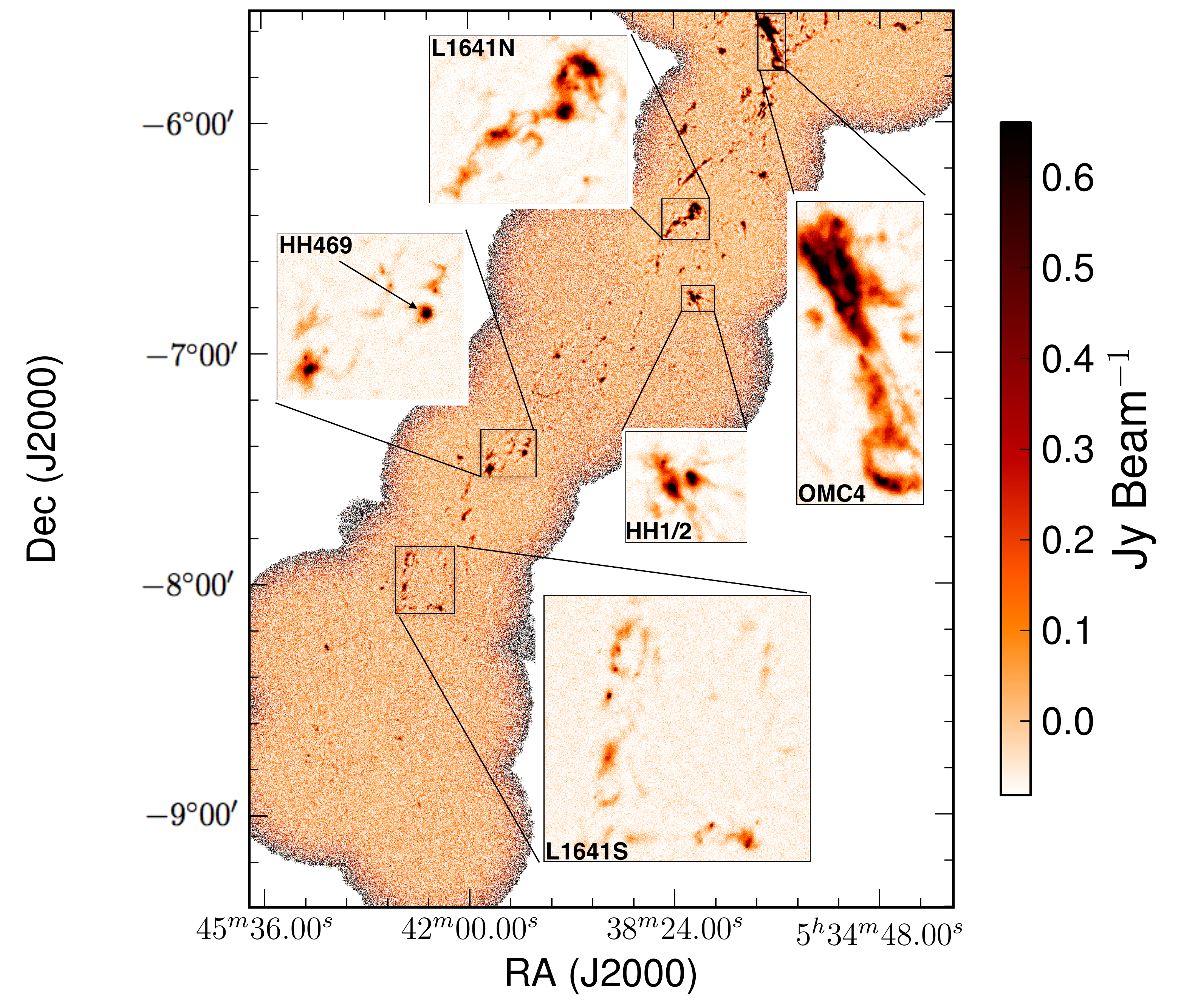}
\caption{The \mbox{450 $\mu$m} SCUBA-2 map of the GBS-defined Southern Orion A region. }
\label{bigmap450}
\end{figure*}

In Figures \ref{bigmap} and \ref{bigmap450}, we present the full \mbox{850 $\mu$m} and \mbox{450 $\mu$m} maps of Southern Orion A, respectively. Note that the northern boundary we have chosen ($\delta = -5\mathrm{:}31\mathrm{:}27.5$) includes the ``V-shaped'' OMC-4. This southern extension of the Orion A Giant Molecular Cloud (GMC) is less confused than its northern locations (e.g. the ISF) but it still shows a diverse set of objects defined by localised emission. It is, therefore, an intriguing location to study the initial stages of star formation at submillimetre wavelengths.   

There are many locations of interest across these maps, several of which are displayed as insets in Figure \ref{bigmap}. Even a cursory glance across the structure reveals a wealth of shapes and sizes of significant emission. Broadly speaking, there are no notable differences in the locations of emission structure between the \mbox{850 $\mu$m} and \mbox{450 $\mu$m} maps. To quantify this structure, several algorithms designed to extract, in an automated manner, structure from a given region are available (for example, see {\sc{GaussClumps}} \citealt{stutzki1990}, 
{\sc{ClumpFind}} \citealt{williams1994}, 
{\sc{Astrodendro}} \citealt{rosolowsky2008}; {\sc{GETSOURCES}} \citealt{getsources2012},
and {\sc{FellWalker}} \citealt{berry2015}). Each method amalgamates locations of significant emission differently based on user supplied criteria. Nevertheless, in maps such as the \mbox{850 $\mu$m} one presented here, structure should always be identified with a goal of answering specific scientific questions. Currently, there is no single technique that is commonly agreed to work well for the broad array of physical analyses possible for these data so different algorithms are used even within the GBS papers (see, for examples, \citealt{salji2015cores}, \citealt{salji2015}, \citealt{pattle2015}, \citealt{kirkfirstlook2016}, Broekhoven-Fiene et al., submitted, and Lane et al. in prep.).

\begin{figure*}
\centering
\subfloat{\label{exampleisland}\includegraphics[width=8.8cm,height=7.3cm]{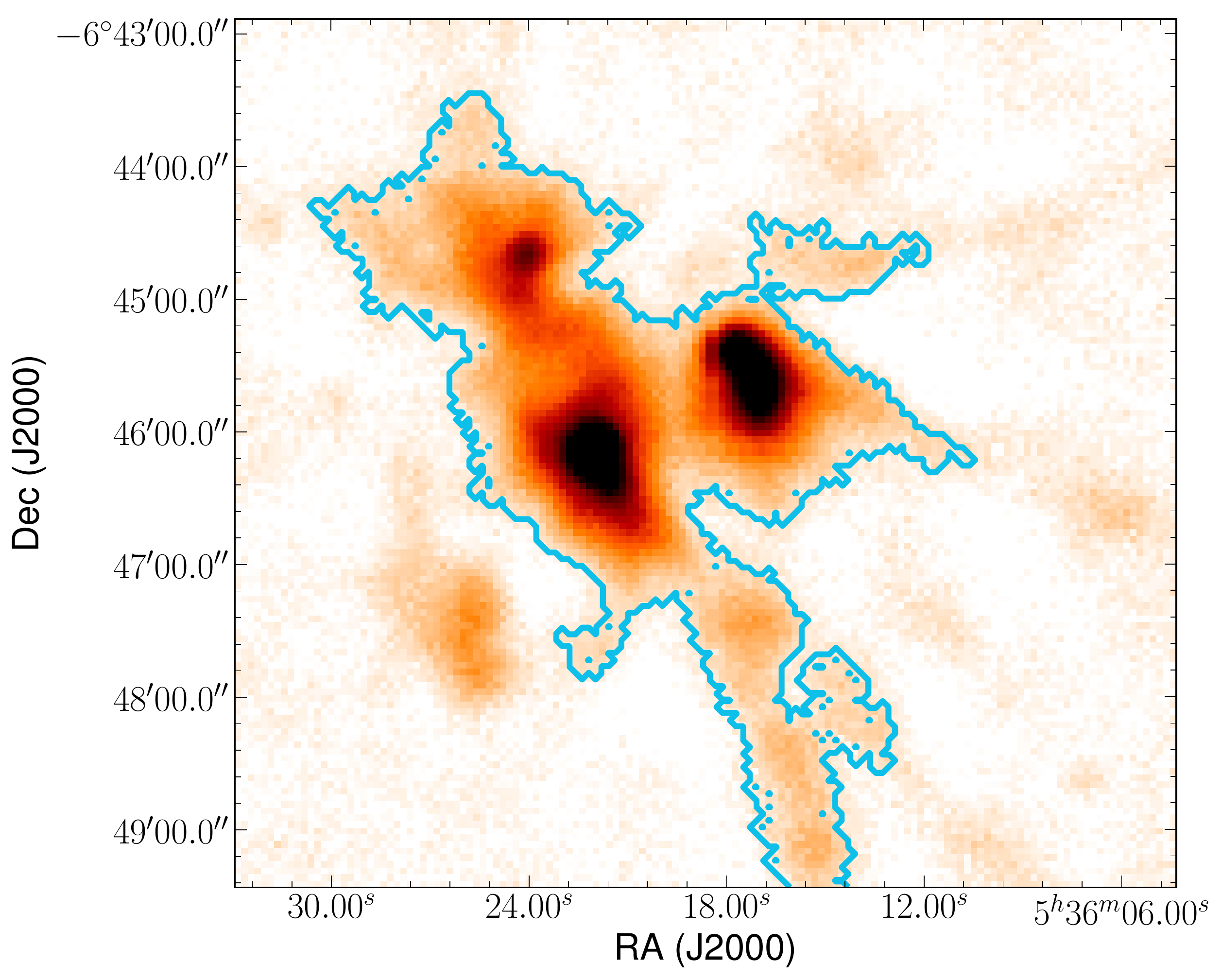}}
\subfloat{\label{examplefrags}\includegraphics[width=7.4cm,height=7.3cm]{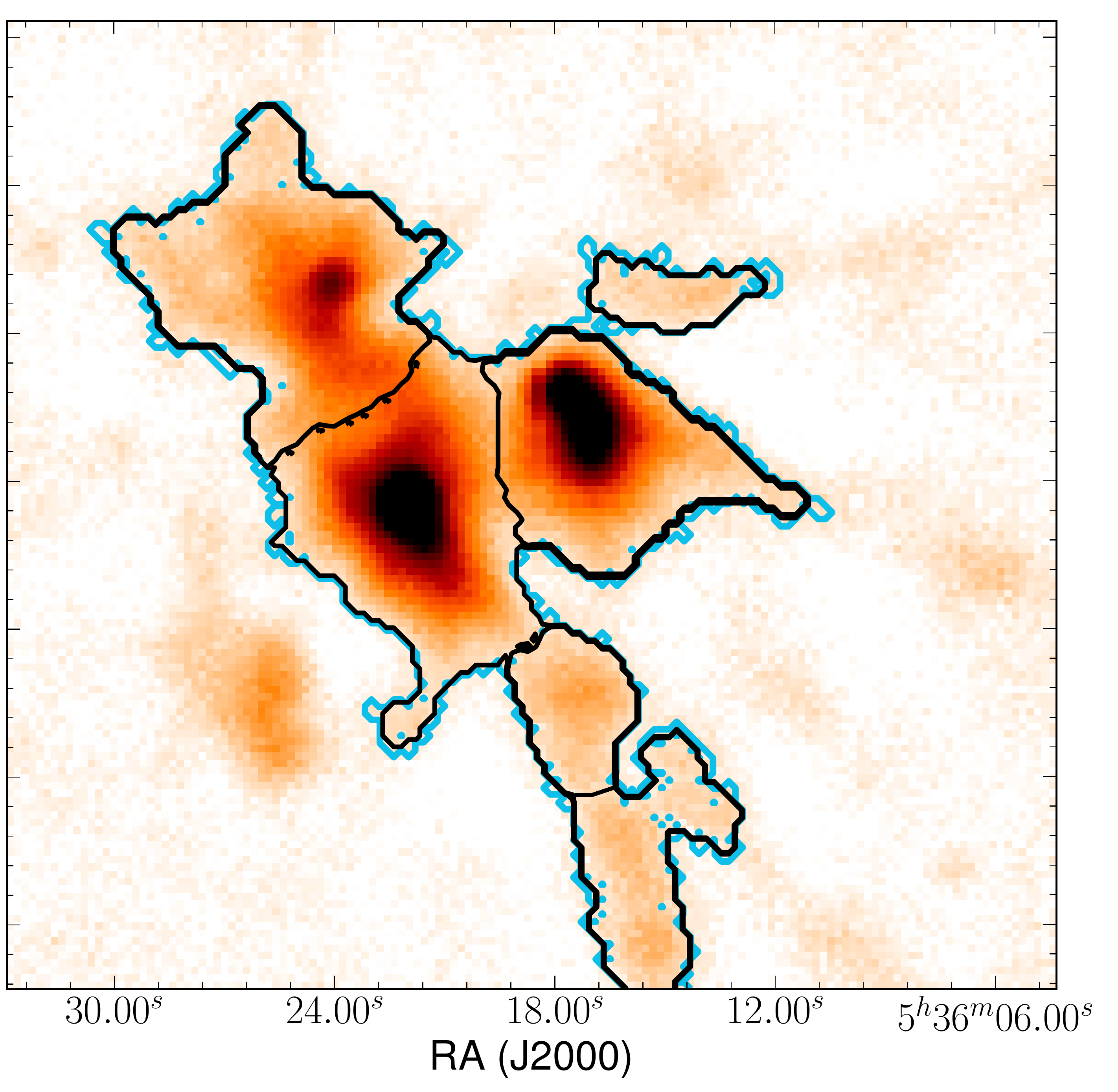}}
\caption{\textit{Left}: An example of an identified island. This blue 3$\sigma_{rms,pix}$ contour contains the Herbig-Haro objects HH 1/2 \citep{johnstone2006}. \textit{Right}: The blue contour again shows the boundaries of the island while the black contours show six individual compact fragments identified by the {\sc{jsa\_catalogue}} algorithm.}
\label{exampleislandandfrags}
\end{figure*}

Our goal here is to characterise both the extended and compact structure present and highlight the connection between the large-scale (up to $\sim$ 7.5$\arcmin$ to 10$\arcmin$) and small-scale components ($<2\arcmin$). 
We define a pixel to be ``significant'' if it has a value of at least 3$\sigma_{rms,pix}$ (\mbox{$\sigma_{rms,pix} = 9.4\mathrm{\:mJy\:beam}^{-1}\mathrm{\:}$}\footnote{This value is higher than what is shown in Table \ref{noisetable} as the flux in a pixel only measures a fraction of the flux in the beam.}) in the CO-subtracted \mbox{850 $\mu$m} map. Thus, we first extract the largest objects studied in this work by simply drawing a contour at 3$\sigma_{rms,pix}$  and retaining all enclosed structures larger than approximately one beam  (15$\arcsec$ in circularly projected diameter). We accomplish this identification using Starlink's version of the algorithm {\sc{ClumpFind}} \citep{williams1994} as implemented in the {\sc{Cupid}} package (\citealt{berry2007}) 
by defining only one flux level over which significant structure is identified. Each non-spurious object detected is referred to as an ``\textit{island}''; any flux present in the map outside of an island is considered to be dominated by noise. The simplicity of this initial step prevents the otherwise sophisticated structure identification algorithms from separating adjoining structures based on more complex criteria. Figure \ref{exampleislandandfrags} (left panel) shows an example island which corresponds to HH 1/2.

In the second step, we employ the JCMT Science Archive algorithm {\sc{jsa\_catalogue}} found in Starlink's {\sc{PICARD}} package \citep{sun265}. This algorithm uses the {\sc{FellWalker}} routine \citep{berry2015}. Briefly, {\sc{FellWalker}} marches through a given image pixel by pixel and identifies the steepest gradient up to an emission peak. After performing tests to ensure that the peak is ``real'' and not just a noise spike, the local maximum is assigned an identifying integer and all the pixels above a user-defined threshold that were included in the path to the peak are given the same identifier. In this way, all of the robust peaks in the image are catalogued and the structure associated with each peak can be analysed. 
The user-defined parameter, \textit{MinDip}, governs the separation of distinct, significant structure. {\sc{FellWalker}} separates structure based on the relative brightness of the region between two areas of peaked emission. If two adjacent structures have peak emission values of $P1$ and $P2$ with $P1<P2$, and the pixels connecting these two peaks have brightnesses larger than $P1-\mathrm{\textit{MinDip}}$, the two emission structures are merged together. For this work, the catalogue produced is focussed on smoothly varying, peaked structure and the \textit{MinDip} parameter was set to $5\times\mathrm{\:local\:noise}$. 

For simplicity in the definition of the largest structures identified, 
we rely on the \textit{islands} described above and we use the ``compact catalogue'' produced by the {\sc{FellWalker}} algorithm to describe the localised, peaked structure visible in the map. These localised peaks are often akin to the individual mountains on an island. {\sc{jsa\_catalogue}} is run independently of the initial contouring, separating emission contained within the larger islands into multiple components. In this way, the compact catalogue generated reveals the substructure present within the context of coincident large-scale emission. For this reason, we refer to the compact components as ``\emph{fragments}''. Fragments are allowed to be somewhat smaller than one beam, their circular projected radius must be at least 5$\arcsec$ (compare this to the JCMT's half width at half maximum of 7.5$\arcsec$). Therefore, they can also exist outside of islands as isolated objects. Note that in many cases, however, the smoothly varying emission structure causes several fragments to be of comparable size to islands, so they should not be directly compared to individual, star-forming cores in all cases. Throughout the rest of this paper, an island which contains at least two fragments will be referred to as a ``complex island'' and an island that contains only one fragment will be referred to as a ``monolithic island''. Note that in the case of the monolithic islands, their corresponding fragments often trace almost the exact same structure. Generally, the total area of a fragment associated with a monolithic island is 80-100\% of the total area of the island. The right panel of Figure \ref{exampleislandandfrags} shows how the HH 1/2 island (blue contours) is separated into six fragments by this technique (black contours). The detected fragments typically trace the islands quite well (to within $\leq$10-20\% in area). Accuracy depends, however, on the morphology of the emission structure.   

\subsection{Calculation of Physical Properties}
\label{physparamsec}

For each island or fragment, we use the associated identification algorithm and the \mbox{850 $\mu$m} SCUBA-2 data to measure the number of pixels associated, the brightest pixel and its location, as well as the total flux density. Table \ref{island450cat} summarises the main observational parameters for each \mbox{850 $\mu$m}-identified island. Note that we align the \mbox{850 $\mu$m} - identified island boundaries with the \mbox{450 $\mu$m} map and we extract the total flux and the peak flux from the latter to include it in Table \ref{island450cat}. We limit the analysis of the \mbox{450 $\mu$m} data to finding the total and peak fluxes of \mbox{850 $\mu$m}-identified island locations as a full comparison between these two datasets goes beyond the scope of this work.  Assuming a constant dust emissivity and temperature, we then calculate the mass ($M$), the peak column density ($N_{peak}$), the radius ($R$; calculated from the circular projection of the given object), the Jeans mass ($M_{J}$, the maximum mass that can be thermally supported in a spherical configuration), and the ``concentration'' (or ``peakiness''). We present this derived information organised in order of the peak brightness of the sources for \mbox{850 $\mu$m} islands and \mbox{850 $\mu$m} fragments in Tables \ref{islandcat}, and \ref{fragcat}, respectively. The \mbox{450 $\mu$m} Orion A data convolved to match the \mbox{850 $\mu$m} data along with temperature maps of all the GBS regions are currently under production and will be released by Rumble et al. (in prep). For a discussion of the determination of source temperatures using \mbox{450 $\mu$m} and \mbox{850 $\mu$m} data in the Ophiuchus Molecular Cloud, see \cite{pattle2015}. 

Assuming the optical depth, $\tau$, is much less than 1, the dust emission observed at \mbox{850 $\mu$m} can be used to derive the mass of a given island or fragment using the following equation
\begin{equation*}  
M_{850} = 2.63\left(\frac{S_{850}}{\mathrm{Jy}}\right)\left(\frac{d}{450\mathrm{\:pc}}\right)^{2}\left(\frac{\kappa_{850}}{0.012\mathrm{\:cm}^2\mathrm{g}^{-1}}\right)^{-1}
\end{equation*}
\begin{equation}
\times \left[\frac{\mathrm{exp}\left(\frac{17\mathrm{\:K}}{T_{d}}\right)-1}{\mathrm{exp}\left(\frac{17\mathrm{\:K}}{15\mathrm{\:K}}\right)-1}\right]M_{\odot},
\label{masseq}
\end{equation}
where $S_{850}$ is the total flux density of the observed emission structure at \mbox{850 $\mu$m}, $d$ is the distance to Southern Orion A, $\kappa_{850}$ is the dust opacity at \mbox{850 $\mu$m}, and $T_{d}$ is the isothermal temperature of the dust, which we assume to be equivalent to the gas temperature. For this work, we choose $d = 450 \mathrm{\:pc}$ \citep{muench2008}, \mbox{$\kappa_{850} = 0.012 \mathrm{\:cm}^{2}\mathrm{g}^{-1}$} (following the parametrization of \citealt{beckwith1990}, $\kappa_{\nu} = 0.1[\nu/10^{12}\mathrm{\:Hz}]^{\beta}\mathrm{\:cm}^{2}\mathrm{\:g}^{-1}$, where \mbox{$\beta = 2.0$}), and $T_{d} = 15\mathrm{\:K}$. Our chosen dust opacity value is consistent with those in other GBS first-look papers such as \cite{pattle2015} and \cite{kirkfirstlook2016}, though, the uncertainty in $\kappa_{850}$ is high (see \citealt{ossenkopf1994}). Preliminary results investigating the temperatures of significant emission regions throughout Orion A by Rumble et al. (in prep) show that temperature values range around 15 K for modest flux values in the CO subtracted \mbox{850 $\mu$m} map. This also agrees with the Orion A temperature map derived by \cite{lombardi2014} using \textit{Herschel Space Observatory} and \textit{Planck Space Observatory} data. 
Thus, we chose an isothermal dust temperature of 15 K for the sources identified in this analysis. Note that recent data from the \textit{Planck Space Observatory} \citep{planck2015} suggests that $\beta \sim 1.8$ for the Orion Molecular Cloud. This small difference in $\beta$ does not affect any of our main conclusions, so we continue to assume a value of $\beta=2$ which is typically assumed in the ISM (see Chen et al., accepted, for a discussion on $\beta$). 

The total uncertainty associated with each term involved in calculating a mass is difficult to precisely quantify. There are uncertainties due to the emission properties of dust grains, temperatures and heating due to YSOs, and distance variations from Northern to Southern Orion A combined with the effects of line of sight projections on the total size of a given source. The dominant contributions to the uncertainty are the temperature and opacity estimates. Temperatures used for similar analyses span \mbox{10-20 K} (see, for example, \citealt{sadavoy2010}) which introduces a factor of $\sim2$ in the mass estimate (see equation \ref{masseq}). Preliminary results from Rumble et al. (in prep) also suggest that while most sources we observe appear to have temperatures of \mbox{$\sim15$ K}, the distribution has a width of \mbox{$\sim\pm5$ K}. In addition, different authors use a range of \mbox{$\kappa_{850}$} values (such as 0.02 \mbox{g cm$^{-2}$}, see \citealt{kirk2007}) introducing another factor of $\sim2$ in uncertainty.  Therefore, an estimate of the total uncertainty in mass is a factor of 3 to 4. Note, however, that most of this is in fundamental properties that are expected to be similar across the cloud (for example, dust opacity, mean temperature, and distance).


The column density of H$_{2}$ molecular hydrogen at \mbox{850 $\mu$m} is given by
\begin{equation*}  
N_{peak} = 1.19\times10^{23}\left(\frac{f_{850,peak}}{\mathrm{Jy\:beam^{-1}}}\right)\left(\frac{\kappa_{850}}{0.012\mathrm{\:cm}^2\mathrm{g}^{-1}}\right)^{-1}
\end{equation*}
\begin{equation}
\times \left[\frac{\mathrm{exp}\left(\frac{17\mathrm{\:K}}{T_{d}}\right)-1}{\mathrm{exp}\left(\frac{17\mathrm{\:K}}{15\mathrm{\:K}}\right)-1}\right]\mathrm{\:cm}^{-2},
\label{Neq}
\end{equation}
assuming a beam width of 14.1$\arcsec$ at \mbox{850 $\mu$m}, where $f_{850,peak}$ is the peak flux density given in \mbox{Jy beam$^{-1}$}.
The Jeans mass can be rewritten in terms of the temperature and the radius of a given island or fragment, $R$ (see \citealt{sadavoystarless})
\begin{equation}
M_{J} = 2.9\left(\frac{T_{d}}{15\mathrm{\:K}}\right)\left(\frac{R}{0.07\mathrm{\:pc}}\right)M_{\odot},
\label{jeanseq}
\end{equation}
where $R$ is the given emission structure's projected circular radius, assuming spherical geometry (the value given in the seventh column of Table \ref{islandcat}). We approximate the aspect ratios of the islands and fragments (tenth column of Table \ref{islandcat} and eleventh column of Table \ref{fragcat}) using flux-weighted horizontal and vertical lengths calculated in the same way by the respective source extraction algorithms, {\sc{ClumpFind}} and {\sc{FellWalker}} (see \citealt{cupidascl2013} for more detailed information). We note that the distributions of aspect ratios (the length of the horizontal dimension divided by the length of the vertical dimension) for both islands and fragments are peaked near 1.0, implying that our assumption of spherical geometry is reasonable. There are, however, sources which deviate by up to a factor of a few. By calculating the ratio between the island or fragment mass and its associated Jeans mass (assuming only thermal pressure support is acting against gravity), we can identify objects that are unstable to gravitational collapse. A gravitationally unstable object has a ratio of \mbox{$M/M_{J}$ $\geq$ 1}. Nevertheless, due to the inherent uncertainties in the measurements described above, 
we define a \textit{significantly} gravitationally unstable island or fragment as one which has $M/M_{J}$ $\geq$ 4.   

The \textit{concentration}, $C$, is a useful metric to quantify whether or not a structure is peaked. The concentration is calculated by comparing the total flux density measured across a given island or fragment to a uniform structure of the same area wherein each pixel is set to the peak brightness, $f_{850,peak}$ (following \citealt{johnstone2001}):

\begin{equation}
C = 1 - \frac{1.13B^{2}S_{850}}{\pi R^{2}\times f_{850,peak}},
\label{concentrationeq}
\end{equation}

\noindent where $B$ is the beam width in arcseconds, $R$ is the radius of the source measured in arcseconds, $S_{850}$ is the total flux of the source measured in Jy, and $f_{850,peak}$ is the peak brightness of the source measured in \mbox{Jy beam$^{-1}$}.  Thus, large islands or fragments which are mostly diffuse will have a low concentration whereas bright, more peaked islands/fragments will have concentration values nearing one. For example, a non self-gravitating, uniform density Bonnor-Ebert sphere has C=0.33 and a critically self-gravitating Bonnor-Ebert sphere has C=0.72 (see \citealt{johnstone2001}).

Peaked structure is often indicative of a higher importance of self-gravity in the observed gas and dust (see \citealt{johnstone2001}, \citealt{kirk2006}, and \cite{kirkfirstlook2016} or heating due to the reprocessing of emission from the presence of young stellar objects. In general, peaked structure is associated with YSOs (see \citealt{jorgensen2007}, \citealt{jorgensen2008}, and \citealt{vankempen2009} for examples), though, \cite{kirkfirstlook2016} found many starless cores with high concentrations (>0.72) in the Orion B Molecular Cloud.

\subsection{Islands}
\label{islandssec}

\begin{table*}
\caption{A sample of the observed parameters corresponding to the \mbox{850 $\mu$m}-identified islands (the full catalogue is available online).}
\label{island450cat}
\begin{tabular}{|c|c|c|c|c|c|c|c|c|}
\hline
\multicolumn{1}{|p{2cm}|}{\centering Source Name$^{\mathrm{a}}$ \\ MJLSG...} & ID & \multicolumn{1}{p{0.7cm}|}{\centering RA$^{\mathrm{\:b}}$ \\ (J2000)} & \multicolumn{1}{p{0.7cm}|}{\centering DEC$^{\mathrm{\:b}}$ \\ (J2000)} & \multicolumn{1}{p{1.2cm}|}{\centering Area$^{\mathrm{\:c}}$ \\ (arcsec$^{2}$)} & \multicolumn{1}{p{1.0cm}|}{\centering S$_{850}$$^{\mathrm{\:d}}$ \\ (Jy)} & \multicolumn{1}{p{1.7cm}|}{\centering f$_{850,peak}$$^{\mathrm{\:e}}$ \\ (Jy beam$^{-1}$)} & \multicolumn{1}{p{1.0cm}|}{\centering S$_{450}$$^{\mathrm{\:f}}$ \\ (Jy)} & \multicolumn{1}{p{1.7cm}|}{\centering f$_{450,peak}$$^{\mathrm{\:g}}$ \\ (Jy beam$^{-1}$)}\\ 
\hline\hline
J053619.0-062212I & 1 & 5:36:18.99 & -6:22:11.88 & 81024.57 & 57.0 & 1.43 & 181.79 & 0.49\\
J053956.2-073027I & 2 & 5:39:56.18 & -7:30:27.31 & 24889.79 & 18.0 & 1.04 & 56.85 & 0.34\\
J053919.9-072611I & 3 & 5:39:19.88 & -7:26:11.05 & 11887.27 & 9.0 & 0.81 & 31.16 & 0.3\\
J053623.1-064608I & 4 & 5:36:23.06 & -6:46:08.20 & 33575.05 & 22.0 & 0.70 & 72.66 & 0.3\\
J053508.8-055551I & 5 & 5:35:08.77 & -5:55:51.43 & 29578.88 & 18.0 & 0.52 & 54.36 & 0.16\\
... & ... & ... & ... & ... & ... & ... & ... & ...\\
J054056.9-081730I & 359 & 5:40:56.87 & -8:17:30.23 & 313.49 & 0.05 & 0.02 & 0.06 & 0.01\\
\hline
\end{tabular}
\begin{flushleft}
a. The source name is based on the coordinates of the peak emission location of each object in right ascension and declination: Jhhmmss.s$\pm$ddmmss. Each source is also designated an ``I'' to signify it is an island as opposed to a fragment. \\ b. The \mbox{850 $\mu$m} map location of the brightest pixel in the island. \\c. The total area of an island. \\d.The total \mbox{850 $\mu$m} flux observed within the island's boundaries. \\e. The maximum \mbox{850 $\mu$m} flux value within the island's boundaries. \\f.The total \mbox{450 $\mu$m} flux observed within the island's boundaries. \\g. The maximum \mbox{450 $\mu$m} flux value within the island's boundaries. 
\end{flushleft}
\end{table*}

\begin{table*}
\caption{A sample of \mbox{850 $\mu$m}-identified islands and their properties (the full catalogue is available online). Islands are ordered from highest to lowest N$_{\mathrm{peak}}$.}
\label{islandcat}
\begin{tabular}{|c|c|c|c|c|c|c|c|c|c|}
\hline
\multicolumn{1}{|p{0.6cm}|}{\centering Island \\ ID} & \multicolumn{1}{p{1cm}|}{\centering $N_{peak}^{\mathrm{\:a}}$ \\ (cm$^{-2}$)} & \multicolumn{1}{p{0.5cm}|}{\centering $M^{\mathrm{\:b}}$ \\ (M$_{\odot}$)} & \multicolumn{1}{p{0.5cm}|}{\centering $R^{\mathrm{\:c}}$ \\ (pc)} & $\frac{M}{M_{J}}^{\mathrm{d}}$  & C$^{\mathrm{\:e}}$ & AR$^{\mathrm{\:f}}$& \multicolumn{1}{p{1cm}|}{\centering $A_{K}^{\mathrm{\:g}}$ \\ (mag)} & Frags$^{\mathrm{\:h}}$ & Protos$^{\mathrm{\:i}}$\\\hline\hline
 1 & 3.66$\times10^{23}$ & 148.61 & 0.35 & 10.42 & 0.95 & 1.14 & 1.51 & 13 & 12\\
 2 & 2.67$\times10^{23}$ & 47.8 & 0.19 & 6.05 & 0.93 & 1.22 & 2.65 & 3 & 6\\
 3 & 2.08$\times10^{23}$ & 23.51 & 0.13 & 4.3 & 0.90 & 1.11 & 1.82 & 2 & 2\\
 4 & 1.79$\times10^{23}$ & 58.17 & 0.23 & 6.33 & 0.90 & 1.0 & 1.29 & 6 & 5\\
 5 & 1.34$\times10^{23}$ & 46.19 & 0.21 & 5.36 & 0.88 & 1.3 & 0.54 & 6 & 4\\
 ... & ... & ... & ... & ... & ... & ... & ... & ... & ...\\
 359 & 4.86$\times10^{21}$ & 0.13 & 0.02 & 0.14 & 0.17 & 4.56 & 1.15 & 0 & 0\\\hline
\end{tabular}
\begin{flushleft}
a. The \textit{peak} column density is calculated by using the flux density of the brightest pixel in the island ($f_{850,peak}$) in Equation \ref{Neq} (using the values shown in the text). \\b.  The mass is calculated by using the total flux of the island ($S_{850}$) in Equation \ref{masseq} (using the standard values shown). \\c.  Effective radius that represents the radius of a circular projection having the same area, A, as the island:  $R = (A/\pi)^{0.5}$. \\d.  The Jeans mass is calculated using the radius of the island in Equation \ref{jeanseq} (using the standard values shown). \\e. The concentration is calculated using Equation \ref{concentrationeq}.\\f. AR is the aspect ratio of the source. It is defined as the length of the horizontal dimension divided by the length of the vertical dimension. \\g. $A_{K}$ is the average value taken directly from the extinction map provided by M. Lombardi (private communication, July 18$^{th}$, 2015) of each source footprint. The extinction can be converted to column density using Equation \ref{aktocol}. \\h. The number of fragments associated with the island. \\i. The number of protostars identified by \cite{megeath2012} and \cite{stutz2013} within the island's boundaries.
\end{flushleft}
\end{table*}

Each identified island is simply defined as a closed, 3$\sigma_{rms,pix}=28\mathrm{\:mJy\:beam^{-1}}$ contour larger than one beam. In Table \ref{islandcat}, we present a small sample of individual island properties derived from the \mbox{850 $\mu$m} data (the full catalogue is available online).  Throughout this section, we give a brief overview of the island population, focussing on the mass and the stability as key observational parameters. In Section \ref{ysosec}, we elaborate on the connections between these structures and the broader physical perspective involving fragmentation and the population of YSOs. There are 359 identified islands in total which comprise 2.2\% of the area of the total map. Out of these, 55 islands were calculated to be Jeans unstable (see Section \ref{physparamsec}) and 75 islands were found to harbour protostars within their boundaries.

\begin{figure*}	
\centering
\subfloat{\label{masshistisland}\includegraphics[width=9.25cm,height=7.5cm]{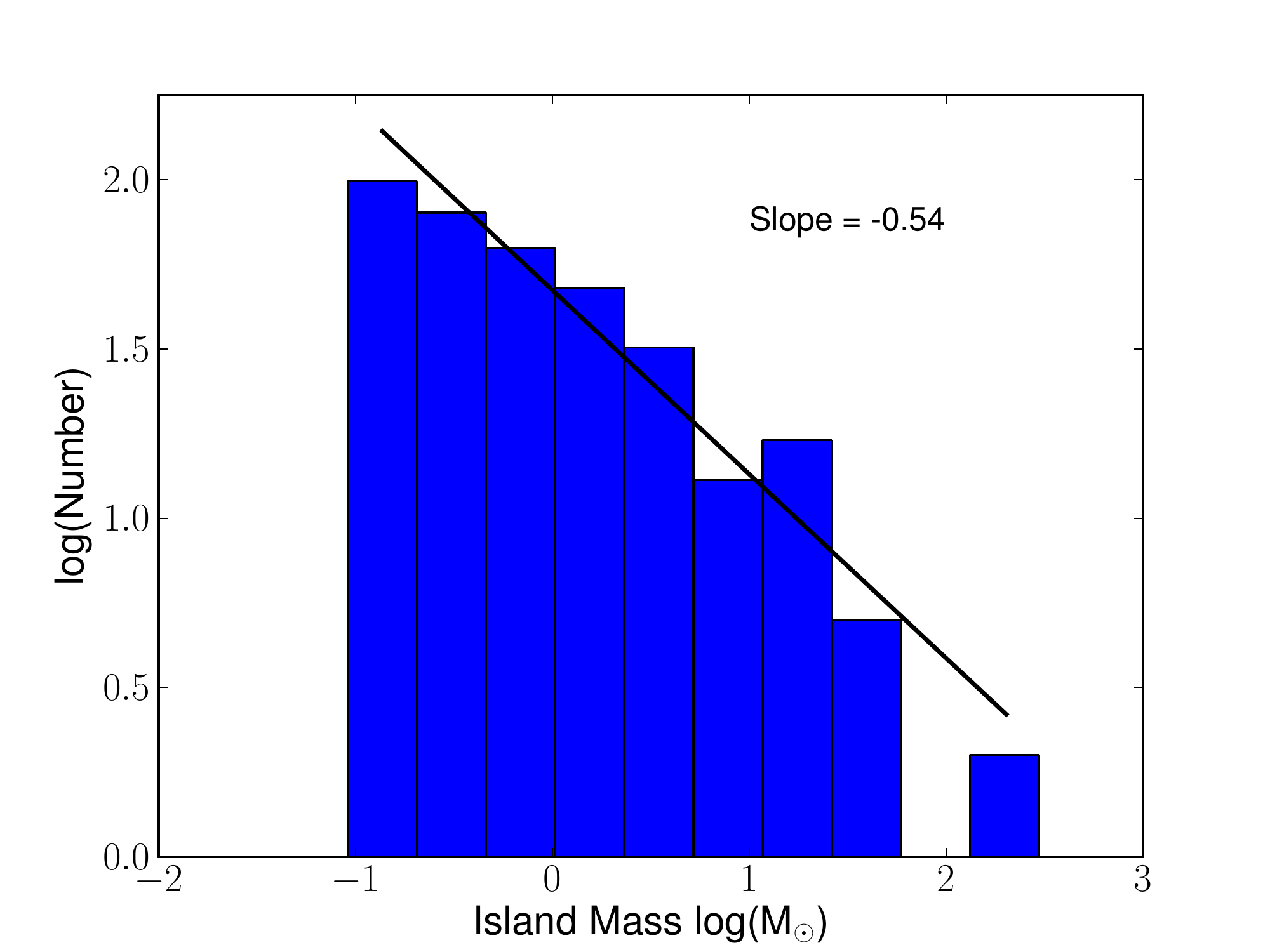}}
\subfloat{\label{stabhist}\includegraphics[width=9.25cm,height=7.5cm]{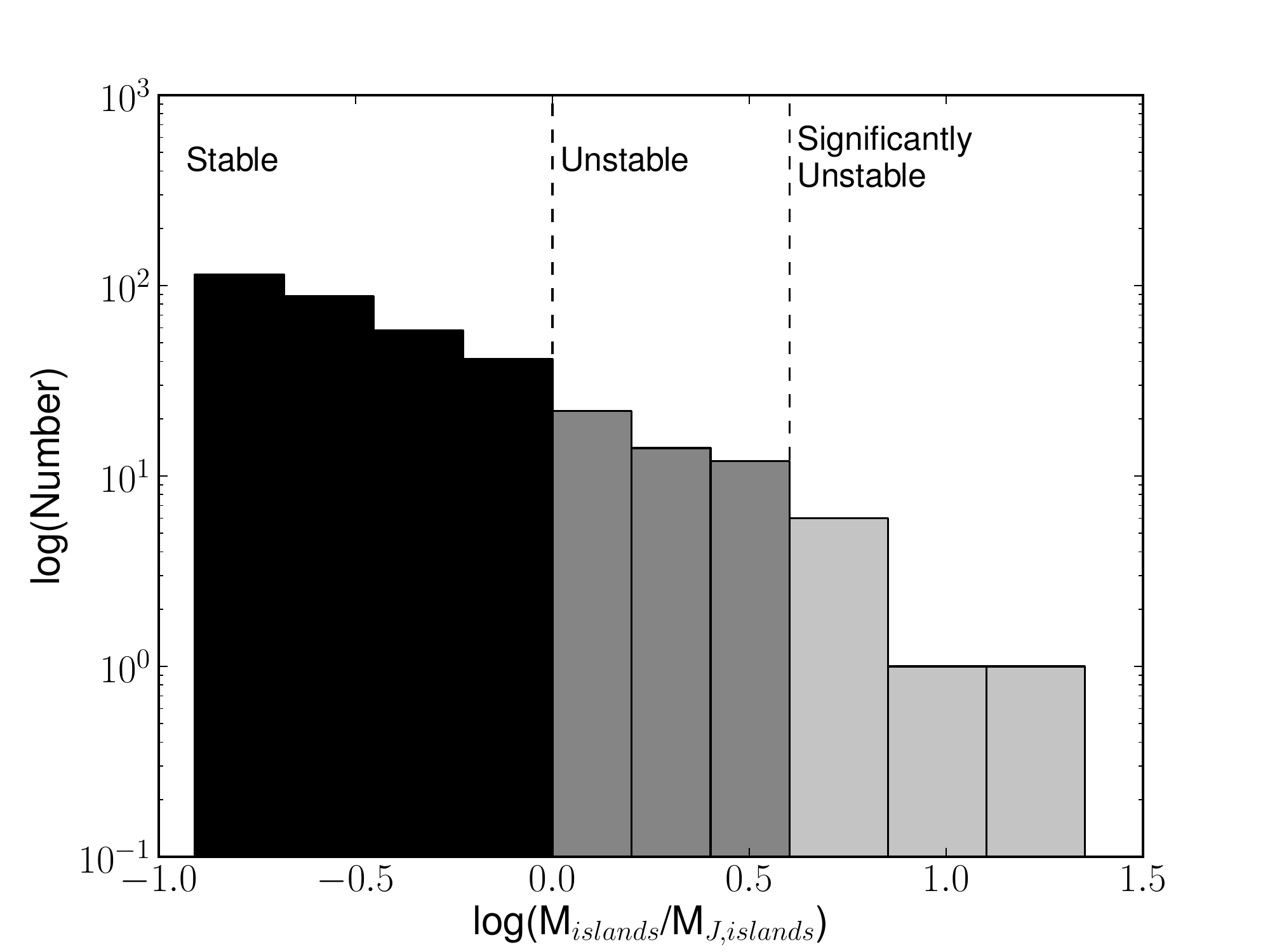}}
\caption{\textit{Left}: Histogram of the masses of the island population. The number of islands decreases with mass following a power law with an exponent of -0.54. 
\textit{Right}: Histogram of the stabilities ($M/M_{J}$) of the island population. Islands with a ratio of $M/M_{J}$ $\geq$ 1 may be gravitationally unstable to collapse, whereas islands with $M/M_{J}$ $\geq$ 4 are defined as \textit{significantly} unstable and are expected to show evidence of gravitational collapse.}
\label{masshist}
\end{figure*}

The left panel of Figure \ref{masshist} shows the mass histogram of the entire island population. 
The masses were calculated using Equation \ref{masseq}, assuming an isothermal temperature of 15 K. As we can clearly see, most island masses are under 10 M$_{\odot}$ with only a few examples of very large, contiguous structures. This situation is to be expected, as large-scale structure is filtered out in SCUBA-2 data and in many cases we only expect to see the brighter components of this underlying emission. 

This histogram does not represent a core mass function as the islands do not uniformly represent pre-stellar objects. Instead, it provides an indication of the largest-scale features to which SCUBA-2 is sensitive. In fact, defining a core mass function from data such as these is inherently difficult due to the broad variety of ways different structure identification algorithms draw borders around adjoining areas of emission \citep{pineda2009}. 


For every island, we calculate the Jeans mass using Equation \ref{jeanseq} and test the stability of the object by comparing it to the observed mass derived from the dust emission. As noted in Section \ref{physparamsec}, an object is theoretically unstable if its $M/M_{J}$ ratio is greater than 1, but we consider a significantly unstable object to have an $M/M_{J}$ ratio greater than or equal to 4 due to the inherent uncertainties in the mass calculation described above (also see Section \ref{fragssec}). We expect large, unstable islands to collapse and fragment on the Jeans length scale (assuming there is only thermal pressure support counteracting gravity in these objects) and small, unstable islands to show some indication of star formation such as high concentration or association with a YSO. Preliminary results from Rumble et al. (in prep) derived from 450/\mbox{850 $\mu$m} flux ratios suggest that a histogram of the median temperature of each island peaks at \mbox{$\sim15\mathrm{\:K}$} within a broad range. The right panel of Figure \ref{masshist}  shows the results on the stability of each island across the map. The two dashed lines show which islands are calculated to be unstable ($M/M_{J}$ $\geq$ 1) and which are significantly unstable ($M/M_{J}$ $\geq$ 4). It is important to note that SCUBA-2 is not sensitive to large-scale structure. As we highlight in Section \ref{extinctionsec}, islands comprise $\sim1.4\%$ of the cloud's mass. 
For the purposes of this analysis, we focus on the smaller-scale star forming sources in the regions of highest column density in the SCUBA-2 \mbox{850 $\mu$m} map and we assume that the mass on the larger scales can be separated out from the more local analysis.  We leave the more thorough stability analysis for the sections below where we combine the island and fragment catalogues, and we can examine individual special cases in the context of fragmentation and YSO association.

\subsection{Fragments}

\begin{figure*}	
\centering
\subfloat{\label{masshistisland}\includegraphics[width=9.25cm,height=7.5cm]{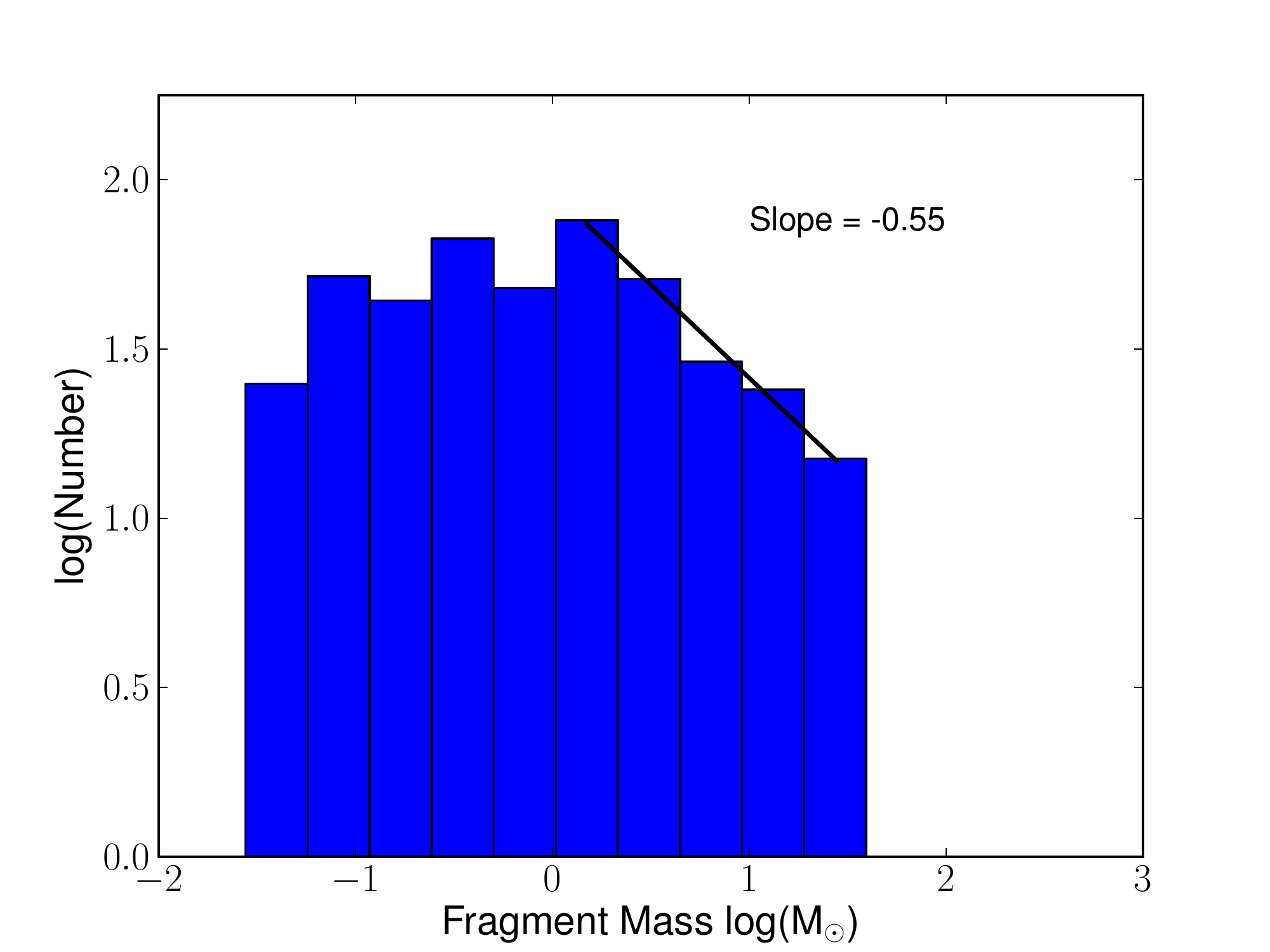}}
\subfloat{\label{stabhist}\includegraphics[width=9.25cm,height=7.5cm]{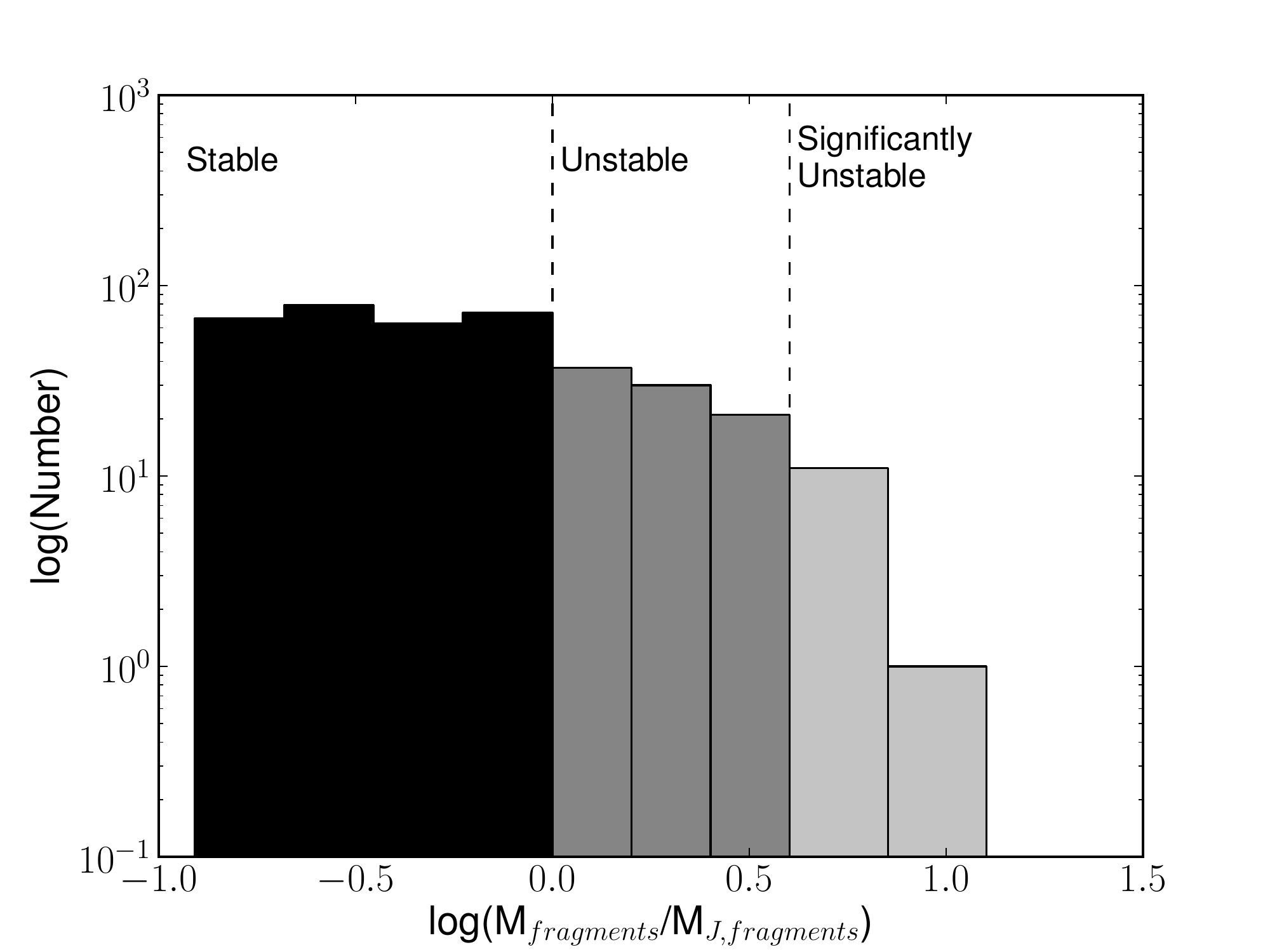}}
\caption{\textit{Left}: Histogram of the masses of the fragment population. 
The high mass slope of the fragment population matches the island high mass slope. \textit{Right}: Histogram of the stabilities ($M/M_{J}$) of the fragment population. Fragments with a ratio of $M/M_{J}$ $\geq$ 1 may be gravitationally unstable to collapse, whereas fragments with $M/M_{J}$ $\geq$ 4 are defined as \textit{significantly} unstable and are expected to show evidence of gravitational collapse.}
\label{fragmasshist}
\end{figure*}

The {\sc{jsa\_catalogue}} algorithm which we use to identify fragments employs the structure identification procedure {\sc{FellWalker}} \citep{berry2015} to detect objects and separate significant emission into individual sources. 
In total, 431 fragments are detected by {\sc{jsa\_catalogue}}, 100 of which are calculated to be Jeans unstable (see Section \ref{physparamsec}) and 103 of which contain at least one protostar within their boundaries. The left panel of Figure \ref{fragmasshist} shows the mass distribution of the observed fragments and the right panel shows the Jeans stability associated with the same population. 
Table \ref{fragcat} shows several examples of fragment properties and the full catalogue is available online.

\begin{table*}
\caption{A sample of \mbox{850 $\mu$m}-identified fragments and their properties (the full catalogue is available online). Fragments are ordered from the highest to lowest N$_{\mathrm{peak}}$ within each parent island.}
\label{fragcat}
\begin{tabular}{|c|c|c|c|c|c|c|c|c|c|c|c|c|}
\hline
\multicolumn{1}{|p{2cm}|}{\centering Source Name$^{a}$ \\ MJLSG...} & \multicolumn{1}{p{0.6cm}|}{\centering Frag \\ ID} & \multicolumn{1}{p{0.5cm}|}{\centering Island \\ ID} & \multicolumn{1}{p{1cm}|}{\centering R.A.$^{\mathrm{\:b}}$ \\ (J2000)} & \multicolumn{1}{p{1cm}|}{\centering Dec$^{\mathrm{\:b}}$ \\ (J2000)} & \multicolumn{1}{p{0.5cm}|}{\centering $N_{peak}$$^{\mathrm{\:c}}$ \\ (cm$^{-2}$)} & \multicolumn{1}{p{0.5cm}|}{\centering $M^{\mathrm{\:d}}$ \\ (M$_{\odot}$)} & \multicolumn{1}{p{0.5cm}|}{\centering $R^{\mathrm{\:e}}$ \\ (pc)} & $\frac{M}{M_{J}}^{\mathrm{\:f}}$  & $C^{\mathrm{\:g}}$ & AR$^{h}$ & \multicolumn{1}{p{1cm}|}{\centering $A_{K}^{\mathrm{\:i}}$ \\ (mag)} & Protos$^{\mathrm{\:j}}$\\
\hline\hline
J053619.0-062212F & 1 & 1 & 5:36:18.99 & -6:22:11.88 & 3.66$\times10^{23}$ & 38.88 & 0.13 & 7.36 & 0.9 & 1.13 & 1.51 & 5\\
J053625.4-062500F & 2 & 1 & 5:36:25.43 & -6:24:59.78 & 9.63$\times10^{22}$ & 27.31 & 0.13 & 4.99 & 0.76 & 1.28 & 1.13 & 5\\
J053641.7-062618F & 3 & 1 & 5:36:41.74 & -6:26:17.59 & 7.15$\times10^{22}$ & 21.1 & 0.13 & 3.9 & 0.74 & 1.63 & 0.31 & 0\\
J053621.0-062151F & 4 & 1 & 5:36:21.00 & -6:21:50.88 & 6.81$\times10^{22}$ & 12.71 & 0.1 & 3.25 & 0.69 & 1.03 & 1.51 & 0\\
J053624.8-062239F & 5 & 1 & 5:36:24.83 & -6:22:38.83 & 6.73$\times10^{22}$ & 19.02 & 0.12 & 3.88 & 0.7 & 1.97 & 1.45 & 1\\
... & ... & ... & ... & ... & ... & ... & ... & ... & ... & ... & ... & ...\\
J054250.0-081209F  & 431 & None & 5:42:49.95 & -8:12:09.16 & 5.91$\times10^{21}$ & 0.05 & 0.01 & 0.09 & 0.28 & 1.8 & 1.06 & 0\\
\hline
\end{tabular}
\begin{flushleft}
a. The source name is based on the coordinates of the peak emission location of each object in right ascension and declination: Jhhmmss.s$\pm$ddmmss. Each source is also designated an ``F'' to signify it is an fragment as opposed to an island. \\ b. The \mbox{850 $\mu$m} map location of the brightest pixel in the fragment. \\c. The \textit{peak} column density is calculated by using the flux density of the brightest pixel in the fragment ($f_{850,peak}$) in Equation \ref{Neq} (using the values shown in the text). \\d. The mass is calculated by using the total flux of the fragment ($S_{850}$) in Equation \ref{masseq} (using the standard values shown). \\e.  Effective radius that represents the radius of a circular projection having the same area, A, as the fragment:  $R = (A/\pi)^{0.5}$. \\f.  The Jeans mass is calculated using the radius of the fragment in Equation \ref{jeanseq} (using the standard values shown). \\g. The concentration is calculated using Equation \ref{concentrationeq}.\\h. AR is the aspect ratio of the source. It is defined as the length of the horizontal dimension divided by the length of the vertical dimension. \\i. $A_{K}$ is the average value taken directly from the extinction map provided by M. Lombardi (private communication, July 18$^{th}$, 2015) of each source footprint. The extinction can be converted to column density using Equation \ref{aktocol}. \\j. The number of protostars identified by \cite{megeath2012} and \cite{stutz2013} within the fragment's boundaries.

\end{flushleft}
\end{table*}

\begin{figure*}	
\centering
\includegraphics[width=17cm,height=15cm]{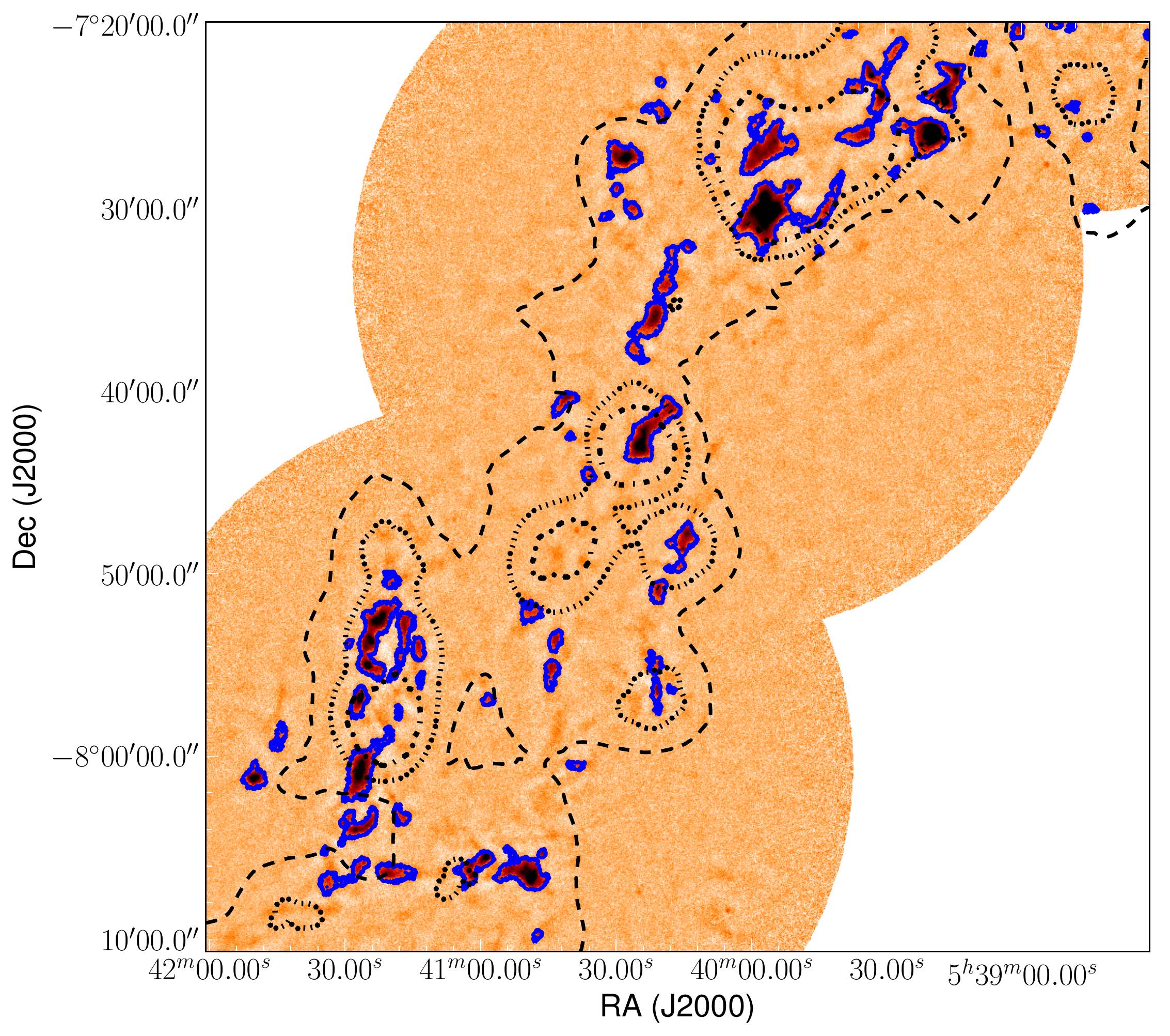}
\caption{A subsection of the \mbox{850 $\mu$m} SCUBA-2 image overlaid with contours from the extinction map obtained from Lombardi (private communication). The solid, blue contours represent islands identified with the SCUBA-2 data while the dashed, dotted, and dash-dot contours represent regions of the extinction map with column densities of 1.67$\times$10$^{22}$ cm$^{-2}$, 3.32$\times$10$^{22}$ cm$^{-2}$, and 5.00$\times$10$^{22}$ cm$^{-2}$, respectively.}
\label{extcontfig}
\end{figure*}

Since each fragment is defined to be associated with a local maximum, these objects often subdivide the larger islands into multiple areas of significant emission. While projection effects are difficult to constrain, the fragments highlight the connection between the larger- and smaller-scale structure in star-forming regions and offer a useful reference for more in-depth studies. Since these fragments are often inherently smaller and less diffuse than their island hosts, it is within the context of fragments that we more thoroughly discuss the connection between dust emission and star formation. There is a wide range in observed fragment masses spanning from 0.03 to 39.3 M$_{\odot}$ with a median mass of \mbox{$\sim$0.7 M$_{\odot}$}. It is interesting to note, however, that there are no detected fragments with masses above $\sim39\mathrm{\:M}_{\odot}$ (Figure \ref{fragmasshist}). Several sources are detected in this high mass regime, but there is a sudden truncation indicating that objects which achieve higher masses are broken into smaller-scale, localised structures. This is obvious when we compare the high mass end of the fragment distribution with the high mass end of the island distribution in Figure \ref{masshist} (left panel). The highest mass islands each contain at least three fragments within their boundaries. Also, note that the slope of the fragment mass histogram is comparable to the island mass histogram at large masses. This indicates that the large fragments are not completely analogous to cores, but represent more extended regions of smoothly varying significant emission. As in the case of the island mass distribution shown in the left panel of Figure \ref{masshist}, this histogram does not represent a core mass function because the fragments do not uniformly represent pre-stellar objects. Note, however, that the {\sc{FellWalker}} algorithm separates objects based on the height of a given emission peak relative to its local surroundings. This means that while many fragments may be large, they only contain one prominently peaked region.

\subsection{Large-scale Structure from Extinction}
\label{extinctionsec}


Here, we analyse the observed islands and associated YSOs from the \cite{megeath2012} and \cite{stutz2013} catalogues in the context of large-scale structure. To this end, we use the extinction data from \cite{lombardi2011} at 1.5$\arcmin$ resolution (Lombardi, M. priv communication, 2015). Figure \ref{extcontfig} shows the Lombardi et al. extinction data as contours overlaid on the SCUBA-2 \mbox{850 $\mu$m} extinction map.  These extinction data were determined using the Near-infrared Color Excess (NICEST) method from \cite{lombardi2009}. In effect, the {\sc{NICEST}} method seeks to remove contamination of foreground stars and inhomogeneities introduced by unresolved structure. The extinction measurements were calculated using near-infrared observations from the Two Micron All Sky Survey (2MASS; \citealt{skrutskie2006}). We note that the difference in the resolution between the SCUBA-2 map and the extinction map results in some small variations in peak emission location as represented in Figure \ref{extcontfig}.  
\begin{figure}	
\centering
\includegraphics[width=9cm,height=7.2cm]{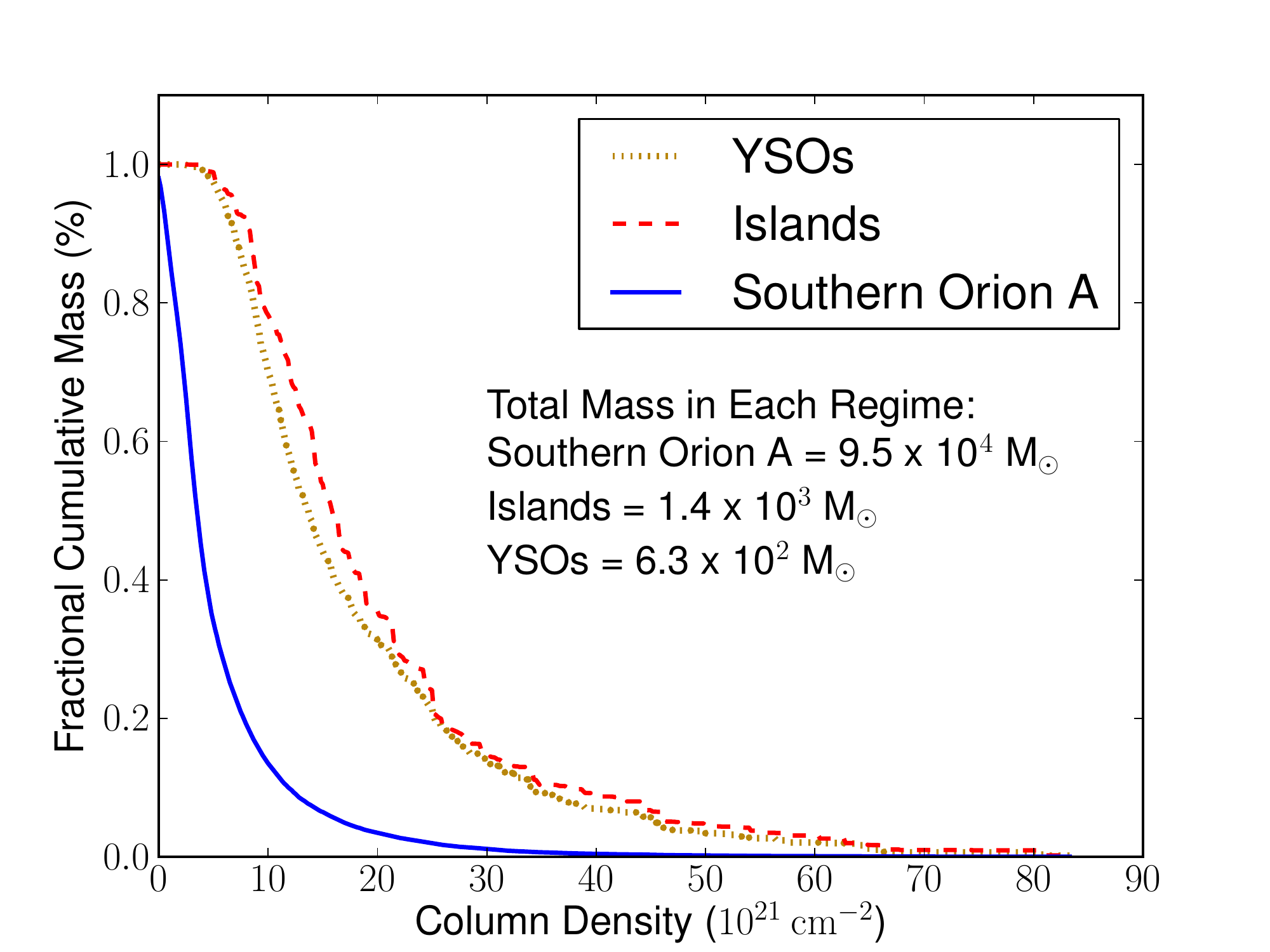}
\caption{Three cumulative mass fractions plotted against the column density: The entire Southern Orion A cloud (NICEST; blue curve), the islands (SCUBA-2; red dashed curve), and the YSOs (\textit{Herschel} and \textit{Spitzer}; dotted curve). The cumulative mass fraction for the whole cloud was derived from the {\sc{NICEST}} extinction map. The cumulative mass fraction of the islands was derived from the SCUBA-2 \mbox{850 $\mu$m} data of all the pixels contained within the boundaries of each sources. The cumulative mass fraction of the YSOs was derived by counting the number of objects in the \protect\cite{megeath2012} and \protect\cite{stutz2013} catalogues and assuming a mass of 0.5 $M_{\odot}$ for each source.} 
\label{cumulativeextfig}
\end{figure}

Following calculations presented in \cite{lombardi2014}, we converted the A$_{\mathrm{K}}$ extinction values to column densities using the conversion
\begin{equation}
\frac{\Sigma}{A_{K}} \simeq 183 \mathrm{\:M}_{\odot} \mathrm{\:pc}^{-2}\mathrm{\:mag}^{-1},
\label{aktocol}
\end{equation}
where $\Sigma$ is the mass surface density. 

Figure \ref{cumulativeextfig} compares the cumulative mass fraction for all of Southern Orion A, the islands, and the YSO population plotted against the column density derived from Lombardi et al's extinction map.  For the cloud distribution, we derive the mass from the extinction map and consider only those data where SCUBA-2 observed.  Similarly, for the islands, we determined the mass associated with the islands from our analysis in Section 3.2 (e.g., contiguous regions with \mbox{850 $\mu$m} emission $>$ 3$\sigma_{rms,pix}$).  Finally, for the YSOs, we use the number of sources in all classes above each column density level, assuming a standard average YSO mass of \mbox{0.5 M$_{\odot}$} (for example, see \citealt{megeath2012} and \citealt{stutz2013}).


Figure \ref{cumulativeextfig} can be compared with a similar analysis performed in Orion B \citep{kirkfirstlook2016} with the caveat that the extinction map used in this paper has much coarser resolution and therefore, on average, much smaller column density values. 
The total mass of the SCUBA-2 observational footprint derived from the extinction map is $9.5\times10^{4}M_{\odot}$. The total mass of all identified islands derived from the \mbox{850 $\mu$m} map is $1.3\times10^{3}M_{\odot}$ and the total mass of the YSOs is $6.6\times10^{2}M_{\odot}$ assuming a typical mass of \mbox{0.5 M$_{\odot}$} for all sources. Clearly, the islands trace the densest material, whereas the broader Southern Orion A cloud includes a significant diffuse component. Also, we see that the YSO population tracks quite well with the islands especially at higher column densities, indicating a connection between the densest gas and the YSO population. The associations between YSOs and observed structure are further explored throughout this paper and especially in Section \ref{ysosec}. 

Note that in Figures \ref{extcontfig} and \ref{cumulativeextfig}, we can see the effect of the large-scale mode subtraction applied to this dataset. The islands we identify are moderate-scale, heavily extincted regions which comprise a small portion of the map in both mass and area (approximately 1.4\% and 2.2\%, respectively). These structures we identify undoubtedly lie within larger-scale, less-dense structures; the material which links our islands to the rest of the cloud. The details of how the largest scales in a molecular cloud connect to localised star-forming regions are complex and not yet well understood. As we explore throughout Section \ref{ysosec}, however, the size scales and mass scales accessible to SCUBA-2 continuum data represent siginificant areas  of star forming material. Throughout this analysis, we assume that the larger-scale modes to which our observations are not sensitive only serve to increase the gravitational instability of islands and fragments and therefore fuel the formation of stars.  



\section{Associations with Young Stellar Objects}
\label{ysosec}


In this section we analyse the SCUBA-2 emission in conjunction with the YSO catalogues presented by \cite{megeath2012} and \cite{stutz2013} in an effort to associate these dense gas structures with evidence of active star formation.   \cite{megeath2012} constructed their catalogue using a large-scale  \textit{Spitzer Space Telescope} survey while the catalogue derived by \cite{stutz2013} targeted more localised regions with the \textit{Herschel Space Observatory} such that their analysis would be sensitive to very deeply embedded protostars. All the figures presented in this section are colour-coded by the given emission structures' individual association with different classes of YSOs. We define an ``association'' between a YSO and an emission structure as the YSO position falling within the boundaries of the object of interest (island or fragment). A ``strong'' protostellar association is when a protostar falls within one beam diameter ($\sim15\arcsec$) of the object's peak emission location. In this work, we make no attempt to determine the class of a given YSO independently and rely on the provided designations of these sources in the catalogues of \citealt{megeath2012} and \citealt{stutz2013}. There are four YSO designations presented by \citealt{megeath2012} which we combine with a ``No YSO'' category to separate our detected emission structures into five main groups.  

\vspace{3mm}

\noindent P: Protostars. These objects have characteristics (such as spectral energy distribution and colour) consistent with Class 0, Class I, or Flat Spectrum sources, i.e., young, embedded protostars. We also include five additional confirmed protostars from \textit{Herschel Space Observatory} observations (see objects with a ``flag'' value of 1, indicating a ``confirmed'' protostar, in Table 3 in \citealt{stutz2013}). We differentiate in the plots here between an island or fragment that simply contains a protostar (denoted by a green outline) and an island or fragment that contains a protostar that lies within one beam of the peak emission position (denoted by a solid green symbol). 

\vspace{3mm}

\noindent FP: Faint Candidate Protostars. These objects have protostar-like colours but \textit{Spitzer} \mbox{MIPS 24 $\mu$m} emission that is too faint (> 7 mag) for them to be considered robust protostar detections (see the \citealt{kryukova2012} criteria and \citealt{megeath2012} for more details). We denote associations with faint candidate protostars by blue outlines.

\vspace{3mm}

\noindent RP: Red Candidate Protostars. These objects have sufficiently bright MIPS 24 $\mu$m emission but lack any detection in \textit{Spitzer}'s shorter wavelength bands. Each source was visually inspected by \cite{megeath2012} to differentiate it from objects such as asteroids or background galaxies. We denote associations with red candidate protostars by red outlines. 

\vspace{3mm}

\noindent D: Discs. These objects have characteristics consistent with Class II sources, i.e., pre-main sequence stars with discs. We denote associations with discs by brown outlines.

\vspace{3mm}

\noindent No YSOs: No Associated YSOs. If none of the above objects lie within the boundaries of a given emission structure, we denote it with a black outline. 

\vspace{3mm}

\noindent We also analysed four protostar candidates which were identified in  \citealt{stutz2013} (objects with a ``flag'' value of 2, indicating a ``candidate'' protostar, in Table 3 in \citealt{stutz2013}). Only one of these, however, is contained within the boundaries of an island or a fragment and it lies a significant distance from the nearest \mbox{850 $\mu$m} dust emission peak. We therefore chose not to include it in this analysis. In total, there are 212 protostars, 1081 disc sources (or, discs), 27 faint candidates, and 2 red candidates within the SCUBA-2 mapped area analysed in this paper.  

\begin{figure*}
\centering
\subfloat{\label{ysobright}\includegraphics[width=12cm,height=10cm]{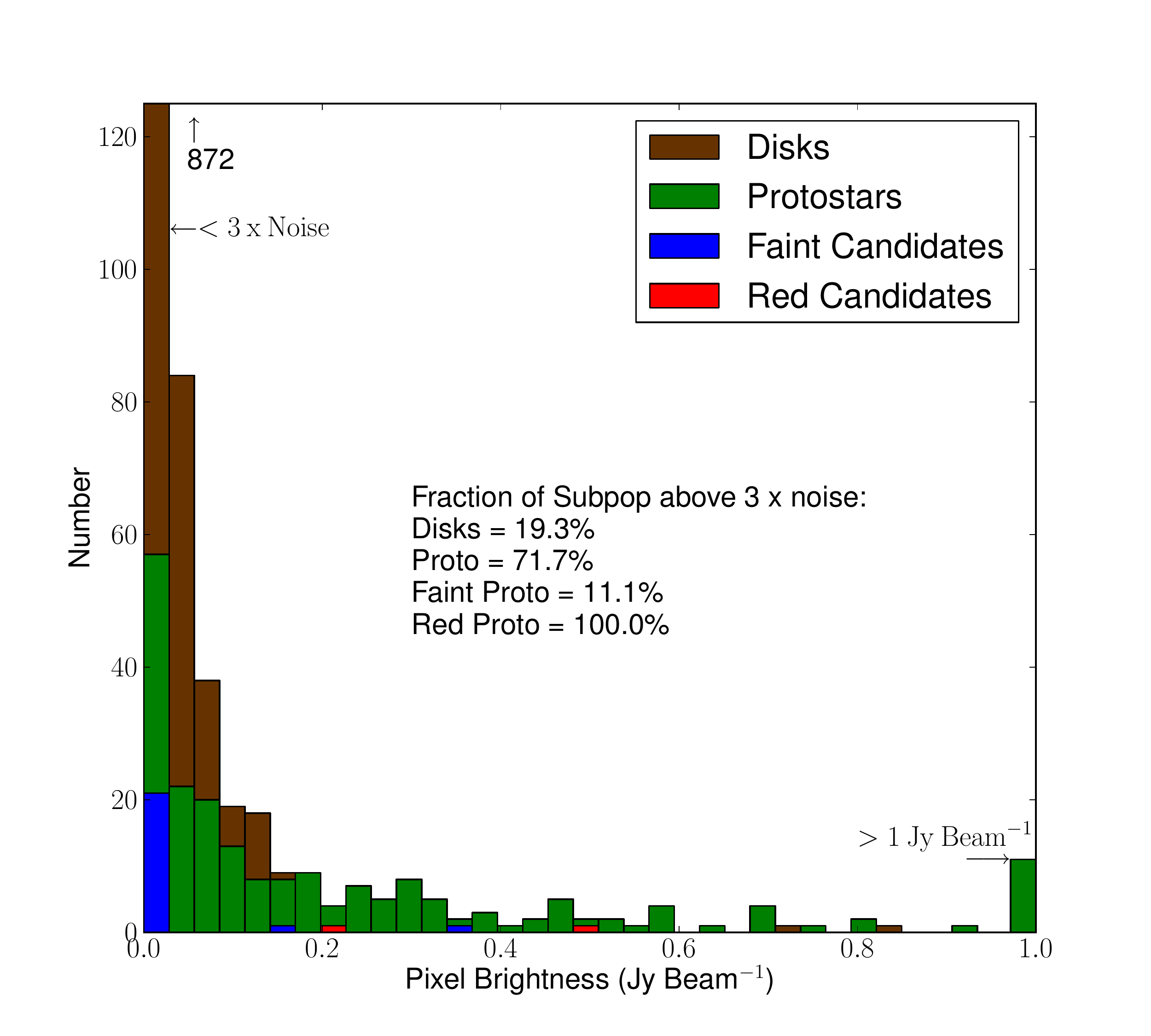}}\\
\subfloat{\label{ysodist}\includegraphics[width=12cm,height=10cm]{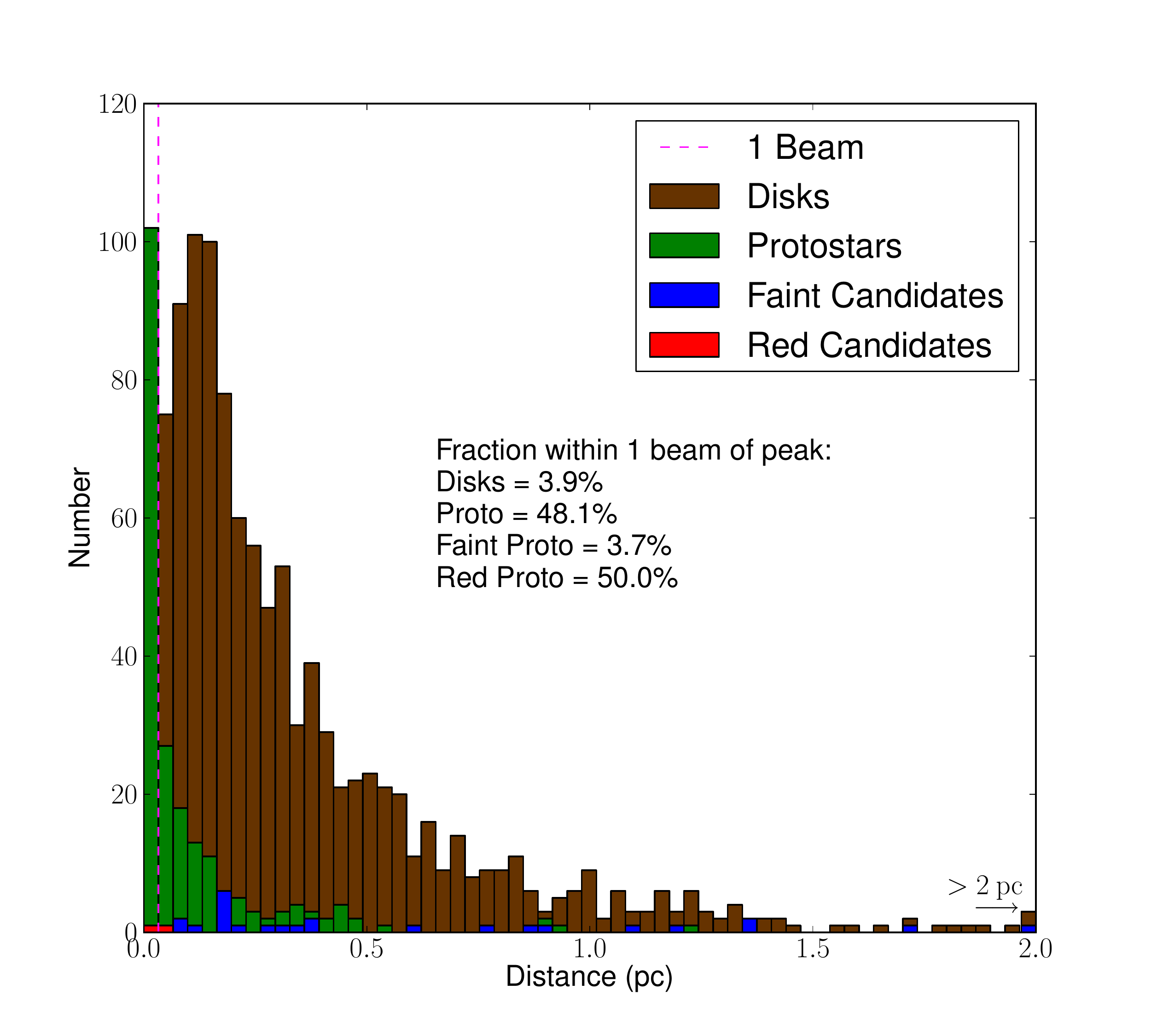}}
\caption{Two metrics to analyse the population of YSOs in the context of their association with fragments. \textit{Top:} A measurement of the \mbox{850 $\mu$m} flux at the location of a YSO in units of \mbox{Jy beam$^{-1}$}. The width of each bin is  \mbox{$3\sigma_{rms,pix} = 0.028\mathrm{\:Jy\:beam^{-1}}$}. The first bin also includes YSOs which are located on negative \mbox{850 $\mu$m} flux pixels; in this bin, there are 872 disc sources. The final bin shows the number of YSOs coincident with pixels that are brighter than \mbox{1.0 $\mathrm{Jy}\:\mathrm{beam}^{-1}$}. \textit{Bottom:} The distance between a given YSO and the location of the nearest fragment's localised emission peak. Each bin has a width of \mbox{$15\arcsec \simeq 1\mathrm{\:beam} = 6750$ AU}. The final bin shows the number of YSOs which lay further than 2.0 pc from the nearest emission peak. The magenta line on the right edge of the first bin highlights objects which are within \mbox{$\sim1\mathrm{\:beam}$} of the nearest localised emission peak.}
\label{ysohists}
\end{figure*}

\subsection{An Overview of the YSO Population in the \mbox{850 $\mu$m} SCUBA-2 map}


In the top panel of Figure \ref{ysohists}, we plot the \mbox{850 $\mu$m} flux measured at each YSO location. The right edge of the first bin represents the threshold flux level for a pixel to be included in an island or a fragment. Each bin has a width of \mbox{3$\sigma_{rms,pix}=28\mathrm{\:mJy\:beam^{-1}}$}. Here, we see that 72\% of protostars lie on pixels with \mbox{850 $\mu$m} fluxes above this adopted threshold value. Since young protostars are deeply embedded objects that are still accreting mass from surrounding material, their correspondence with bright \mbox{850 $\mu$m} emission is expected. More-evolved protostars eventually disperse this surrounding material and should have lower associated \mbox{850 $\mu$m} fluxes than their younger counterparts. Due to their still young ages, however, even the more-evolved protostars have not had time to move a significant distance away from their parent emission structure or for this structure to have dispersed and thus still reside within islands (see Section \ref{diskssec}, \citealt{stutz2015}, and \citealt{megeath2016} for further discussion). 

The remaining 28\% of protostars which do not appear within islands represent an interesting population. In some cases, protostars lie just beyond island boundaries by $\sim$ 3$\arcsec$ to $10\arcsec$ and these could well be more-evolved objects that formed in the nearest island but shed enough local material or were gravitationally ejected such that they now lie outside its boundaries. In other cases, the protostars may simply be misclassified. Interestingly, \cite{heiderman2015}  recently found the same percentage of protostars which appear to be misclassified using an independent data set: the Gould Belt ``MISFITS'' survey. In their survey, \cite{heiderman2015} observed \mbox{HCO$^{+}$(J=3-2)} toward all the Class 0/I and Flat spectral sources identified by Spitzer (from \citealt{megeath2012}) and distinguished protostars from discs following \cite{vankempen2009}.   Similar to our results with SCUBA-2 at \mbox{850 $\mu$m}, \cite{heiderman2015} found that only 72\% of their sample met the line criteria for protostellar classification.  Thus, a significant fraction of protostars may be misclassified based on their SEDs.  In addition, line-of-sight coincidences between more-evolved Class II/III sources and dense gas could result in additional misclassifications (e.g., from underestimated extinction corrections in the near-infrared bands).  Since the Orion cloud has a large and dense YSO population, such coincidences are more likely.


The top panel of Figure \ref{ysohists} shows that disc sources, as expected, are found generally at locations of low emission. These more-evolved objects have had time to migrate away from their parent structures and by definition they should not have a dense envelope. Emission we detect around isolated disc sources is presumably due to the remnant of the dispersed natal envelope or excess material finishing its collapse. Of course, we expect some discs to also align with bright emission locations simply because of projection effects. The majority of faint protostar candidates also seem to lie at lower levels of \mbox{850 $\mu$m} flux, indicating that they are likely not young protostellar objects. The two red protostar candidates which fall into our Southern Orion A map, however, do have significant associated flux which strengthens the evidence of their classification. 

The bottom panel of Figure \ref{ysohists} shows the distribution of distance between a given YSO and its nearest fragment's peak emission location. 
Fragment peak emission locations were chosen as opposed to islands as it is the former objects that are more likely the formation sites of an individual to a few protostars. We only include YSOs which lie on pixels within the SCUBA-2 footprint of Southern Orion A. Here, we find similar results to the top panel, i.e., disc sources appear to be more scattered about the map than protostars (see Section \ref{diskssec}). In contrast, approximately half of the protostars lie within one beam of the nearest peak flux location and the population as a whole is peaked toward closer distances. Moreover, the red protostar candidates seem to have strong associations with potential star-forming sites whereas the faint protostars can lie quite separated from these regions, indicating the latter may be misclassified background galaxies.

\subsection{Star Formation in Fragments}
\label{fragssec}

\begin{figure*}	
\centering
\includegraphics[width=16cm,height=13cm]{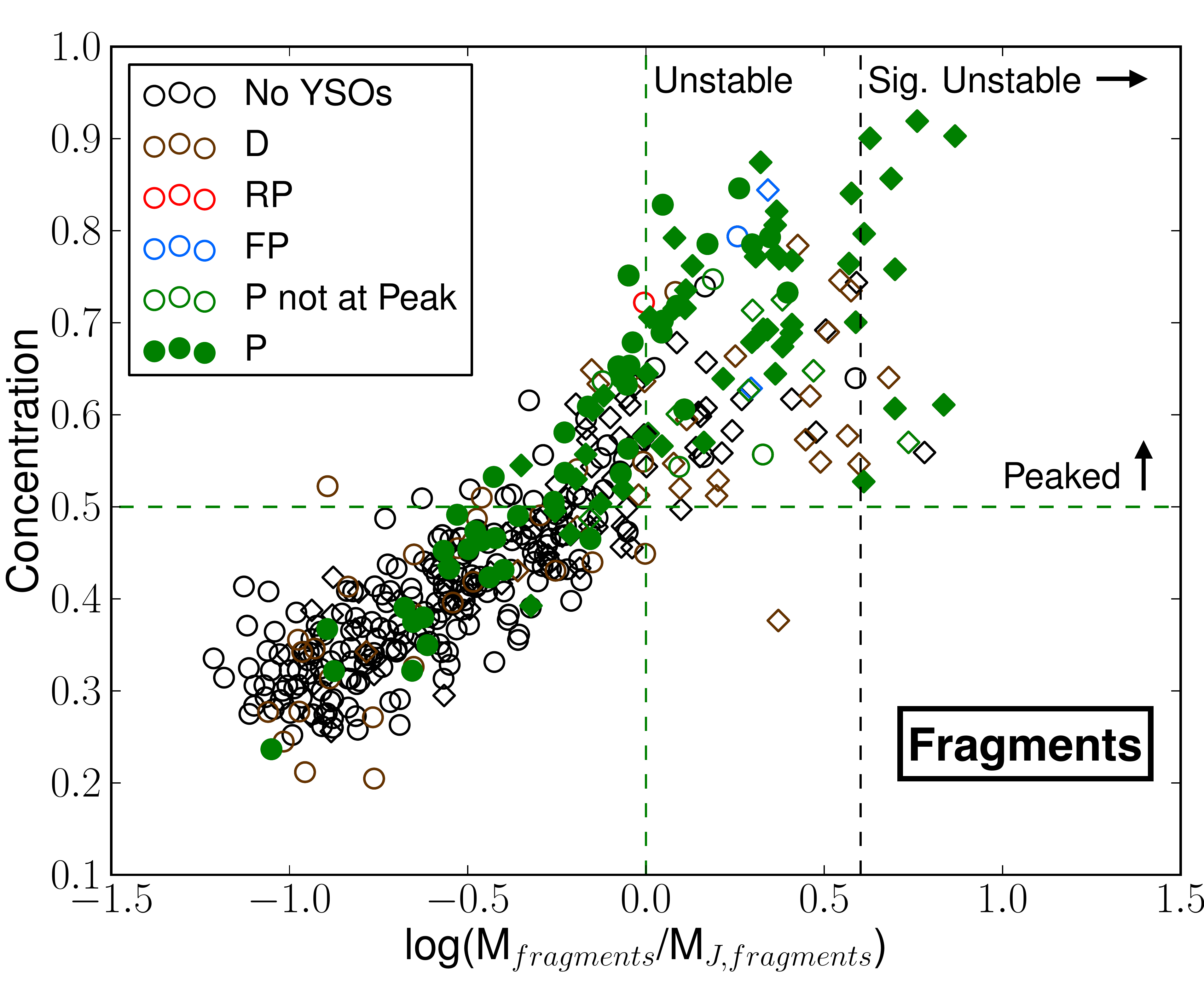}
\caption{Fragment concentration versus fragment stability. The dashed green lines show a concentration of 0.5 on the ordinate and the gravitational instability line on the abscissa. The vertical dashed black line represents an $M/M_{J}$ ratio of 4 where we define sources to be significantly unstable. Colours represent associations between the identified fragment and several classes of YSOs as denoted in the legend. Diamonds represent a fragment which belongs to a complex island and a circle represents a fragment which traces isolated, monolithic structure.}
\label{concenstab}
\end{figure*}

More so than islands, it is the compact, localised fragments for which we expect Jeans unstable cases to be forming (or to eventually go on to form) stars. Thus, in Figure \ref{concenstab}, we compare fragment concentrations with their Jeans stabilities. Highly concentrated sources are expected to have a higher degree of self-gravity, eventually collapsing and forming one to a few stellar systems. As discussed in Section \ref{physparamsec}, the concentration is a measure of the spatial distribution of emission. Using Equation \ref{concentrationeq}, we determine which fragments are concentrated (values nearer to 1) or more uniform (values nearer to 0). 
In Figure \ref{concenstab}, green dashed lines indicate the nominal gravitational instability line $M/M_{J} \geq $ 1 (horizontal) and  $C = 0.5$ (vertical). $C = 0.5$ is chosen because it represents a relatively concentrated core approximately half way between a uniform density (0.33) and self gravitating Bonnor Ebert sphere (0.72) (see \citealt{johnstone2001}). Note that the fragments fall broadly into two regimes: 1.) gravitationally stable and with uniform emission and 2.) gravitationally unstable and with peaked emission. 
We note as well that the diamond symbols in Figure \ref{concenstab} represent a fragment which belongs to a complex island (an island containing at least two fragments) and a circle represents a fragment which traces isolated, monolithic structure.

We would expect the gravitationally unstable, peaked fragments to be the population which is associated with protostars. In general, we see this is the case. In Figure \ref{concenstab}, only 8\% of the fragments without discernible signs of YSOs appear unstable and concentrated. Of those, the fragments which were extracted from monolithic islands (or have no island associations) are outnumbered by those which were extracted from complex islands (21\% and 79\%, respectively). Conversely, we would expect the gravitationally stable, less peaked fragments to be the population which is not actively forming stars. Indeed, only 23\% of the  stable and uniform fragments appear to have YSOs. Almost all of these fragments are associated with monolithic islands (83\%); that is, they do not have ``siblings'' within the same island. 

\begin{figure*}
\centering
\subfloat{\label{frag398}\includegraphics[width=9cm,height=7.5cm]{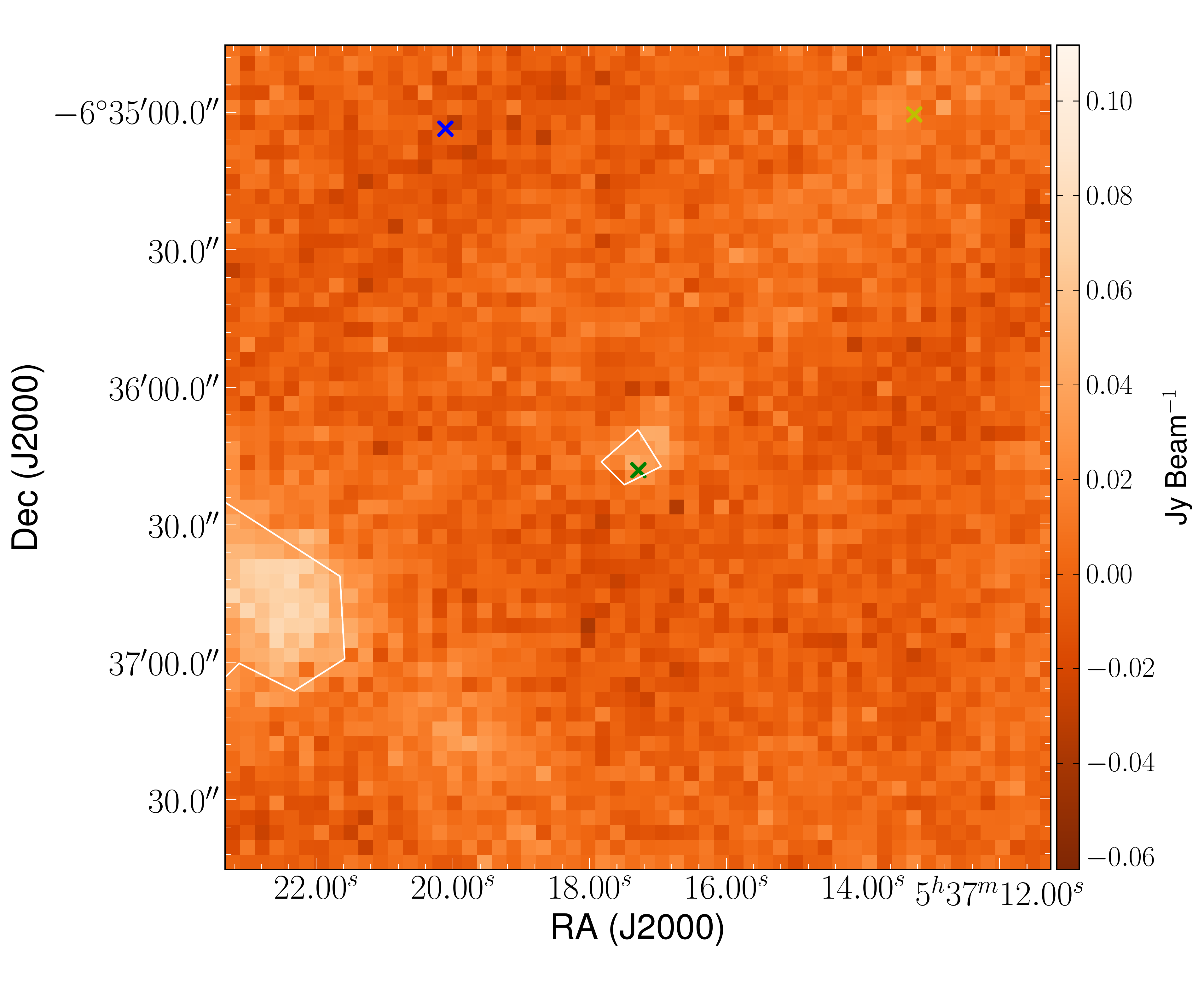}}
\subfloat{\label{isl63}\includegraphics[width=9cm,height=7.5cm]{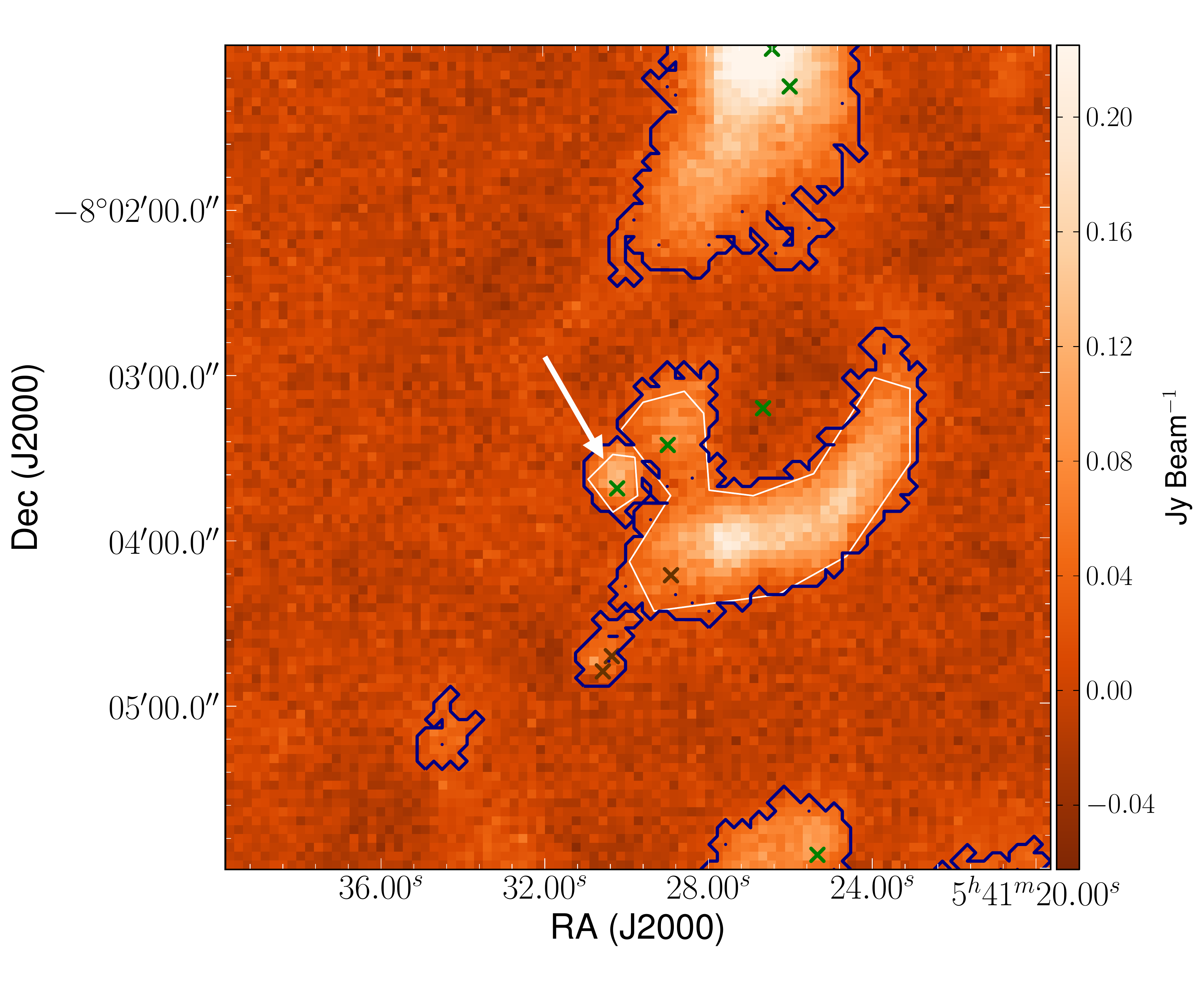}}
\caption{Typical examples of fragments calculated to be gravitationally stable to collapse yet having a strong association with a confirmed protostar. In general, it is the lack of large-scale structure in the SCUBA-2 map which leads to these non-intuitive detections. White contours show the boundaries of selected fragments. The crosses show the locations of YSOs following the same colour scheme as outlined in previous figures and the text. \textit{Left}: The isolated monolithic case. This particular fragment of interest (center) has no associated island. \textit{Right}: A case where the fragment is extracted from an island with multiple areas of significant emission. The blue contours show the boundaries of islands in the field of view (part of L1641S). The fragment of interest is highlighted by the white arrow.}
\label{lssubexamples}
\end{figure*}

There are two main possibilities for explaining the fragment population in the bottom left quadrant of Figure \ref{concenstab} that are associated with protostars. First, during the formation of the protostar, the mass reservoir around the central, bright object has been depleted by accretion to the extent that the now diffuse gas and dust falls below our detection limit. Similar situations were noted by \cite{mairs2014} through the synthetic observations of a numerical simulation. Thus, these objects are more-evolved Class I protostars. Second, our data may be insensitive to some mass due to the large-scale structure subtraction discussed previously (see \citealt{chapin2013} and \citealt{mairs2015}). In at least some cases (see below) this can cause structure identification algorithms to detect multiple individual sources instead of one larger source, leading to an underestimate of the true stability.  
Also note that the dust continuum traces the envelope and disc and not the mass associated with the central protostar itself (which is optically thick but slightly beam diluted at these wavelengths). The actual mass of the system, therefore, is greater than the measured mass (see \citealt{mairs2014} for a discussion on including protostellar masses in stability calculations based on synthetic observations of numerical simulations).  

Figure \ref{lssubexamples} shows two examples of the types of fragments we identify with protostellar associations in the purportedly ``stable regime'' of Figure \ref{concenstab}. In the left panel, 
we see bright, dense regions which may sit on top of a more uniform, large-scale background to which the SCUBA-2 instrument is less sensitive. 
In the data reduction procedure, if we were to filter out less of the large scale structure, the boundaries of isolated sources would broaden further into the diffuse structure and this may result in the blending of multiple islands and fragments. Relaxing the filtering constraints, however, leads to less confidence in the robustness of the detected diffuse structure (see \citealt{chapin2013}).

The right panel of Figure \ref{lssubexamples} shows one of the two ``stable'' fragments associated with protostars extracted from complex (not monolithic) islands. The difference between this structure and the monolithic, stable structure in the left panel which harbours a protostar, however, is that the smaller fragment was close enough to a larger structure to have been included in the boundaries of the same island rather than being identified as an isolated object. Both stability as well as concentration of course will depend on how boundaries are drawn between the significant areas of emission. This example shows why performing source extraction in crowded areas is a difficult process, especially when lacking the entire large-scale component. The unstable, low concentration fragment associated with a disc source in the lower right quadrant of Figure \ref{concenstab} is a similar object to the small fragment with the protostar near its peak  presented in the right panel of Figure \ref{lssubexamples}. As described in Section \ref{cataloguessec}, the {\sc{FellWalker}} algorithm has chosen the boundaries of these individual fragments based on the minimum value between localised peaks, i.e., the valleys between the mountains. If a sufficiently low value is achieved, the algorithm will separate structure accordingly (see \citealt{berry2015}). 
Again, a robust recovery of the large-scale background structure may prove useful in identifying how each fragmented area is related, depending on the morphology of that structure. Any algorithm designed to extract structure will have uncertainties in object boundaries based on the user's specific input parameters, culling processes, and end goals. Similarly, without spectroscopic information, any algorithm  will also be subject to projection effects, i.e., the possibility of more than one source in the same line of sight. In terms of associations with YSOs, however, we expect projection effects to be a larger factor when associating dust-emission regions with disc sources as opposed to protostars as the latter tend to be embedded in their parent material. A further discussion of the distribution of  disc sources and protostars is provided in Section \ref{diskssec}. 


While there is an intrinsic uncertainty in the opacity (by assuming a fixed dust grain size) and the distance of each object, we assume these two values are fairly consistent across the entire Southern Orion A map. What may change in different areas, however, is the temperature. We have calculated each fragment's $M/M_{J}$ stability ratio based on the assumption of an isothermal temperature of 15 K. If this temperature was higher by 5 K, the stability ratio would decrease by approximately a factor of 2 and each object will be found to be ``more stable'', assuming only thermal support is counteracting the force of gravity. This difference comes from a combination of the lower fragment mass as well as the higher Jeans mass arising from assuming a hotter temperature. Potential sources of heating include nearby high-mass stars,  the embedded YSOs themselves, and cosmic rays. If the temperature was 5 K colder, however, the calculated stability ratio would increase by a factor of 3 and objects would be more unstable.

\subsection{Island Fragmentation}
\label{fragsec}

\begin{figure*}	
\centering
\subfloat{\label{islandstability}\includegraphics[width=13.75cm,height=10.5cm]{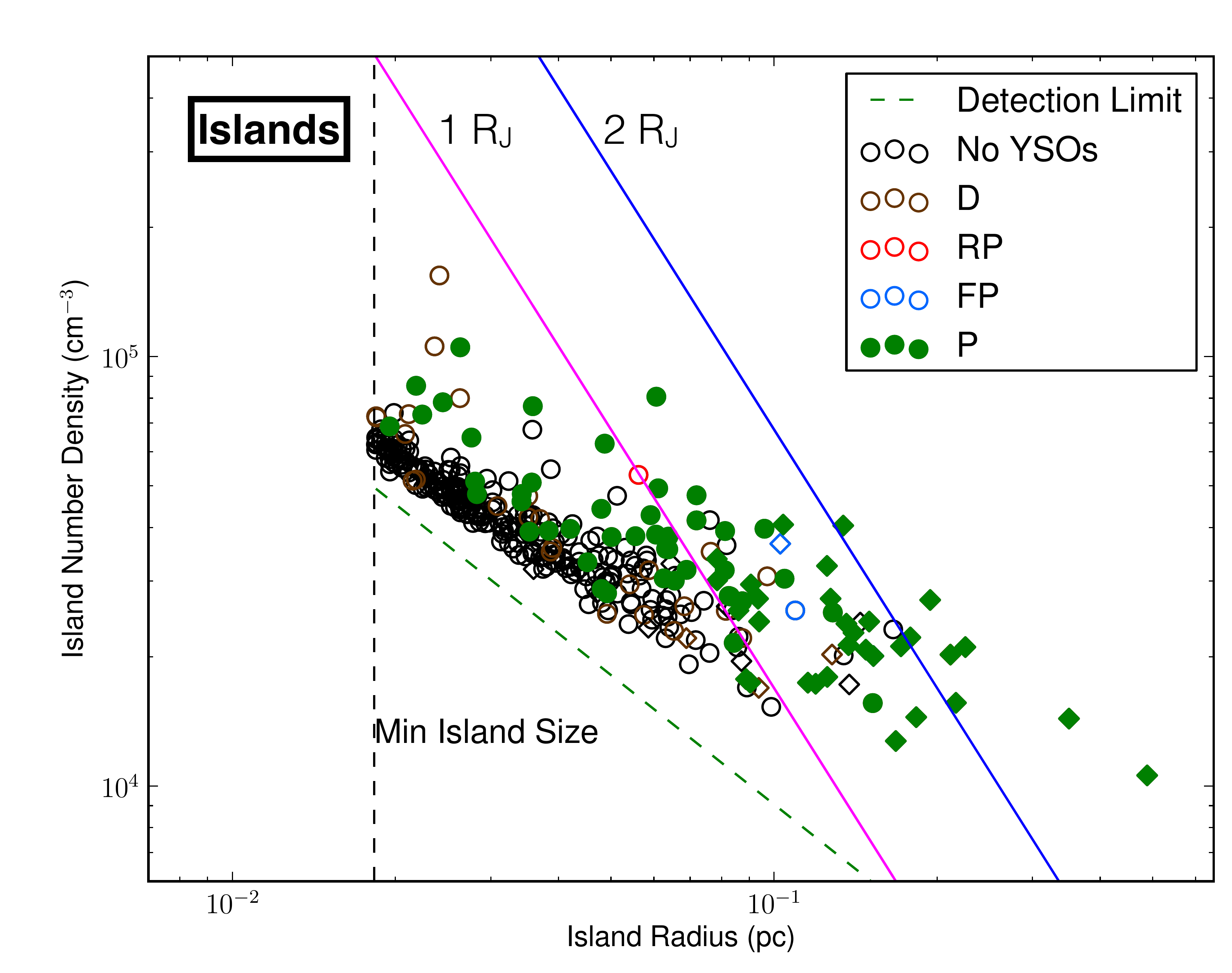}}\\
\subfloat{\label{fragstability}\includegraphics[width=13.75cm,height=10.5cm]{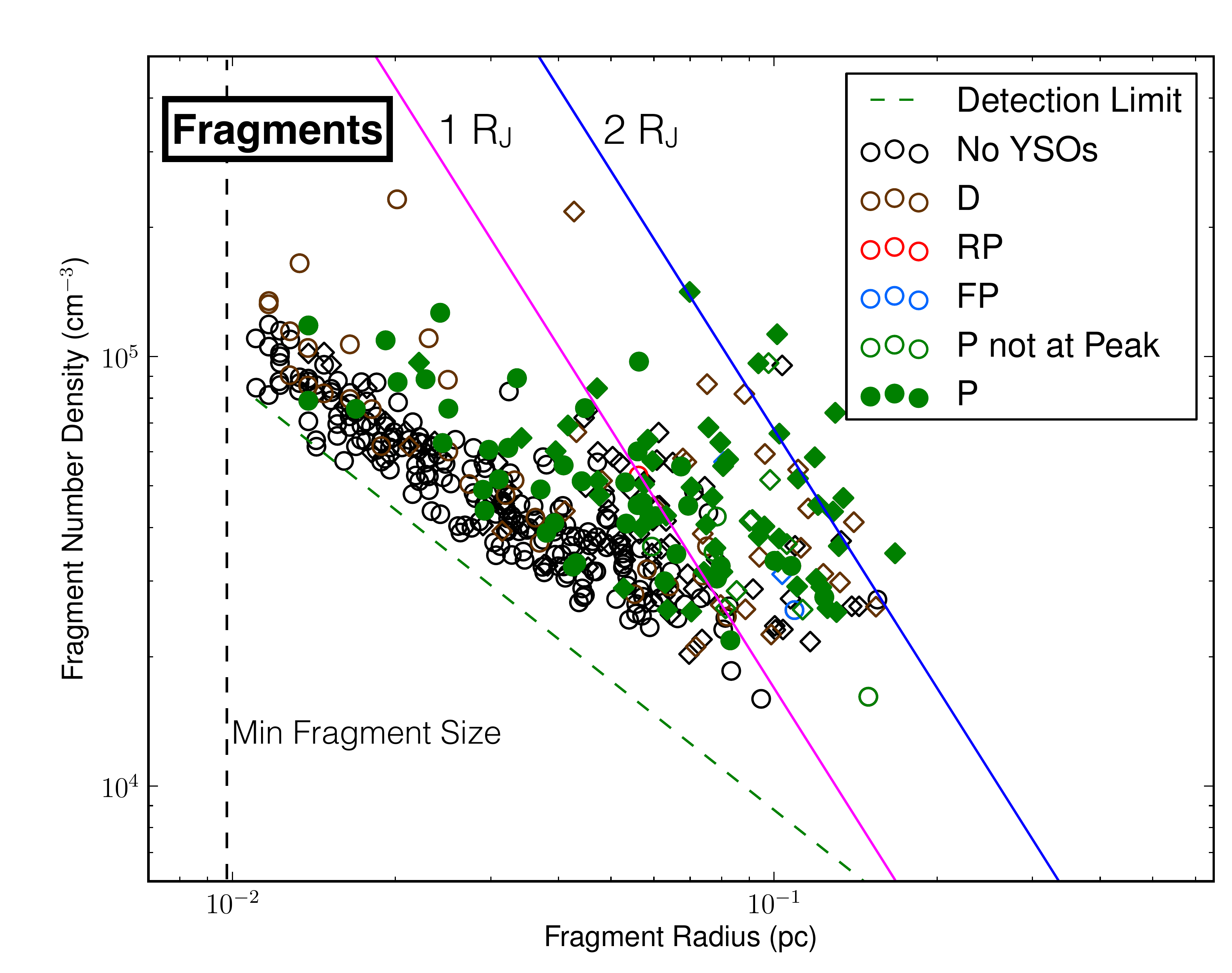}}
\caption{The number density of a given object assuming a spherical configuration versus the radius of the object's circular projection. The colour scheme follows Figure \ref{concenstab}. \textit{Top:} Islands; diamonds represent complex islands and circles represent monolithic islands. The green dashed line shows the detection limit. We chose the minimum island size such that every object had at least some measurable structure. \textit{Bottom:} Fragments; diamonds represent fragments extracted from complex islands and circles represent fragments extracted from monolithic islands. Note that the smallest fragments were allowed to be smaller than the minimum island size. The magenta and blue lines show 1 Jeans radius and 2 Jeans radii, respectively.}
\label{stabilityfigs}
\end{figure*}

We now turn our discussion to the connection between islands and fragments in the context of fragmentation and star formation. We remind the reader that a ``complex island'' is defined to contain at least two fragments whereas an island that displays only one area of significant emission is referred to as ``monolithic''. In Figure \ref{stabilityfigs}, we compare the mean gas number density with effective radius for islands (top panel) and fragments (bottom panel). The colour scheme of symbols remains the same for the YSO associations but there is a subtle difference in the symbols themselves. For the islands, a diamond represents a complex island and a circle represents a monolithic island whereas for the fragments, a diamond represents an object extracted from a complex island and a circle represents an object extracted from a monolithic island. The number densities were calculated assuming spherical symmetry using the effective radii (see Table \ref{islandcat}). Two lines of instability are shown representing one Jeans radius (beyond which we expect an object to be unstable to collapse) and two Jeans radii (beyond which we observe all objects to be fragmented). A third, dashed green, line represents the detection lower limit for an island (\mbox{$3\sigma_{rms,pix} = 28 \mathrm{\:mJy\:beam}^{-1} = 3.73\times10^{21}\mathrm{\:cm}^{-2}$}). The reason there is a gap between larger structures and this detection limit is because the data reduction process filters out uniform, extended emission. An area of the sky with significant emission will only be recovered if it has some slope, otherwise it will be filtered out with the signal attributed to the sky. Thus, a uniform 3$\sigma_{rms,pix}$ flux across an island's area is a conservative, rather than realistic, lower limit. The Jeans radius is calculated by inverting Equation \ref{jeanseq}, assuming the observed mass is the Jeans mass.

In Figure \ref{stabilityfigs}, larger objects are generally less dense, but more unstable, as expected. The majority of the 43 complex islands (79\%) lie beyond the Jeans instability line (R$_{\mathrm{J}}$) and all the islands beyond the second instability line (2R$_{\mathrm{J}}$) are complex (19\% of the complex sample). An object can be unstable to collapse and not fragment when it is only slightly too large (between R$_{\mathrm{J}}$ and 2R$_{\mathrm{J}}$), but for an island to remain monolithic above two Jeans radii, a non-thermal pressure support would be needed in addition to thermal energy to counteract gravity\footnote{Note that if the molecular gas is indeed cooler than 15 K, each object will be shifted upward to higher densities (recall that assuming a temperature of 10 K results in masses which are a factor of three larger) and more of the complex islands would lie beyond the lines of instability. The lines of instability will also vertically shift as they vary linearly with temperature (e.g., a factor of 1.5 downward assuming 10 K as opposed to 15 K), but to a lesser degree than the density. We note that the majority of the large islands show signs of star formation via associations with YSOs. There are, however, a few special cases which will be explored in more detail in Section \ref{starlesssec}, below.}.

The bottom panel of Figure \ref{stabilityfigs} shows the break-up of the larger islands into significant, individual fragments. 
We see several cases where the individual fragments drawn from complex islands are larger than two Jeans radii. The small-scale monolithic objects in both panels are fairly consistent with one another, indicating isolated regions have similar properties whether we lay a simple contour around the emission region (as we did for islands) or we employ the {\sc{FellWalker}} algorithm (as we did for fragments). Between the two instability lines we see several cases of fragments that do not have associated YSOs but which can be found in complex islands. There are also many interesting areas of seemingly unstable, starless dust emission (see Section \ref{starlesssec}), though new, deeper surveys such as {\sc{Vision}}\footnote{\url{http://homepage.univie.ac.at/stefan.meingast/vision.html}}, however, may uncover previously undetected embedded protostars.

\begin{figure*}	
\centering
\subfloat{\label{monolithichist}\includegraphics[width=15cm,height=10.8cm]{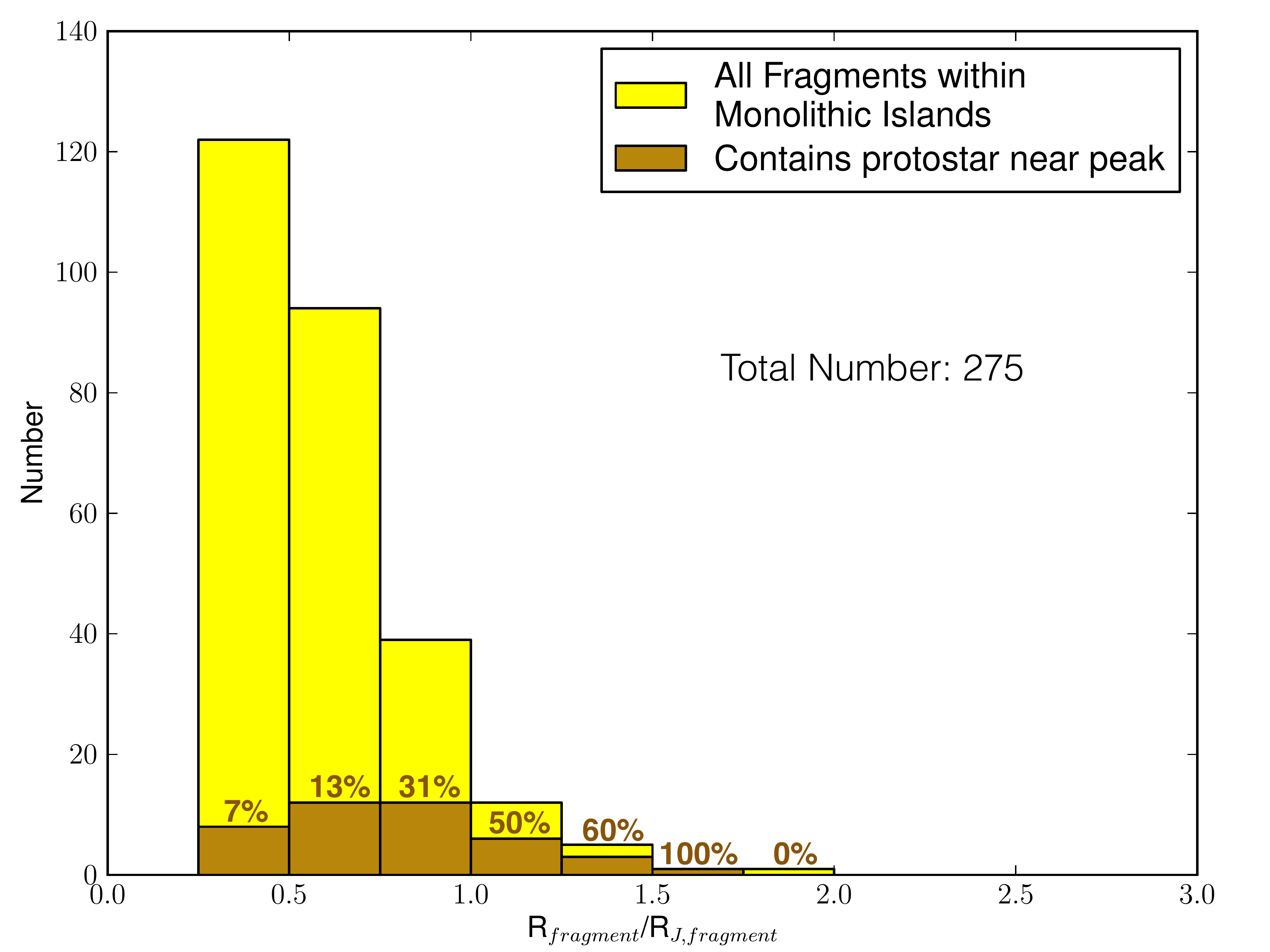}}\\
\subfloat{\label{fragmentedhist}\includegraphics[width=15cm,height=10.8cm]{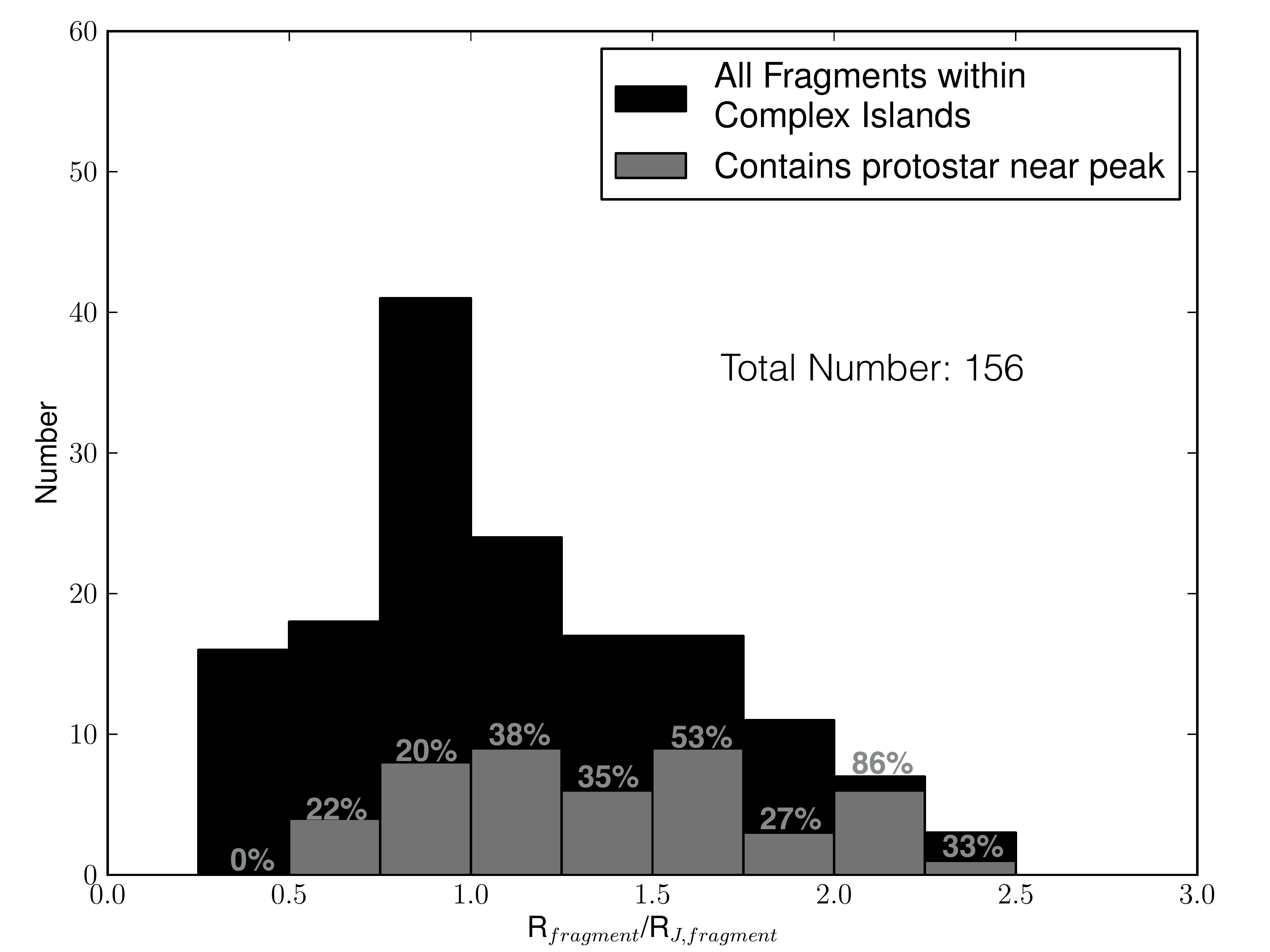}}
\caption{Histograms showing the total population of fragments extracted from monolithic islands (275 in total, 23 of which have no island association; \textit{top}) and fragments extracted from complex islands (156 in total; \textit{bottom}) in the context of each object's Jeans radius. The main histograms (light yellow in the top panel and black in the bottom panel) show all fragments within each classification whereas the secondary histograms (dark yellow in the top panel and grey in the bottom panel) show the fraction of fragments which contain a confirmed protostar within one beam width of the peak location. The percentages written are the fraction of the subpopulation which contains a protostar near the peak in each bin.}
\label{stabilityhists}
\end{figure*}

To investigate further the connection between fragmentation and star formation, Figure \ref{stabilityhists} shows histograms of Jeans radii for the fragments in monolithic islands (top) and complex islands (bottom), with separate distributions for all sources and for those sources with protostars residing less than 15$"$ from the fragment peak flux position. 
In both panels, the percentages above each bar show the fraction of fragments within a particular bin that have a strong association with a known protostar. The top panel percentages reveal that more unstable ($R>R_{J}$) monolithic structures indeed show increasingly more evidence of star formation (except in the final bin which represents one curious object discussed further in Section \ref{starlesssec}). The bottom panel percentages, however, reveal the same cannot be said for fragments in more complicated, clustered environments. Here, it appears that more unstable structures within complex islands do not necessarily show more evidence of star formation. Although their parent islands may have protostars within their boundaries, there are still some significantly dense, unstable emission peaks which have no associations with YSOs. Such examples could indicate on-going collapse across a time longer than the collapse of a single core (i.e.  clustered star formation may be more drawn out). Similar objects were noted in models by \cite{mairs2014} (also see \citealt{offner2010} for more information on the simulations used in that study and a further analysis on fragmentation). 

\subsection{Starless Super-Jeans Islands}
\label{starlesssec}

\begin{table*}
\caption{A list of gravitationally unstable, starless islands. These objects are good candidates for follow-up studies.}
\label{interestingtable}
\begin{tabular}{|c|c|c|c|c|c|}
\hline
Source Name (MJLSG) & Island ID & $\frac{\mathrm{M}}{\mathrm{M}_{J}}$ & Concentration & Aspect Ratio & Monolithic/Complex\\
\hline\hline
J053700.5-063711I & 20 & 1.8 & 0.79 & 3.14 & Monolithic\\
J053228.4-053420I & 25 & 1.4 & 0.76 & 1.36 & Monolithic\\
J053511.0-061400I & 29 & 3.7 & 0.67 & 1.25 & Monolithic\\
J053509.8-053754I & 45 & 1.4 & 0.59 & 1.10 & Monolithic\\
J053550.8-054142I & 33 & 2.9 & 0.67 & 1.15 & Complex$^{a}$\\
J053622.8-055618I & 38 & 1.9 & 0.73 & 1.01 & Complex$^{b}$\\
J053403.9-053412I & 63 & 1.9 & 0.60 & 5.29 & Complex$^{c}$\\
\hline
\end{tabular}
\begin{flushleft}
a. Both fragments are also gravitationally unstable with $\frac{M}{M_{J}}$ ratios of $\sim$2 and $\sim$3. \\b. Both fragments are also gravitationally unstable with $\frac{M}{M_{J}}$ ratios of $\sim$7 and $\sim$5. \\ c. Both fragments are nearly gravitationally unstable with $\frac{M}{M_{J}}$ ratios of $\sim$1.
\end{flushleft}
\end{table*}

\begin{figure*}	
\centering
\subfloat{\label{island27}\includegraphics[width=9cm,height=7.5cm]{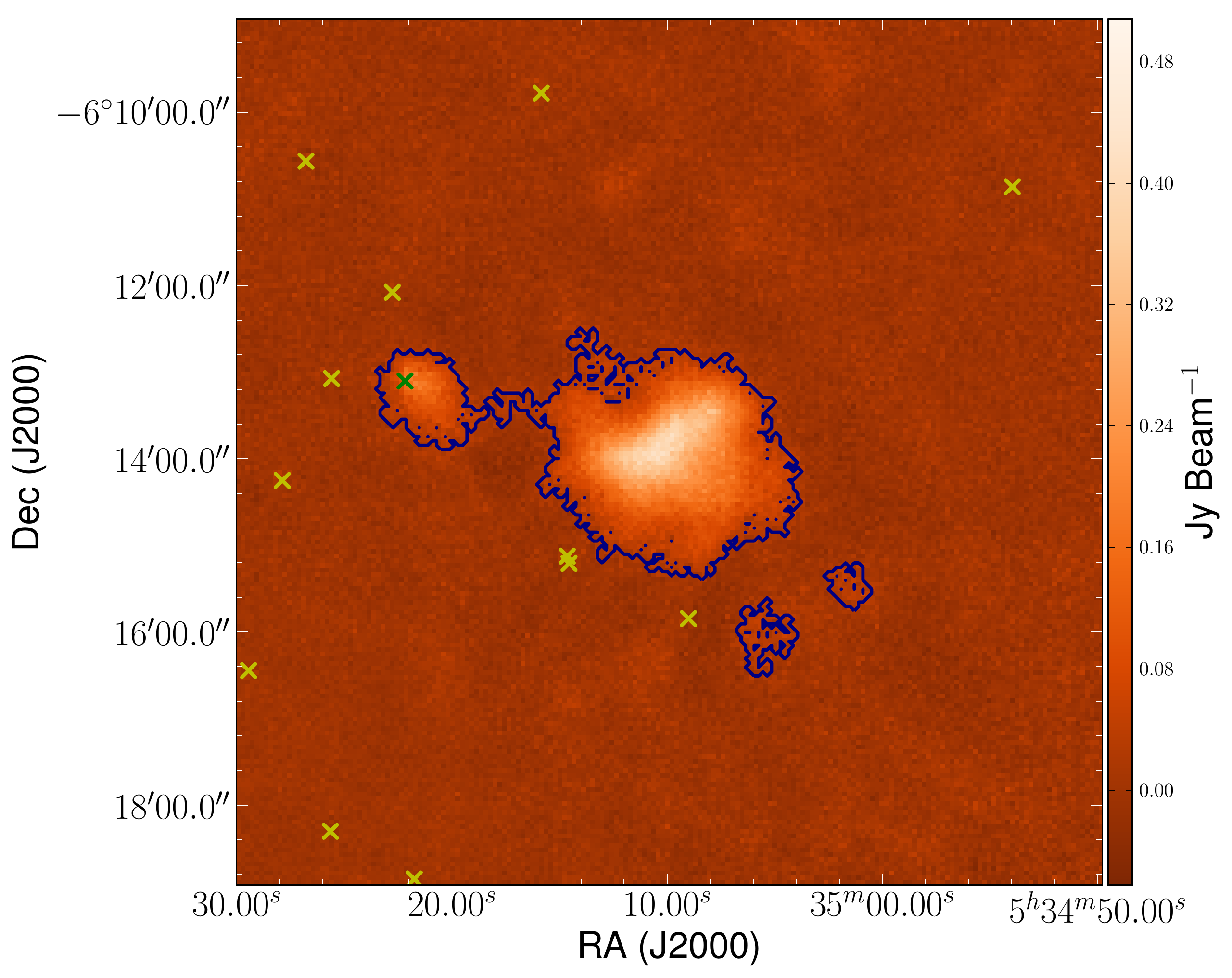}}
\subfloat{\label{island31}\includegraphics[width=9cm,height=7.5cm]{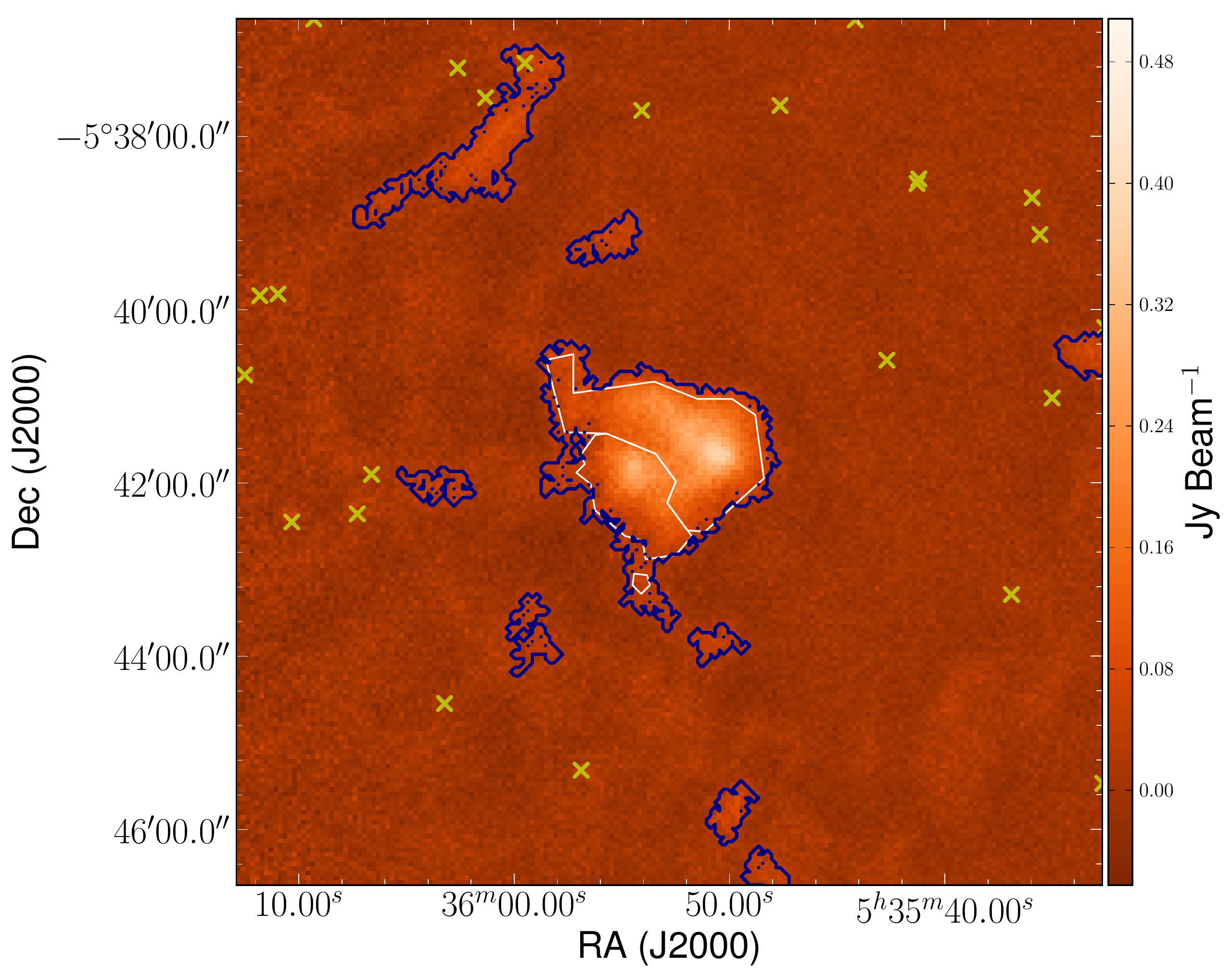}}
\caption{Islands which are calculated to be unstable to gravitational collapse yet harbour no evidence of associated YSOs of any class. The blue contours indicate the boundaries of the island and white contours indicate the boundaries of selected fragments. Note that we do not show the singular fragment in the main island in the left panel to emphasise that it is monolithic. Crosses denote YSOs colour coded as in previous figures and outlined in the text (protostars appear in green; disc sources, however, have been shown in yellow so that they are more visible). The colour scale has been chosen to accentuate the main islands of interest. \textit{Left}: A monolithic island with an $M/M_{J}$ ratio of $\sim$4. The secondary structure to the left of centre is its own island, separate from the main emission region. \textit{Right}: A complex island wherein the two main fragments have $M/M_{J}$ ratios of $\sim$2 and $\sim$3 from left to right, respectively.}
\label{starlesssuperJeans}
\end{figure*}

In Table \ref{interestingtable}, we present a list of starless islands in Southern Orion A which are good candidates for follow-up studies. Throughout this section, we highlight two islands which appear to be significantly gravitationally unstable, yet harbour no YSOs of any class (see Figure \ref{starlesssuperJeans}). For these two objects, there is no evidence from the existing \textit{Spitzer} and \textit{Herschel} catalogues that star formation is taking place. In the left panel of Figure \ref{starlesssuperJeans}, the central island (island index = 29) appears to be entirely monolithic with no sign of fragmentation, e.g., we calculate the $M/M_{J}$ ratio of this object to be $\sim$4 with a concentration of 0.67.  If there are no projection effects making this object appear larger and brighter than it truly is due to line-of-sight superposition, there are four scenarios which could explain its existence.

\begin{enumerate}[labelindent=0pt,labelwidth=\widthof{\ref{last-item}},label=\arabic*.,itemindent=1em,leftmargin=!]

\item There are indeed deeply embedded protostars which cannot be detected by \textit{Spitzer} because the optical depth is too high or the protostars are too faint. The Orion A Molecular Cloud has a lot of bright, diffuse infrared emission which can obscure faint protostellar sources (see \citealt{sadavoystarless} for a further discussion).  Note that the \cite{stutz2013} \textit{Herschel} catalogue does not  cover this particular island. 

\vspace{3mm}

\item The $M/M_{J}$ ratio is slightly overestimated because the gas in this region is hotter than 15 K. Even with a 5 K difference, however, the island would still have an $M/M_{J}$ ratio of $\sim$2. We also note that preliminary results from temperature maps derived by Rumble et al. (in prep) suggest that the temperature for this specific island is 14 K.  

\vspace{3mm}

\item The mass in the island has been assembled using a non-thermal support mechanism such as turbulent or magnetic pressure and it is out of thermal equilibrium.

\vspace{3mm}

\item The island is still very young and has not had the time to form protostars yet. This island may be a good follow-up location for a first hydrostatic core, an early stage of star formation which has long been theorised in the literature \citep{larson1969}.

\vspace{3mm}

\item This island is not associated with the Orion Molecular Cloud. If this object lies in the foreground of Orion by a significant distance, the Jeans mass ratio would be overestimated.

\end{enumerate}

In the right panel of Figure \ref{starlesssuperJeans}, we see a similar island (island index = 33) but in this case, the object is complex. In total, the island has an $M/M_{J}$ ratio of $\sim$3 so the observed multiple fragments are consistent with our expectation. Each of the two main fragments, however, are themselves Jeans unstable with $M/M_{J}$ ratios of 2 and 3, respectively (from left to right), and concentrations of 0.65 and 0.61, respectively. Thus, even if the island is the result of line of sight coincidence, each individual object is both super-Jeans and starless.


\subsection{A Toy Model for the Spatial Distribution of Young Stellar Objects}
\label{diskssec}

\vspace{3mm}

\begin{figure*}	
\centering
\includegraphics[width=15cm,height=15cm]{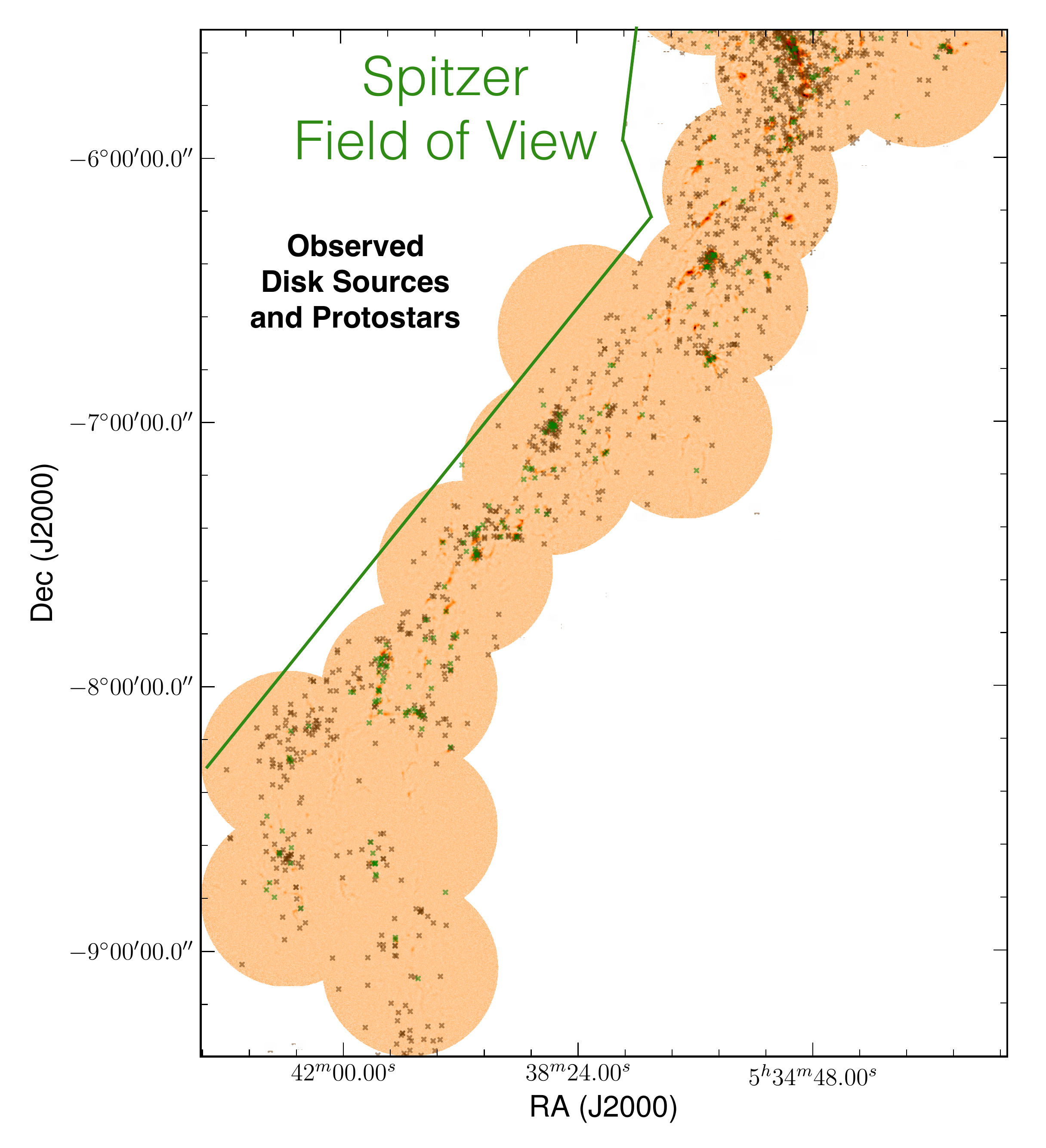}
\caption{The observed spatial distributions of discs (brown) and protostars (green) plotted over the map of Southern Orion A. The positions of these sources have been taken from the \protect\cite{megeath2012} and \protect\cite{stutz2013} catalogues.}
\label{disksdata}
\end{figure*}

In this section, we characterise the observed spatial distribution of disc sources and protostars from the \cite{megeath2012} \textit{Spitzer} catalogue with a toy model based on the locations of each YSO with respect to the fragments calculated to be Jeans unstable. Figure \ref{disksdata} shows the SCUBA-2 \mbox{850 $\mu$m} map with the locations of the discs and protostars overplotted. It is clear from the Figure that the surface densities of these sources can be separated into two populations and we label as ``clustered'' (away from the edges of the map and close to fragments) and ``distributed'' (the sporadic sources at larger distances from the clustered objects around fragments). Recently, \cite{megeath2016} studied the spatial distribution of YSOs in Orion A and found that the distributed population has a much lower fraction of protostars than the clustered population, suggesting that this is an older generation of YSOs. \cite{stutz2015} found evidence that the Orion A filament may be oscillating, so this distributed population of YSOs may have no association with the dense gas observed as the gas itself has moved away from this generation of forming stars, creating a ``slingshot'' mechanism. 

With a simple model, we attempt to recreate simultaneously both the clustered and distributed populations of YSOs using a few assumptions.

\begin{enumerate}[labelindent=0pt,labelwidth=\widthof{\ref{last-item}},label=\arabic*.,itemindent=1em,leftmargin=!]

\item The lifetimes of large-scale emission structures are much longer than those of individual discs and protostars such that the currently observed structures are linked to the formation of young stars and their present distribution.

\vspace{3mm}

\item All observed YSOs formed in fragments which are calculated to be Jeans unstable and every Jeans unstable fragment has the same probability of producing a YSO.

\vspace{3mm}

\item The half-life age of discs is estimated to be $t_{0.5}= 2 \mathrm{\:Myr}$ (\citealt{mamajek2009}; also see \citealt{ppvihalflife} for a discussion of disc dispersal)  and we detect no discs older than 10 Myr. We choose 10 Myr as a hard limit for two reasons. First, it is unlikely that a YSO older than 10 Myr would have enough surrounding material to achieve a suitable signal to noise ratio to be visible in our \mbox{850 $\mu$m} map (see \citealt{dunham2015} for a review of YSO lifetimes).  Second, a YSO moving at a reasonable velocity has a high probability of being ejected from the SCUBA-2 footprint of Southern Orion A within 10 Myr. 

\vspace{3mm}

\item We define protostars to have an age $\leq 0.5$ Myr \citep{dunham2015}.

\vspace{3mm}

\item Discs and protostars are ejected in a random (3D) direction from their parent fragment (see \citealt{stutz2015} for an alternative model). 

\vspace{3mm}

\item The space velocities of the observed YSOs follow a Maxwell-Boltzmann distribution with a fixed most probable speed, $v_{p}$ (see below).

\end{enumerate}

\begin{figure*}	
\centering
\includegraphics[width=16cm,height=13cm]{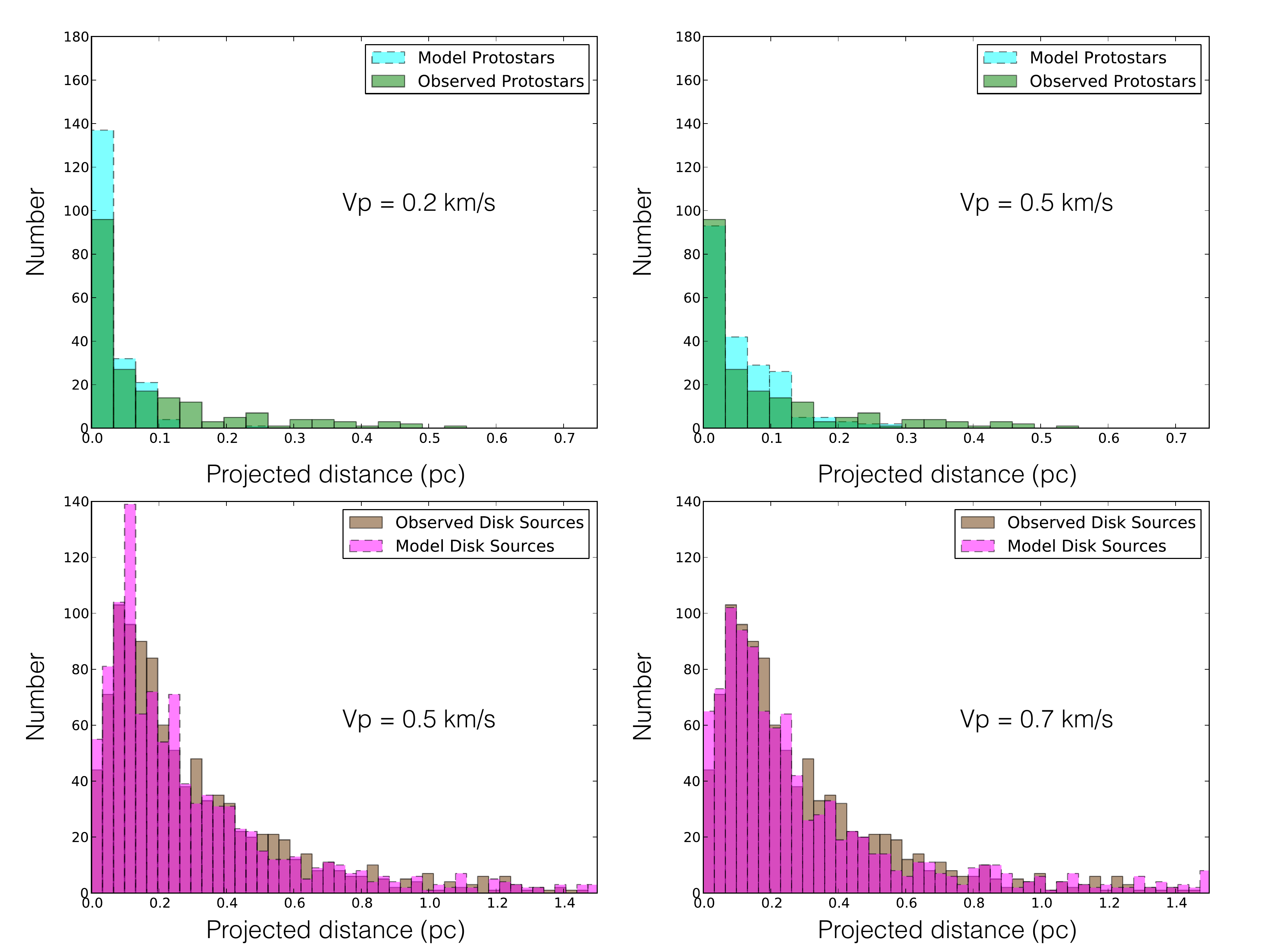}
\caption{\textit{Top Left:} The calculated projected distance between model protostar locations and the nearest fragment peak brightness location assuming $v_{p} = 0.2 \mathrm{\:km\:s}^{-1}$ in Equation \ref{maxboltzeq} (cyan, dashed lines) plotted along with the observed distribution (green, solid lines). We only include YSOs which lie on pixels within the SCUBA-2 footprint of Southern Orion A. \textit{Top Right:} Same as top left, but with a $v_{p}$ value of 0.5 km s$^{-1}$. \textit{Bottom Left:} The calculated projected distance between model disc source locations and the nearest fragment peak brightness location assuming $v_{p} = 0.5 \mathrm{\:km\:s}^{-1}$ in Equation \ref{maxboltzeq} (magenta, dashed lines) plotted along with the observed distribution (brown, solid lines). \textit{Bottom Right:} Same as bottom left, but with a $v_{p}$ value of 0.7 km s$^{-1}$.}
\label{projdistfig}
\end{figure*}

\begin{figure*}	
\centering
\includegraphics[width=15cm,height=15cm]{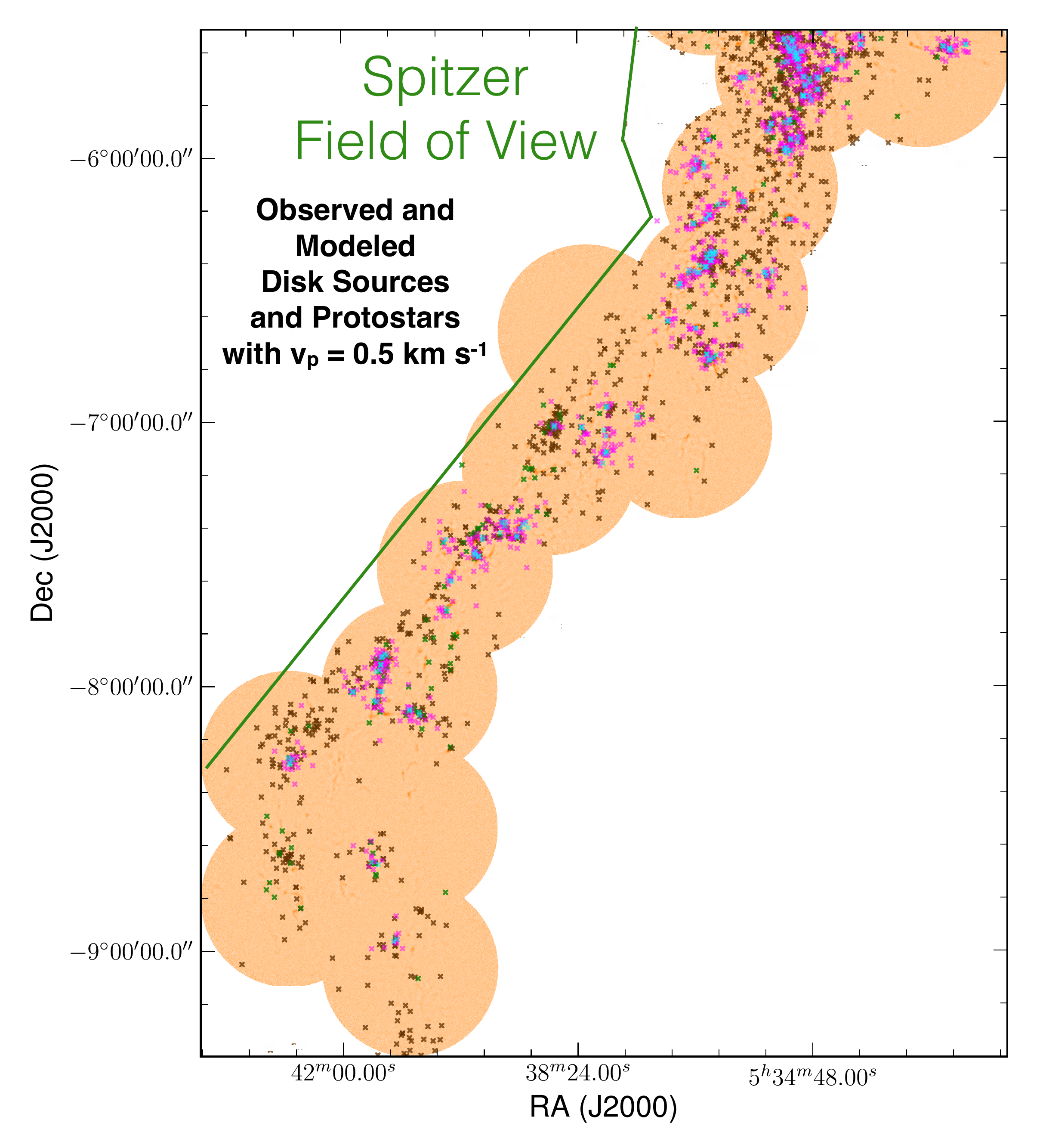}
\caption{The distributions of observed discs (brown), observed protostars (green), model discs (magenta), and model protostars (cyan) plotted over the \mbox{850 $\mu$m} map of Southern Orion A for a $v_{p}$ value of 0.5 km s$^{-1}$.}
\label{disksmodel}
\end{figure*}

For the combined number of discs and protostars present in the map, we first assign for each object a random age between 0 Myr and 10 Myr (assuming a uniform distribution). Then, we determine whether that age corresponds to a protostar or a disc source (see Assumption 4, above) and randomly determine the likelihood, $l$, that a YSO of that age is detectable based on the half-life age of a disc source ($l = 0.5^{\left[\frac{\mathrm{age}}{t_{0.5}}\right]}$). If the disc source is ``not detected'', we do not calculate a speed or direction for it, we simply start the code again until we detect the same total number of discs and protostars present in the observed map. The ratio of the numbers of protostars to discs derived through this sampling remains relatively constant and reflects the observed populations to within 7\%. 

Next, to determine the locations of the protostars and discs which are detected, we select a random speed, $v$, from a Maxwell-Boltzmann distribution with a specified most probable speed, $v_{p}$: 
\begin{equation}
\mathrm{PDF}_{Max-Boltz} = \left(\frac{1}{v_{p}\sqrt{\pi}}\right)^{3} 4\pi v^{2}e^{-\left(\frac{v}{v_{p}}\right)^{2}}. 
\label{maxboltzeq}
\end{equation}
Ten $v_{p}$ values were tested from 0.1 km s$^{-1}$ to 1.0 km s$^{-1}$. Finally, a random 3D direction is selected from an origin representing the location of the peak emission pixel within a selected Jeans unstable fragment, and we calculate the projected distance traveled during the lifetime of the YSO at the constant velocity drawn from Equation \ref{maxboltzeq}. 

The top panels of Figure \ref{projdistfig} show histograms of the new, ``detected'' protostar locations based on chosen $v_{p}$ values of 0.2 km s$^{-1}$ and 0.5 km s$^{-1}$, respectively, along with the observed distribution. It is important to note that fitting the model protostar projected distances to the observed data in the first few bins is more important than the extended tail. The observed protostars in the extended tail are unlikely to be true Class 0+I or flat-spectrum objects as they would need to have particularly high velocities or advanced ages to have travelled projected distances of more than \mbox{0.1 - 0.2 pc}. Recall from Section \ref{ysosec} that based on the brightness underlying the protostars in the \mbox{850 $\mu$m} continuum map, we expect up to 28\% of these objects to be misclassified. 
We find that 40 to 45 of the 209 total modelled protostars (20 to 22\%) lie beyond 0.1 pc in projected distance from their parent fragment (also see the bottom panel of Figure \ref{ysohists}). 

 As Figure \ref{projdistfig} (top left) shows, selecting 0.2 km s$^{-1}$ as the most probable speed somewhat overestimates the number of protostars that are very close to Jeans unstable fragments. Selecting \mbox{0.5 km s$^{-1}$} (Figure \ref{projdistfig}, top right), however, overestimates the number of protostars in the second, third, and fourth bins where we expect to find only a few ``real'' protostars. Thus, the $v_{p}$  value which best fits the observed projected distances between protostars and fragment peaks is between 0.2 km s$^{-1}$ and \mbox{0.5 km s$^{-1}$} assuming the same $v_{p}$ value applies to the entire Orion A filament. To test the accuracy of this toy model, each set of protostar projected distances produced 
 within this range reasonably fit the data. If the speed is decreased below 0.2 km s$^{-1}$, the model protostars are too clustered near their parent fragments compared to the observed data while an increased speed above \mbox{0.5 km s$^{-1}$} does not recreate this clustered population accurately. 

Note that \cite{jorgensen2007} observe the young, Class 0 protostellar population of the Perseus Molecular Cloud to have a velocity dispersion that is comparable to the sound speed $c_{s} \simeq 0.2 \mathrm{\:km\:s}^{-1}$ in this region. 
More recently, \cite{frimann2016} synthetically observed the distribution of Class 0 protostars within the MHD simulation {\sc{RAMSES}} and noted that the young, protostellar population has a 2D velocity dispersion of $\sim$0.15 km s$^{-1}$. In the same analysis, \cite{frimann2016} also note that the protostellar velocity distribution resembles a log-normal function in the simulations as opposed to our assumed Maxwell-Boltzmann distribution. A log-normal distribution will have a much higher fraction of high velocity sources that would travel further distances from their places of origin in the same amount of time. \cite{frimann2016} also assumed the age of a Class 0 protostar to be $\leq$0.1 Myr. Note, however, that we follow the protostellar definition from \cite{megeath2012} that also includes Class I and Flat-spectrum sources in addition to Class 0 objects. Thus, we expect our best fitting $v_{p}$ values to be somewhat higher in comparison (see below).   

In Figure \ref{disksmodel}, we plot the same observed objects as in Figure \ref{disksdata} but we now include the model protostars (cyan) and discs (magenta) produced assuming a $v_{p}$ value of \mbox{0.5 km s$^{-1}$}. In this figure, we see that the overall spatial distribution of model protostars is well matched to that of the observations. Note, however, that Figure \ref{disksmodel} shows isolated observed protostars (green crosses) that do not lie near the model protostar positions (cyan crosses). This difference results from the fact that we do not consider every fragment to be producing YSOs, only those which we calculate to be Jeans unstable. While this assumption holds true in many cases, clearly there are other fragments which we do not calculate to be Jeans unstable that are also associated with protostars. Our interpretation is that these objects are either more-evolved protostars (e.g. Class I/flat spectrum sources) and thus have had time to blow away much of their outer material (i.e. remnants of what were unstable islands/fragments) or are simply misclassified objects. 

The bottom left panel of Figure \ref{projdistfig} shows the projected distance between the model disc sources and fragment peaks, assuming $v_{p} = 0.5$ km s$^{-1}$. This histogram is too peaked relative to the observations. This difference can be seen more clearly in Figure \ref{disksmodel} where the more distributed population of observed disc sources is not well matched to the positions of the model discs. For the observed discs to have formed in the currently observed emission structure and then migrated to their present locations they either have to live longer (half-life > 2 Myr) or be moving at faster speeds than we are assuming. Possibly, these older objects have undergone a velocity evolution due to a more complicated gravitational interaction history than their younger counterparts. Thus, it is reasonable to assume that they may have a higher $v_{p}$ than the protostars. We suggest a $v_{p}$ value of 0.7 km s$^{-1}$ to represent better the observed, distributed disc sources (see the bottom right panel of Figure \ref{projdistfig}).

Note that there appears to be a trend in the velocity with YSO class. \cite{jorgensen2007}, \cite{frimann2016} and references therein find the velocity dispersion of Class 0 objects to be $\sim$0.1-0.2 km s$^{-1}$. In this analysis, we find that the population of Class 0, Class I, and flat spectrum sources (the protostars) can be fit reasonably with most probable velocities in the range of $\sim$0.2-0.5 km s$^{-1}$.
We find that the Class II objects (the disc sources) have a most probable velocity of $\sim$0.7 km s$^{-1}$. These velocities, however, are highly dependent on the lifetimes of each type of object.

Direct measurements of the velocity dispersion of young (\mbox{1 - 2 Myr}; Class II) stars were measured by \cite{foster2015} in the NGC 1333 star-forming region as part of the INfrared Spectra of Young Nebulous Clusters (IN-SYNC) project  \citep{cottaar2014}. In this region, \cite{foster2015} find that their sample of young stars have a velocity dispersion of $0.92 \pm 0.12\mathrm{\:km\:s^{-1}}$. This is significantly higher than \citealt{offner2009}'s predicted velocity dispersion based on a turbulent star-forming simulation as well as the velocity dispersion of dense cores in the region of $0.51\pm0.05\mathrm{\:km\:s^{-1}}$ as measured by \cite{kirk2007} using N$_{2}$H$^{+}$(1-0) observations. Evidently, the velocity dispersion of YSOs in NGC 1333 appears to increase quickly after their formation \citep{foster2015} which is consistent with our results in Southern Orion A.

\section{Conclusions}
\label{conclusionsec}
 
In this paper, we present the first-look analysis for the Southern Orion A (south of $\delta=-5\mathrm{:}31\mathrm{:}27.5$) SCUBA-2 continuum maps observed by the JCMT Gould Belt Survey, concentrating on the \mbox{850 $\mu$m} results. At a distance of \mbox{450 pc}, the Orion A Molecular Cloud is a nearby laboratory for examining active star-formation sites with the relatively less-studied southern extent offering a wealth of objects to aid in a better understanding of the dominant physical processes present in the region. We identify structures in the map using two-step procedure to find islands and fragments, the former based on a simple flux threshold and the latter defined using the algorithm {\sc{jsa\_catalogue}} (see Section \ref{cataloguessec}). We then examine the column-density map derived from 2MASS extinctions (Lombardi, M. private communication) for the whole of Southern Orion A, the islands, and the YSOs to characterise the large-scale context to which our SCUBA-2 map is not sensitive (see Section \ref{extinctionsec}). We show the mass distributions and comment on the concentration of fragments in terms of their Jeans stability (see Sections \ref{islandssec} and \ref{fragssec}). We then discuss the number density of the identified emission structures in terms of their Jeans radii and highlight two examples of starless, super-Jeans objects which merit a follow-up study with kinematic information (see Section \ref{fragsec}). 

Using the \cite{megeath2012} and \cite{stutz2013} \textit{Spitzer} and \textit{Herschel} YSO catalogues, we associate protostars, protostar candidates, and disc sources with the detected islands and fragments. We also discuss the YSO population itself by measuring the \mbox{850 $\mu$m} intensities at the locations of each object (as well as the column densities derived from the extinction map) and the distances between each object and its nearest localised SCUBA-2 emission peak. To extend this analysis further, we  examine the spatial distributions of disc sources and protostars in more detail by constructing a toy model of their locations based on simple assumptions and compare them with those of observations (see Section \ref{ysosec}). The \mbox{450 $\mu$m} and \mbox{850 $\mu$m} maps, their associated variance maps, and the island and fragment catalogues are all publicly available at: \url{https://doi.org/10.11570/16.0007}. 

\vspace{2mm}

Our main results are enumerated below.

\begin{enumerate}[labelindent=0pt,labelwidth=\widthof{\ref{last-item}},label=\arabic*.,itemindent=1em,leftmargin=!]

\item There are emission structures with a variety of sizes, flux levels, and morphologies present in Southern Orion A (see Figure \ref{bigmap}). As expected from local Jeans lengths, many large islands are often subdivided into multiple localised fragments (see Figures \ref{masshist} and \ref{stabilityfigs}). There are, however, several objects which require further study (see Table \ref{interestingtable} and Section \ref{starlesssec}).

\vspace{3mm}

\item Fragments are significant sites of star formation (see Figures \ref{concenstab} and \ref{stabilityfigs}). We find that those fragments that are Jeans unstable tend to have higher concentrations than those fragments that appear stable.

\vspace{3mm}

\item  The most Jeans unstable, monolithic structures show the most evidence for ongoing star formation due to their associations with protostars near the peak brightness location (see Figure \ref{stabilityhists}). This is in contrast to fragments extracted from complex islands (i.e., they have siblings in their parent cloud). Starting at an $M/M_{J}$ ratio of 1, these latter objects do not necessarily show more evidence of star formation at higher degrees of instability (see Figure \ref{stabilityhists}) implying clustered star formation may be more drawn out.

\vspace{3mm}

\item Class 0+I and flat-spectrum sources have higher associated \mbox{850 $\mu$m} brightness values and are closer to the nearest fragment's peak emission than their more-evolved disc counterparts. We find a similar result as \cite{heiderman2015} in that only $\sim$72\% of the objects defined as Class 0+I and flat-spectrum protostars are above a significant flux threshold, suggesting that some of the protostars identified in previous surveys may be misclassified.

\vspace{3mm}

\item The observed spatial distribution of disc sources across Southern Orion A has a ``clustered'' population and a ``distributed'' population. We can reproduce the projected distances between protostars and their nearest fragment reasonably well by using a simple toy model. 
Assuming a Maxwell-Boltzmann velocity distribution for these objects, we derive a range of most probable velocity values, $v_{p} = 0.2 - 0.5$ km s$^{-1}$, which reasonably fit the spatial distribution of protostars observed by \cite{megeath2012} and \cite{stutz2013}. The model disc source locations, however, do not recreate the distributed population in Southern Orion A using the same $v_{p}$ values. 
There appears to be a trend in velocity with respect to YSO classes. 
We find the Class II objects (the disc sources) require a $v_{p}$ value of 0.7 km s$^{-1}$ (see Figure \ref{projdistfig}). 

\end{enumerate}

\section*{Acknowledgements}

We thank our anonymous referee for their invaluable comments that have significantly
strengthened and added clarity to this paper. Steve Mairs was partially supported by the Natural Sciences and
Engineering Research Council (NSERC) of Canada graduate scholarship
program. Doug Johnstone is supported by the National Research Council
of Canada and by an NSERC Discovery Grant. 

\vspace{3mm}

The authors wish to recognise and acknowledge the very significant cultural role and
reverence that the summit of Maunakea has always had within the indigenous Hawaiian
community. We are most fortunate to have the opportunity to conduct observations from
this mountain. The James Clerk Maxwell Telescope has historically been operated by the Joint Astronomy Centre on behalf of the Science and Technology Facilities Council of the United Kingdom, the National Research Council of Canada and the Netherlands Organisation for Scientific Research. Additional funds for the construction of SCUBA-2 were provided by the Canada Foundation for Innovation.  The identification number for the programme under which the SCUBA-2
data used in this paper is MJLSG31. The authors thank the JCMT staff for their support
of the GBS team in data collection and reduction efforts. This research has made use of NASA's Astrophysics Data System and the facilities of the Canadian Astronomy Data Centre operated by the National Research Council of Canada with the support of the Canadian Space Agency. This research used the services
of the Canadian Advanced Network for Astronomy Research (CANFAR) which in turn is
supported by CANARIE, Compute Canada, University of Victoria, the National Research
Council of Canada, and the Canadian Space Agency. This research made use of {\sc{APLpy}}, an open-source plotting package for Python hosted at http://aplpy.github.com, and {\sc{matplotlib}}, a 2D plotting library for Python \citep{matplotlib}. This research used the TAPAS header archive of the IRAM-30m telescope, which was created in collaboration with the Instituto de Astrof\'isica de Andaluc\'ia - CSIC, partially supported by Spanish MICINN DGI grant AYA2005-07516-C02.

\bibliography{orionasouth_2}


$^{1}$Department of Physics and Astronomy, University of Victoria, Victoria, BC, V8P 1A1, Canada\\
$^{2}$NRC Herzberg Astronomy and Astrophysics, 5071 West Saanich Rd, Victoria, BC, V9E 2E7, Canada\\
$^{3}$Astrophysics Group, Cavendish Laboratory, J J Thomson Avenue, Cambridge, CB3 0HE, UK\\
$^{4}$Kavli Institute for Cosmology, Institute of Astronomy, University of Cambridge, Madingley Road, Cambridge, CB3 0HA, UK\\
$^{5}$Joint Astronomy Centre, 660 North A`oh\={o}k\={u} Place, University Park, Hilo, Hawaii 96720, USA\\
$^{6}$Department of Physics and Astronomy, University of Waterloo, Waterloo, Ontario, N2L 3G1, Canada \\
$^{7}$East Asian Observatory, 660 North A`oh\={o}k\={u} Place, University Park, Hilo, Hawaii 96720, USA\\
$^{8}$Physics and Astronomy, University of Exeter, Stocker Road, Exeter EX4 4QL, UK\\
$^{9}$Large Synoptic Survey Telescope Project Office, 933 N. Cherry Ave, Tucson, Arizona 85721, USA\\
$^{10}$Leiden Observatory, Leiden University, PO Box 9513, 2300 RA Leiden, The Netherlands\\
$^{11}$Max Planck Institute for Astronomy, K{\"o}nigstuhl 17, D-69117 Heidelberg, Germany\\
$^{12}$School of Physics and Astronomy, Cardiff University, The Parade, Cardiff, CF24 3AA, UK\\
$^{13}$Jeremiah Horrocks Institute, University of Central Lancashire, Preston, Lancashire, PR1 2HE, UK\\
$^{14}$European Southern Observatory (ESO), Garching, Germany\\
$^{15}$Jodrell Bank Centre for Astrophysics, Alan Turing Building, School of Physics and Astronomy, University of Manchester, Oxford Road, Manchester, M13 9PL, UK\\
$^{16}$Current address: Max Planck Institute for Extraterrestrial Physics, Giessenbachstrasse 1, 85748 Garching, Germany\\
$^{17}$Universit\'e de Montr\'eal, Centre de Recherche en Astrophysique du Qu\'ebec et d\'epartement de physique,
C.P. 6128, succ. centre-ville, Montr\'eal, QC, H3C 3J7, Canada\\
$^{18}$James Madison University, Harrisonburg, Virginia 22807, USA\\
$^{19}$School of Physics, Astronomy \& Mathematics, University of Hertfordshire, College Lane, Hatfield,
Herts, AL10 9AB, UK\\
$^{20}$Astrophysics Research Institute, Liverpool John Moores University, Egerton Warf, Birkenhead, CH41
1LD, UK\\
$^{21}$Imperial College London, Blackett Laboratory, Prince Consort Rd, London SW7 2BB, UK\\
$^{22}$Dept of Physics \& Astronomy, University of Manitoba, Winnipeg, Manitoba, R3T 2N2 Canada\\
$^{23}$Dunlap Institute for Astronomy \& Astrophysics, University of Toronto, 50 St. George St., Toronto ON
M5S 3H4 Canada\\
$^{24}$Physics \& Astronomy, University of St Andrews, North Haugh, St Andrews, Fife KY16 9SS, UK\\
$^{25}$Dept. of Physical Sciences, The Open University, Milton Keynes MK7 6AA, UK\\
$^{26}$The Rutherford Appleton Laboratory, Chilton, Didcot, OX11 0NL, UK.\\
$^{27}$UK Astronomy Technology Centre, Royal Observatory, Blackford Hill, Edinburgh EH9 3HJ, UK\\
$^{28}$Institute for Astronomy, Royal Observatory, University of Edinburgh, Blackford Hill, Edinburgh EH9
3HJ, UK\\
$^{29}$Centre de recherche en astrophysique du Qu\'ebec et D\'epartement de physique, de g\'enie physique et
d'optique, Universit\'e Laval, 1045 avenue de la m\'edecine, Qu\'ebec, G1V 0A6, Canada\\
$^{30}$Department of Physics and Astronomy, UCL, Gower St, London, WC1E 6BT, UK\\
$^{31}$Department of Physics and Astronomy, McMaster University, Hamilton, ON, L8S 4M1, Canada\\
$^{32}$Department of Physics, University of Alberta, Edmonton, AB T6G 2E1, Canada\\
$^{33}$University of Western Sydney, Locked Bag 1797, Penrith NSW 2751, Australia\\
$^{34}$National Astronomical Observatory of China, 20A Datun Road, Chaoyang District, Beijing 100012,
China


\newpage

\appendix

\counterwithin{figure}{section}

\renewcommand\thefigure{\thesection.\arabic{figure}}

\section{}

\begin{figure*}
\centering
\includegraphics[width=16cm,height=14cm]{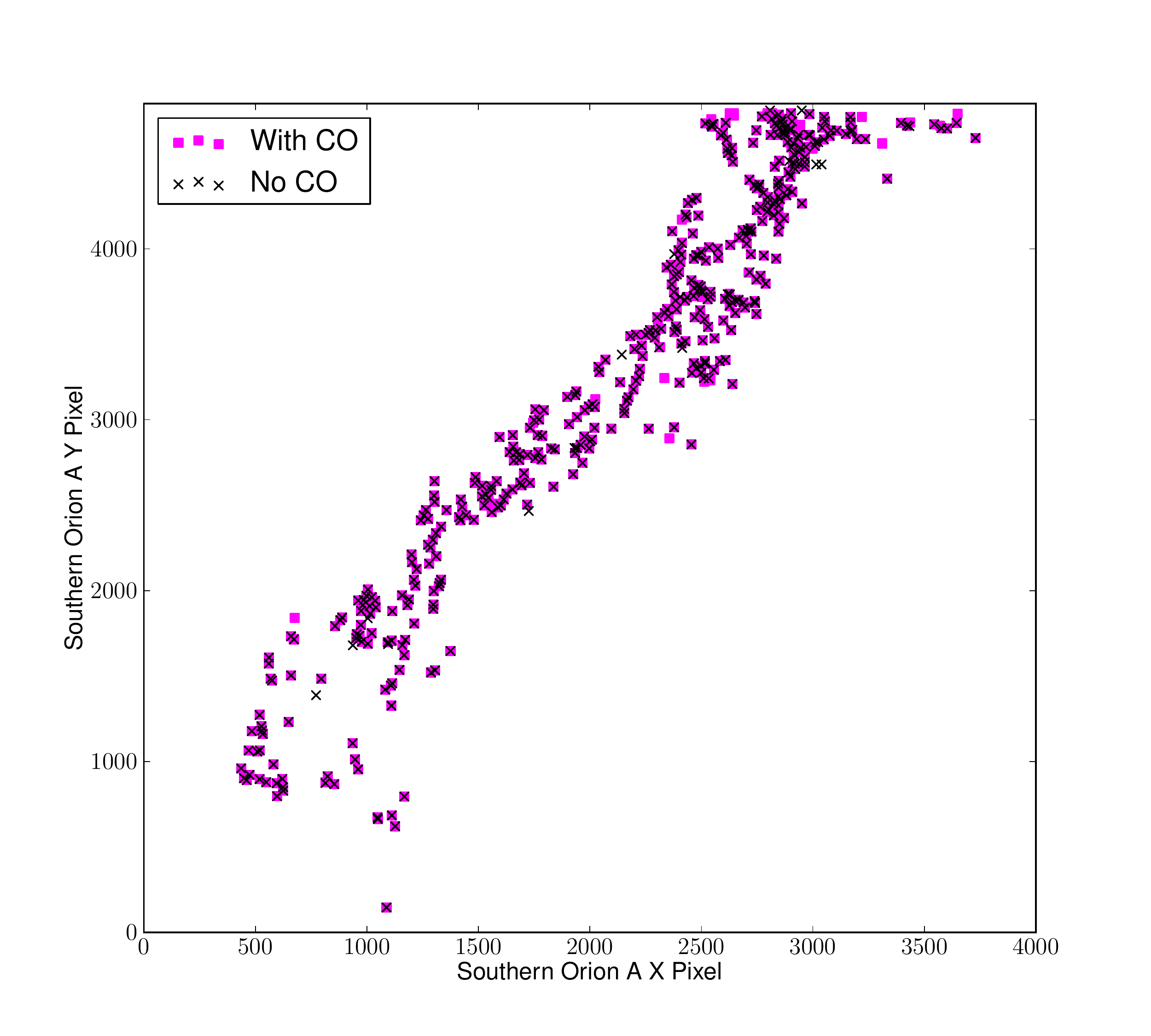}
\caption{The relative locations of detected  fragments detected using the {\sc{jsa\_catalogue}} algorithm in the Southern Orion A map where the \mbox{CO(J=3-2)} emission has been subtracted (magenta squares) and the map which includes the \mbox{CO(J=3-2)} emission (black crosses).}
\label{noco-withco-locations}
\end{figure*}

In this section, we perform a brief comparison between the Southern Orion A \mbox{850 $\mu$m} map which has had the \mbox{CO(J=3-2)} emission removed and the \mbox{850 $\mu$m} map where it has not been removed. Figure \ref{noco-withco-locations} shows the relative locations of the fragments detected in each map using the {\sc{jsa\_catalogue}} algorithm. The magenta squares represent the peak locations of fragments that were found in the map containing no \mbox{CO(J=3-2)} emission and the black crosses denote peak locations of fragments found in the map containing \mbox{CO(J=3-2)} emission. It is clear that the subtraction of this broad emission line has a minimal effect on the detected structure throughout the entire Southern Orion A region. The occasional ($\sim2\%$) fragments which have no counterpart are small areas of low-level emission that do not have any significant bearing on the final results. 
   
\begin{figure*}	
\centering
\includegraphics[width=12cm,height=10cm]{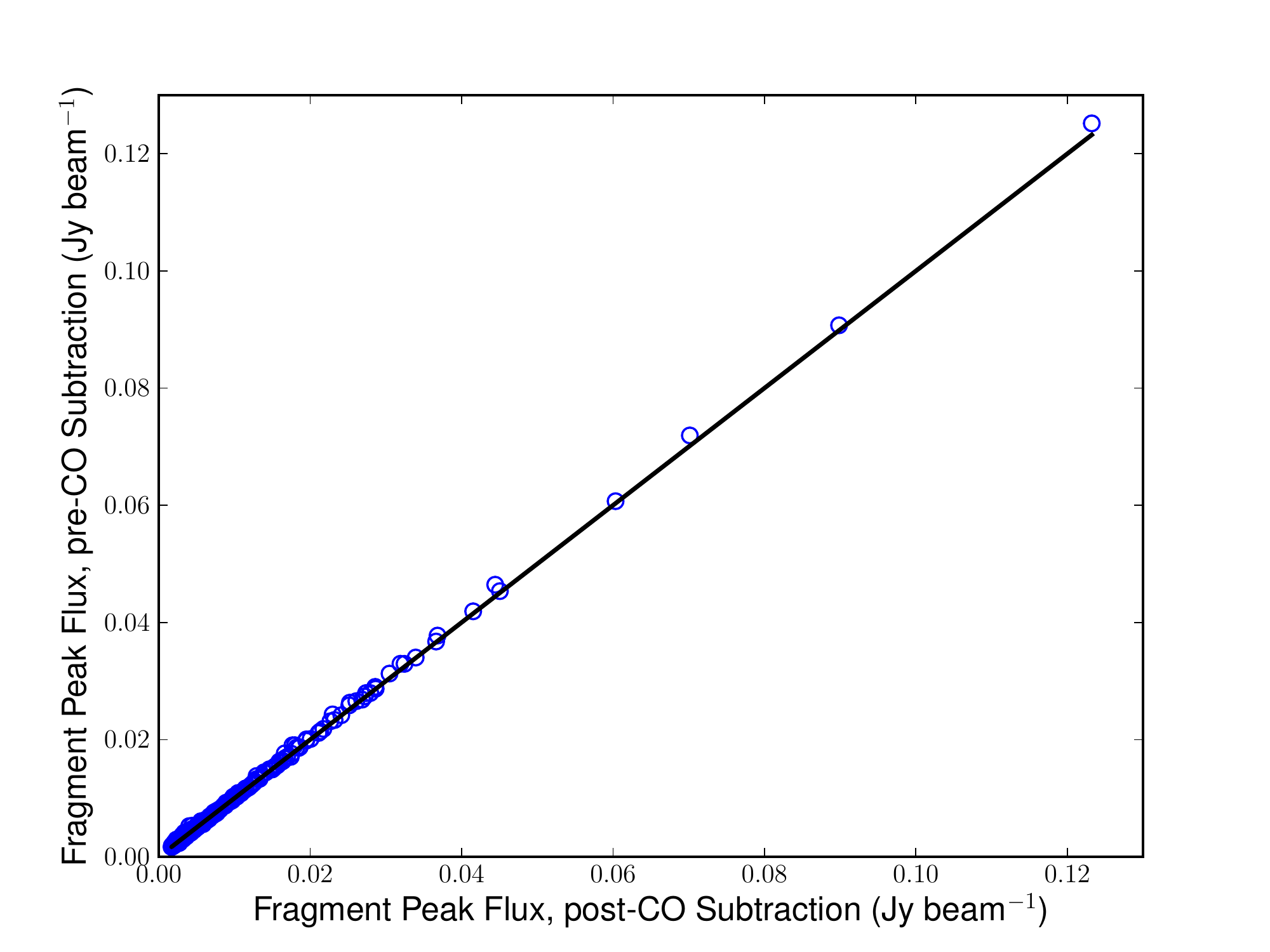}
\caption{The peak flux values of the fragments detected in each map (with and without the \mbox{CO(J=3-2)} emission). The solid, black line is a 1:1 ratio.}
\label{noco-withco-peakflux}
\end{figure*}

In Figure \ref{noco-withco-peakflux}, we plot the peak flux values of the fragments which coincide in each of the two maps. If a fragment in one map has a peak pixel location within one beam diameter (15$\arcsec$) of the peak pixel location of a fragment in the other map, it is included in the plot. 408 out of 431 fragments met this condition. The solid black line in the Figure shows a 1:1 ratio. Evidently, even the faintest peak brightness values are not significantly altered when the \mbox{CO(J=3-2)} emission line is subtracted from the \mbox{850 $\mu$m} continuum emission.

\begin{figure*}	
\centering
\includegraphics[width=12cm,height=10cm]{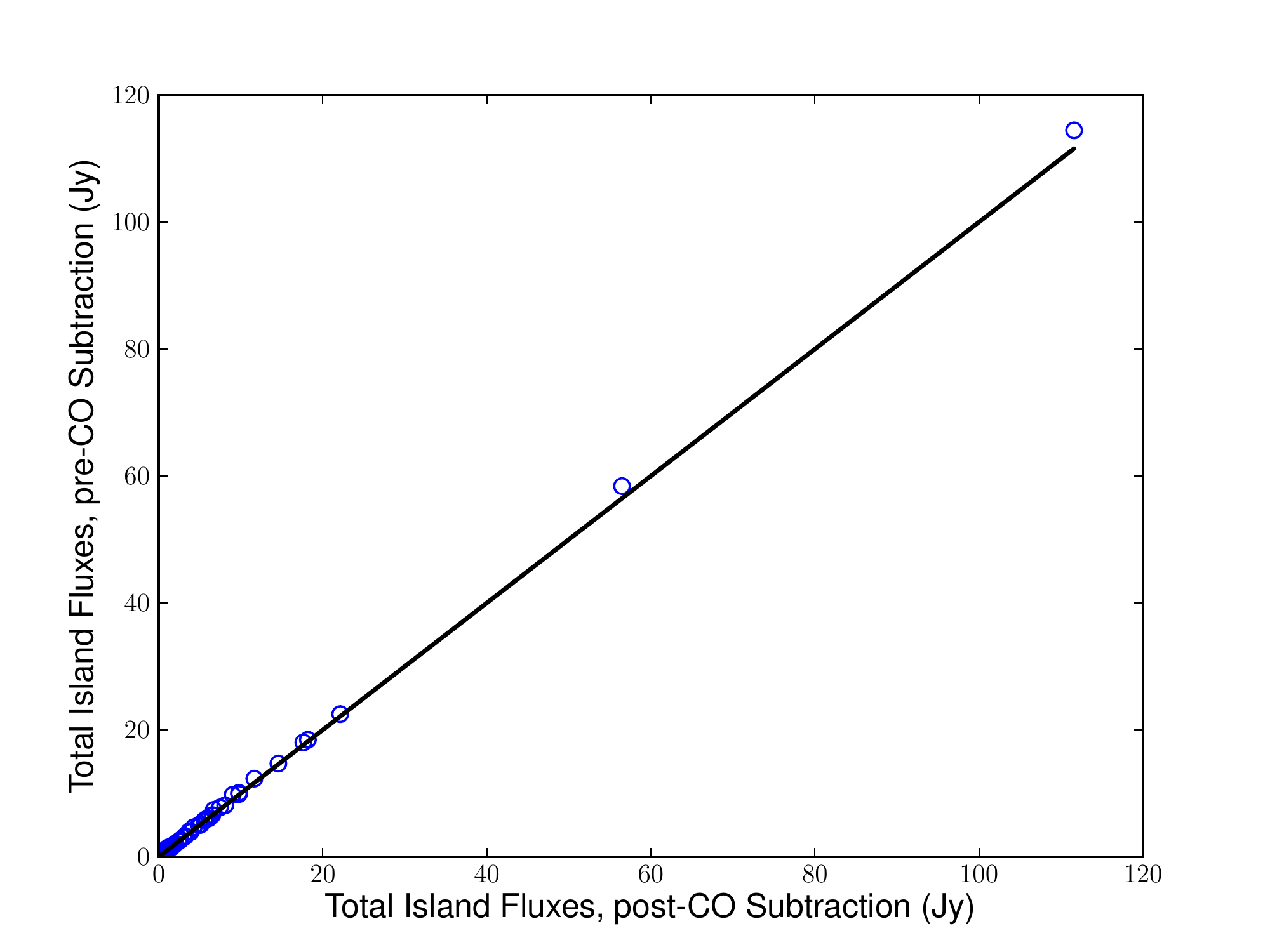}
\caption{The total flux values of the islands detected in each map (with and without the \mbox{CO(J=3-2)} emission). The solid, black line is a 1:1 ratio.}
\label{noco-withco-totalflux}
\end{figure*}

\begin{figure*}	
\centering
\includegraphics[width=17cm,height=14cm]{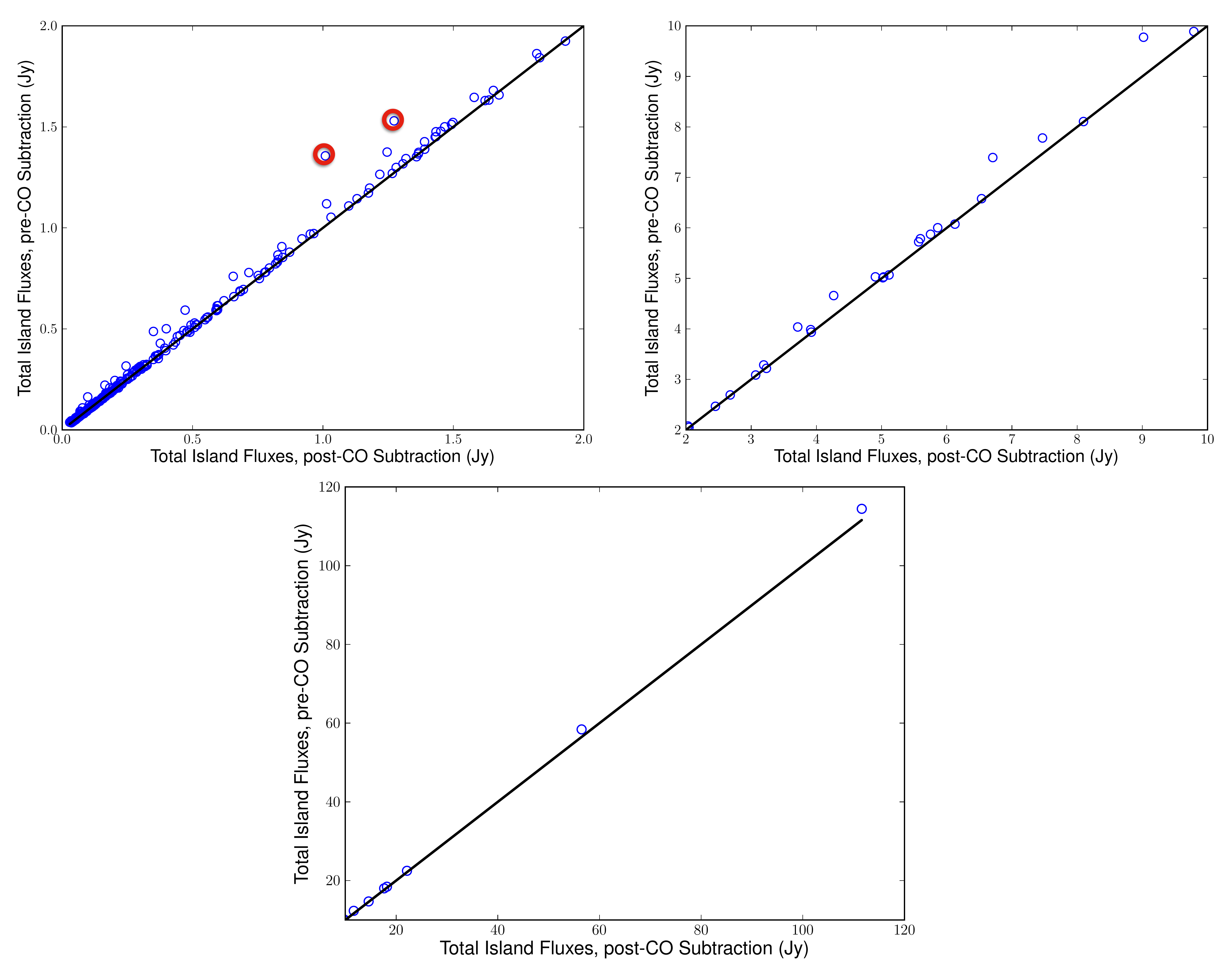}
\caption{Same as Figure \ref{noco-withco-totalflux}, but zoomed in to three sections for clarity. The solid, black line is a 1:1 ratio. \textit{Top left}: Low total flux. The two red circled islands are the sources which were most affected by the subtraction of the CO line emission.   \textit{Top Right}: Medium total flux. \textit{bottom}: High total flux.}
\label{noco-withco-totalflux-zoom}
\end{figure*}

To see how structure is affected on larger scales before and after the CO subtraction, we first identify islands in the \mbox{850 $\mu$m} map that includes emission from the \mbox{CO(J=3-2)} line and measure their total fluxes. We then compare these values to the total fluxes measured within the same boundaries using the map which has had the CO subtracted. Figure \ref{noco-withco-totalflux} shows the results for all these islands, and in Figure \ref{noco-withco-totalflux-zoom}, we split the results into three sections that highlight low total flux, medium total flux, and high total flux, zooming in for clarity.  

In general, we find that the CO-subtracted islands match well the emission from the non-subtracted islands, suggesting that the CO emission is a minor contribution to the total flux. We see in the medium and high total flux regimes, the islands follow a 1:1 relationship and they do not vary more than 10\%. This is approximately the error associated with flux calibration of the images. In the low total flux regime, however, we see more scatter. The lowest flux objects are not of any particular concern as they will have little bearing on the results and they are clustered quite close to the 1:1 line. We highlight the two sources which are the most affected by the CO subtraction using red circles (although they are both below the 3$\sigma$ level of the scatter). These two islands are found near the northern border of the map. They are both $\sim$0.2 pc in diameter assuming a circular projection, starless, and faint. Therefore, including or subtracting the CO emission from these small sources will not affect any of the main conclusions in this analysis. In the map that has undergone CO subtraction, one of these islands breaks up into two components (J053318-053421I and J053313-053506I)  while the other remains one single structure (J053556-053418I). None of these small features have been analysed in previous literature and they were not included in the SCUBA catalogue constructed by \cite{difrancesco2008}. 

In summary, it does not appear that \mbox{CO(J=3-2)} line contamination has any significant effect on the Southern Orion A \mbox{850 $\mu$m} continuum data. In the analysis performed in this paper, we used the CO subtracted SCUBA-2 maps. Both, however, are available online. 

\end{document}